% ...
\documentclass[1p]{elsarticle}

\usepackage{ulem}

\usepackage{graphicx,float,amsmath,amssymb,mathrsfs}

\usepackage[english,francais]{babel}

\usepackage{url,hyperref}
\usepackage[usenames,dvipsnames]{color}

\renewcommand{\em}{\it}

  \definecolor{blue}{rgb}{0,0,1}
  \definecolor{green}{rgb}{0,.6,0}
  \definecolor{red}{rgb}{1,0,0}
  \definecolor{vio}{rgb}{1,0,1}
  \definecolor{uv}{rgb}{0.5,0,0.5}
  \definecolor{ama}{rgb}{0.3,0.3,0.3}

\definecolor{M_Beige}         {rgb}{0.96 , 0.96 , 0.86}

\definecolor{M_Brown}         {rgb}{0.65 , 0.16 , 0.16}

\definecolor{M_Gold}          {rgb}{1.00 , 0.84 , 0.00}

\definecolor{M_LemonChiffon}  {rgb}{1.00 , 0.98 , 0.80}

\definecolor{M_Orange}        {rgb}{1.00 , 0.60 , 0.00}

\definecolor{M_Pink}          {rgb}{0.80 , 0.55 , 0.60}

\definecolor{M_Violet}          {rgb}{0.83 , 0.21 , 0.93}

\definecolor{M_Green}          {rgb}{0.2 , 0.6 , 0.2}

\definecolor{M_Gray}          {rgb}{0.7 , 0.7 , 0.7}

\definecolor{M_BluPal}          {rgb}{0.7 , 0.7 , 0.9}

\renewcommand{\leq}{\leqslant}
\renewcommand{\geq}{\geqslant}

\newcommand{\ket}[1]{|\kern.3ex#1\kern.3ex\rangle}
\newcommand{\bra}[1]{\langle\kern.3ex #1 \kern.3ex|}
\newcommand{\scalar}[2]{\langle\kern.3ex{#1}\kern.3ex|\kern.3ex{#2}\kern.3ex\rangle}
\newcommand{\mean}[1]{\left\langle #1\right\rangle}
\newcommand{\smean}[1]{\langle #1\rangle}
\newcommand{\EXP}[1]{\mathrm{e}^{#1}}         % exponentielle

\newcommand{\re}{\mathop{\mathrm{Re}}\nolimits}      % partie reelle
\def\I{\mathrm{i}}                  % le i mathematique
\def\D{\mathrm{d}}                  % la differenciation
\newcommand{\deriv}[2]{\frac{\mathrm{d}#1}{\mathrm{d}#2}}
\newcommand{\derivp}[2]{\frac{\partial #1}{\partial #2}}

\def\XXint#1#2#3{{\setbox0=\hbox{$#1{#2#3}{\int}$}
\vcenter{\hbox{$#2#3$}}\kern-.5\wd0}}

\def\IDoS{N(E)}
\def\rr{\zeta} % Rescaled Riccati
\def\potRic{\mathscr{U}}

\def\identity{\mathbf{1}}

\def\rate{r}

\def\rf{v_p}  % correlation length for the potential (r_f in previous papers)

\def\LamNoA{\widehat{\Lambda}}

\def\rmass{\mu} % rescaled mass m^2  \mu = (L_c/L_m)^2 \propto m^2

\def\reshe{\theta} % old q of Pierre
\def\Hem{H_{em}}  % old H_{e_m} of Pierre

\def\LatNb{M}
\def\N{\mathbb{N}} % ???
\newcommand\bigo[1]{\mathcal{O}(#1)}
\begin{document}

\selectlanguage{english}

%%%%%%%%%%%%%%%%%%%%%%%%%%%%%%%%%%%%%%%%%%%%%%%%%%%%%%%%%%%%%%%%%%%%%%%%%%%%%
\renewcommand{\labelitemi}{$\bullet$}
\renewcommand{\labelitemii}{$\star$}
%%%%%%%%%%%%%%%%%%%%%%%%%%%%%%%%%%%%%%%%%%%%%%%%%%%%%%%%%%%%%%%%%%%%%%%%%%%%%

\title{Exponential number of equilibria and depinning threshold for a directed polymer in a random potential}

\author{Yan V. Fyodorov} 
\ead{yan.fyodorov@kcl.ac.uk}
\address{King's College London, Department of Mathematics, London  WC2R 2LS, United Kingdom}

\author{Pierre Le Doussal}
\ead{ledou@lpt.ens.fr}
\address{CNRS-Laboratoire de Physique Th\'eorique de l'\'Ecole Normale Sup\'erieure, 24 rue Lhomond, 75231 Paris, France}

\author{Alberto Rosso} 
\ead{alberto.rosso@u-psud.fr}
\author{Christophe Texier}
\ead{christophe.texier@u-psud.fr}
\address{LPTMS, CNRS, Univ. Paris-Sud, Universit\'e Paris-Saclay, 91405 Orsay, France}

%\author[kcl]{Yan V. Fyodorov} 
%\ead{yan.fyodorov@kcl.ac.uk}
%\author[lptens]{Pierre Le Doussal}
%\ead{ledou@lpt.ens.fr}
%\author[lptms]{Alberto Rosso} 
%\ead{alberto.rosso@u-psud.fr}
%\author[lptms]{Christophe Texier}
%\ead{christophe.texier@u-psud.fr}
%
%\address[kcl]{King's College London, Department of Mathematics, London  WC2R 2LS, United Kingdom}
%\address[lptens]{CNRS-Laboratoire de Physique Th\'eorique de l'\'Ecole Normale Sup\'erieure, 24 rue Lhomond, 75231 Paris, France}
%\address[lptms]{LPTMS, CNRS, Univ. Paris-Sud, Universit\'e Paris-Saclay, 91405 Orsay, France}

%\date{\today}
%\date{January 30, 2018}

\date{August 13, 2018}

\begin{abstract}
By extending the Kac-Rice approach to manifolds of finite internal dimension, 
we show that the mean number $\left\langle\mathcal{N}_\mathrm{tot}\right\rangle$ of all possible equilibria (i.e. force-free configurations, a.k.a. equilibrium points) 
of an elastic line (directed polymer), confined in a harmonic well and submitted to a quenched random Gaussian potential in dimension $d=1+1$, grows exponentially $\left\langle\mathcal{N}_\mathrm{tot}\right\rangle\sim\exp{(r\,L)}$ with its length $L$. The growth rate $r$ is found to be directly related to the generalized Lyapunov exponent (GLE) which is a moment-generating function characterizing the large-deviation type fluctuations of the solution to the initial value problem associated with the random Schr\"odinger operator of the 1D Anderson localization problem. For strong confinement, the rate $r$ is small and given by a non-perturbative (instanton, Lifshitz tail-like) contribution to GLE. For weak confinement, the rate $r$ is found to be proportional to the inverse Larkin length of the pinning theory.
As an application, identifying the depinning with a landscape ``topology trivialization'' phenomenon, we obtain an upper bound for the depinning threshold $f_c$, in the presence of an applied force, for 
elastic lines and $d$-dimensional manifolds, expressed through the mean 
modulus of the spectral determinant
of the Laplace operators with a random potential. 
We also discuss the question of counting of stable equilibria.
Finally, we extend the method to calculate the asymptotic number of equilibria at
fixed energy (elastic, potential and total), and obtain 
the (annealed) distribution of the energy density over 
these equilibria (i.e. force-free configurations). 
Some connections with the Larkin model are also established.
\end{abstract}

%\pacs{05.60.Gg ; 03.65.Nk ; 05.45.Mt}

%\pacs{73.20.Fz}{Weak or Anderson localisation}

% 02.50.-r Probability theory, stochastic processes, and statistics
% 02.50.Cw 	Probability theory
% 02.50.Ey 	Stochastic processes
% 03.65.Nk    Scattering theory
% 05.10.Gg 	Stochastic analysis methods (Fokker-Planck, Langevin, etc.)
% 05.40.-a 	Fluctuation phenomena, random processes, noise, and
% 05.40.Jc Brownian motion
% 05.45.Mt    Quantum chaos ; semiclassical methods
% 05.60.Gg    Quantum transport
% 05.60.-k 	Transport processes
% 05.70.Np  Interface and surface thermodynamics
%72.   Electronic transport in condensed matter
% 72.10.-d   Theory of electronic transport; scattering mechanisms
% 72.10.Bg   General formulation of transport theory
% 72.15.Rn Localization effects (Anderson or weak localization)

%73.   Electronic structure and electrical properties of surfaces, interfaces,
%      thin films, and low-dimensional structures
%73.23.-b     Electronic transport in mesoscopic systems
%73.20.Fz Weak or Anderson localization

\begin{keyword} 
\PACS 05.40.-a \sep 75.10.Nr
\end{keyword} 

\maketitle

{\small
\setcounter{tocdepth}{2}
\tableofcontents
}

%%%%%%%%%%%%%%%%%%%%%%%%%%%%%%%%%%%%%%%%%%%%%%%%%%%%%%%%%%%%%%%%%%%%%%%%%%%%%%%%%%%%%%%%%%
%%%%%%%%%%%%%%%%%%%%%%%%%%%%%%%%%%%%%%%%%%%%%%%%%%%%%%%%%%%%%%%%%%%%%%%%%%%%%%%%%%%%%%%%%%

\section{Introduction}
\label{sec:Introduction}

Various aspects of the behaviour of a directed polymer, i.e. an elastic line, in a quenched random potential keep attracting permanent research efforts of both physicists and mathematicians for more than three decades. Among other applications,  it was at the center of attention as a model for vortex lines in superconductors, leading to important developments in the physics of pinning
(see Refs.~\cite{BlaFeiGesLarVin94,LeD11} for reviews). Its connection to the Kardar-Parisi-Zhang growth (see Ref.~\cite{HalZha95} for review of earlier works) led to a recent outburst of interest, and it was shown that the probability density of the free energy for a long polymer converges to the famous Tracy-Widom distribution \cite{CalLeDRos10,Dot10a,Dot10b,SasSpo10,AmiCorQua11}, extending the result for the ground state energy~\cite{Joh00}.

In this article we address a somewhat different aspect %of models of that type
and consider the problem of counting the total number of {\it equilibria} for a directed polymer (DP), in general
harmonically confined, immersed
in a random potential.
Those are defined as the stationary points (minima, maxima, or saddles) of an energy functional %for the DP
(see below). From a broader perspective, %understanding
describing the statistical structure of the stationary points of random landscapes and fields of various types is a rich problem of intrinsic current interest in various areas of pure and applied mathematics~\cite{AzaWsc09,Fyo15,AufBenCer13,AufBen13,Nic14,SubZei17,CamWig17,FyoKho16}.
It also keeps attracting steady interest in the theoretical physics community, and this over more than fifty years~\cite{Lon60,HalLax66,WeiHal82,Fre95,Fyo04,BraDea07,FyoWil07,FyoNad12}, with recent applications to statistical physics \cite{Par05,AnnCavGiaPar03,FyoLeD14,FyoWil07,FyoNad12}, neural networks and complex dynamics \cite{WaiTou13,FyoKho16,Fyo16}, string theory~\cite{DouShiZel04,DouShiZel06} and cosmology~\cite{EasGutMas16}.

Note, however, that all the previous works considered  only the case of zero internal dimension,
equivalent to dealing with a single-particle embedded in a random potential of arbitrary dimension.
 In such a  setting the counting of equilibria (a.k.a, stationary points or force-free configurations)
can be placed in a framework of the standard Random Matrix Theory, see Refs.~\cite{Fyo15,Fyo05}.
In contrast we aim here to address the counting problem of manifolds of finite internal dimension.
As was already anticipated in Ref.~\cite{Fyo05}, in the latter case the problem turns out to be
intimately related to properties of random Schr\"odinger operators appearing
in the problems of Anderson localization.  To the best of our knowledge, this aspect
of the counting problem  was never investigated before. 
Its treatment calls for a quite different
technique and requires understanding of less studied properties of random Schr\"odinger operators, such as
the modulus of its determinant and generalized Lyapunov exponents.
We develop the corresponding approaches, mainly for the 1D case,  in the present article.

%%%%%%%%%%%%%%%%%%%%%%%%%%%%%%%%%%%%%%%%%%%%%%%%%%%%%%%%%%%%%%%%%%%%%%%%%%%%%%%%%%%%%%%%%%
%%%%%%%%%%%%%%%%%%%%%%%%%%%%%%%%%%%%%%%%%%%%%%%%%%%%%%%%%%%%%%%%%%%%%%%%%%%%%%%%%%%%%%%%%%

\section{Model and main results}

\subsection{The continuous model}

\begin{figure}[!ht]
\centering
\includegraphics[width=6cm]{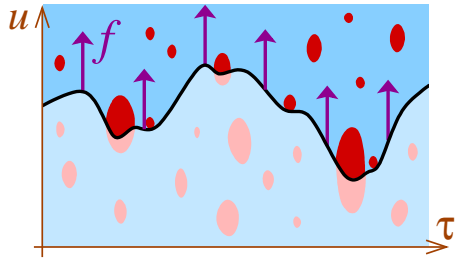}
\caption{\it Elastic line in a disordered medium submitted to a uniform force field. }
  \label{fig:ElasticLine}
\end{figure}

We consider the following energy functional
\begin{equation}
  \label{eq:H}
  \mathcal{H}[u(\tau)]
  =\int_0^L
   \D\tau
  \left[
      \frac{\kappa}{2}
      \left(
         \derivp{u(\tau)}{\tau}
       \right)^2
      + \frac{m^2}{2}u^2(\tau)
      +V(u(\tau),\tau)
  \right]
\end{equation}
where $u(\tau), \, \tau\in [0,L]$ describes the polymer configuration trajectory and $\kappa\geq 0$ is the elastic energy coefficient (cf. Fig.~\ref{fig:ElasticLine}).
Unless stated otherwise, {in the main text of the paper} we assume the fixed ends configuration $u(0)=u(L)=0$ {for simplicity, other types of boundary conditions are briefly discussed in the \ref{app:bc}.}
The random potential $V(u(\tau),\tau)$ is chosen to be Gaussian with zero mean and with a translationally-invariant covariance
\begin{equation}
  \label{eq:CorrV}
  \left\langle V(u,\tau)\,V(u',\tau')\right\rangle=\delta(\tau-\tau')\, R(u-u')
  \:,
\end{equation}
where we assume the symmetric function $R(u)$ to be at least four times differentiable at $u=0$.
{To have a better defined problem}, the polymer is considered to be confined inside a harmonic well
of curvature $m^2 \geq 0$, called the mass parameter,
which flattens the line beyond an infrared length, defined as 
\begin{equation}
  \label{eq:DefLm}
  L_m :=\sqrt{\kappa}/m
\end{equation}
The limit $m \to 0^+$ is of special interest, as the system becomes critical, with
a non trivial roughness exponent in the $L/L_m \gg 1$ limit \cite{BlaFeiGesLarVin94,LeD11}.

As is well known \cite{BlaFeiGesLarVin94,LeD11,LeDMulWie08,BalBouMez96}
the zero temperature problem is characterized by the Larkin length $L_c$, which gives the
order of magnitude of the scale above which metastability (i.e. multiple extrema) sets in.
Below that scale, there is typically a unique minimum and the system responds elastically (i.e. linearly) to perturbations,
such as an external force $f(\tau)$, i.e. adding $- \int_0^L \D\tau\, f(\tau)\, u(\tau)$ in (\ref{eq:H}).
In the absence of a mass, standard estimates at weak disorder \cite{BlaFeiGesLarVin94,LeD11} yield
\begin{equation}
  \label{LL}
  L_c := (\kappa^2/R''''(0))^{1/3}
  \:,
\end{equation}
taken here as a definition. It is also often expressed as
$L_c =  (\kappa^2 \rf ^2/|R''(0)|)^{1/3}$, where $\rf=(|R''(0)|/R''''(0))^{1/2}$
is a characteristic scale of variation of the random potential.
$L_c(m)$ %(or equivalently $m_c(L_c)$)
in the presence of a mass, was also estimated: in the simplest (Gaussian replica variational) approximation, it corresponds to the scale beyond which replica symmetry breaking (RSB) occurs \cite{MezPar91,LeDMulWie08}
while in the more elaborate Functional Renormalization Group (FRG) treatment \cite{Fis86,LeD09AnnPhys,LeD11}, it is associated
with a cusp (non-analyticity)
%to a cusp non-analyticity
$R'''_\mathrm{ren}(0^+)>0$ which develops in the renormalized force correlator,
signalling the appearance of shocks and avalanches
\cite{LeDWie09b,LeDMulWie08}. %Finally,
It also provides the so-called Larkin-Ovchinikov (LO) estimate
for the depinning threshold $f_c$
under the action of a uniform force $f(\tau)=f$, as $f_c \sim \kappa \rf /L_c^2=\rf\,R''''(0)^{2/3}\kappa^{-1/3}$
\cite{BlaFeiGesLarVin94,LeD11}, such that for
$f>f_c$ there are no more barriers to motion
\cite{Fis85,BlaFeiGesLarVin94,LeD11,RosKra02,KolBusFerRos13,DemLecRos14,DemRosPon14,LeDWie09a,FedLeDWie06,RosLeDWie09,LeDWieCha02}.
These estimates are either dimensional (balancing elastic and pinning energy), or valid in large embedding dimension (RSB) or within an (internal) dimensional expansion (FRG), and at present there are no exact
result for the number of minima, or even of equilibria, for a
model of a directed polymer in one dimension.

%%%%%%%%%%%%%%%%%%%%%%%%%%%%%%%%%%%%%%%%%%%%%%%%%%%%%%%%%%%%%%%%%%%%%%%%%%%%%%%%%%%%%%%%%%%%%%%%%%%%%%%%%%%%%

\subsection{The discrete model}

Although most calculations will be performed within the continuous model, {technically} it will be often more easy to start from a discrete version of the model, {passing to the continuous limit in the end of calculations. In this way we replace}
the continuous variable $\tau$  by a discrete {lattice index} $i=1,\ldots, K$ with $K=L/a$ (for simplicity we choose units such that the lattice spacing is $a=1$). The energy {of the polymer and the correlations of the random potential in such setting}  are given by
\begin{align}
  \label{eq:Hdiscrete}
  &\mathcal{H}( {\bf u} )
  =
  \frac{\kappa}{2}\sum_{i=0}^{K}\left(u_i-u_{i+1}\right)^2
  +
  \sum_{i=1}^K
    \left[\frac{m^2}{2} u_i^2+V_i(u_i) \right]
    \\
  \label{corr2}
  &\left\langle V_i(u)V_j(u')\right\rangle=\delta_{ij}\, R(u-u')
  \:.
\end{align}
Configurations of the polymer are described by vectors of transverse coordinates ${\bf u}=(u_1, \ldots,u_K)$, with $u_i\in \mathbb{R}$ for $i=1,\,2,\ldots,\,K$.
The fixed ends condition now reads $u_0=u_{K+1}=0$ (see~\ref{app:bc} for a discussion of other boundary conditions).
We will also consider the effect of an external force field, which requires to add a term $-{\bf f}^\mathrm{T}{\bf u}=\sum_{i=1}^Kf_iu_i$ to the energy~\eqref{eq:Hdiscrete} (the superscript $^\mathrm{T}$ denotes transposition).

%%%%%%%%%%%%%%%%%%%%%%%%%%%%%%%%%%%%%%%%%%%%%%%%%%%%%%%%%%%%%%%%%%%%%%%%%%%%%%%%%%%%%%%%%%%%%%%%%%%%%%%%%%%%%

\subsection{Main results}

{In the present article,} we provide some exact results for this problem. Namely, for model (\ref{eq:H}) we show that the mean
total number of equilibria grows exponentially
at large $L$, and that the rate $\rate$ is given in terms of the two length scales
$L_m$ and $L_c$ defined above as
\begin{equation}
  \label{res1}
  \mean{ \mathcal{N}_\mathrm{tot} }
   \sim \EXP{\rate L}
   \quad \mbox{with} \quad
   \rate = \frac{1}{L_m} \: g\!\left(\frac{L_m}{L_c}\right)
   \:,
\end{equation}
where $g(x)$ is a function calculated below.
Although this result is derived for the continuous model (\ref{eq:H}) and Gaussian disorder, we argue that the function $g(x)$ is universal for a broader class of models, in the limit of weak disorder $L_c \gg a$ and small mass $L_m \gg a$ (where $a$ is a UV cutoff, such as the lattice spacing).
We obtain the asymptotic behaviours
\begin{equation}
  \label{eq:MainResultG}
  g(x) \simeq
  \begin{cases}
    c_0\, x^{3} \exp\{ -8/(3x^3)\} & \mbox{for } x\ll1
    \\[0.125cm]
    C\,x                  & \mbox{for } x\gg1
  \end{cases}
\end{equation}
where the constant $c_0 = 1/(8 \pi)$ has been calculated analytically (see \ref{app:Saykin} written by David Saykin), and we also obtained numerically both  $c_0\simeq0.04$ and
\begin{equation}
  C \simeq 0.46 
%  C=0.461\pm0.001
  \:.
\end{equation}
Correspondingly, we get
\begin{equation}
  \label{eq:RateAtZeroMass}
  \rate = C/L_c
\end{equation}
for the rate in the zero mass limit {$m=0$ (i.e. $L_m=\infty$)}~:
the rate of growth of the number of equilibria increases with the disorder strength as
$\rate\sim R''''(0)^{1/3}$.
In the other limit of large confinement, the rate is exponentially suppressed
$\rate \sim \exp\big[-8m^3\sqrt{\kappa}/\big(3R''''(0)\big)\big]$, which shows that as long as the elastic line is shorter than the exponentially large scale, $L\lesssim(L_c^3/L_m^2)\,\exp\big[8m^3\sqrt{\kappa}/\big(3R''''(0)\big)\big]$, it is typically in a unique equilibrium configuration. This reflects exponentially
rare metastable states of a strongly confined polymer induced by the rare events in the 
Gaussian tail of the disorder.

We can ask the question about the universality of our result Eq.~\eqref{res1}. 
First it is immediate that (\ref{res1}) can be applied to the discrete model (\ref{eq:Hdiscrete}) in the limit $L_m ,\, L_c \gg a$ with parameters $\kappa$ and the same $R(u)$ function (in units such that $a=1$).
Our result \eqref{res1} is based on the asymptotic analysis of the mean number of equilibria, expressed in terms of a ratio of determinants, see Eqs.~\eqref{eq:DetRepres} and \eqref{eq:DetRepres2} below. Below, the analysis is performed in the continuum limit, which leads to evaluating some functional determinants with the Gelfand-Yaglom method.
In fact, although we will not pursue it here, one can extend the Gelfand-Yaglom method to calculate the ratio of discrete determinants \eqref{eq:DetRepres}, see~\ref{app:bc}. 
% Old sentence:
%In fact, although we will not pursue it here, one can extend the Gelfand-Yaglom method  (which we use here to obtain our results for the continuum model, see below) to calculate the ratio of discrete determinants \eqref{eq:DetRepres}, see~\ref{app:bc}.
It is a particular
model since it has uncorrelated, Gaussian disorder, and quadratic elastic energy. 
First it is clear from our derivation that the precise shape of the correlator function
$R(u)$ (within the Gaussian class) does not matter for fixed values of the second
and fourth derivative at zero. We expect furthermore that the scaling form
\eqref{res1}, up to two non-universal scales, $L_c$ and
$L_m$ (which, in some cases, can be independently measured),
extends to a broader class of elastic line models 
 (e.g. with short-range correlations in the $\tau$ direction, and with non Gaussian disorder).
A remaining question for future work is to which extent the function $g(x)$ is universal.

Furthermore, the method of calculation developed here is interesting in itself as it reveals connections to other stochastic problems such as, multifractality-like scaling of moments for the solution of the initial value problem associated with the 1D Schr\"odinger operator with a white noise random potential, reflecting anomalous fluctuations of finite-size Lyapunov exponents, thermally activated dynamics of a particle near depinning, and probably more.~\footnote{For example the problem of calculating large deviation function addressed here is 
closely related to recent work in chaos theory~\cite{HubPraPumWil18}, to be explored.}

\begin{figure}[!ht]
\centering
\includegraphics[width=5.5cm]{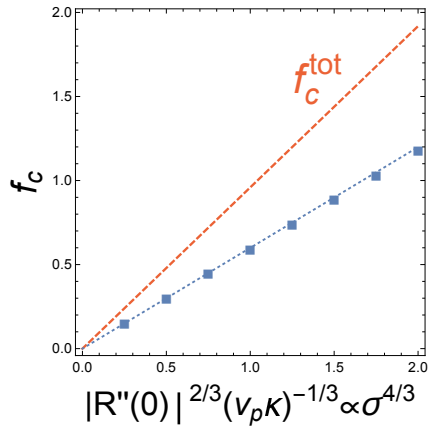}
\caption{\it Depinning threshold as a function of the disorder strength $|R''(0)|=\sigma^2$ for the case of a random periodic disorder~;
we compare numerics (squares) and the upper bound \eqref{fclast0} in the continuum limit (dashed line)~; numerical calculations are performed with averaging over $100$ realizations with chains of $K=512$ monomers.
For details, cf.~Section~\ref{sec:NumericsAlberto}.
}
  \label{fig:DataAlbe}
\end{figure}

As an application of our result we consider the depinning transition in the presence of a uniform applied force $f$.
Predicting the value of $f_c$ is a difficult problem in the theory of pinning.	
The value $f_c$ of the depinning threshold is identified \cite{RosKra02,KolBusFerRos13} as the force $f$ beyond which no metastable state survive.
Depinning can thus be put in the broader framework of the so-called {\it topology trivialization} phenomenon, which
has been studied recently in the context of spin glasses and other complex dynamical systems  \cite{FyoLeD14,WaiTou13,FyoKho16,Fyo16}.
 Here we calculate $\mean{ \mathcal{N}_\mathrm{tot} }$ and obtain the value, $f_c^\mathrm{tot}$ at which it
drops below unity. As we argue below, this provides an upper bound
\begin{equation}
  \label{fclast0}
  f_c \leq f_c^\mathrm{tot} =  \frac{\sqrt{2 C|R''(0)| R''''(0)^{1/3}}}{\kappa^{1/3} }
   =\sqrt{2C}\,\frac{|R''(0)|^{2/3}}{(\rf\kappa)^{1/3}}
  \:,
\end{equation}
with the constant $C$ given above. The formula for $f_c^\mathrm{tot}$ is compatible with the LO estimate, and we confirm from numerical simulations that \eqref{fclast0} is a strict inequality, as shown in Fig.~\ref{fig:DataAlbe}.
For the random periodic disordered introduced in the numerical simulations, although the bound seems far from the numerical data, one should keep in mind that $f_c^\mathrm{tot}$ provides an upper bound depending only on $R''(0)$ and $R''''(0)$ but not the complete shape of the function $R(u)$, in particular irrespectively of its range.

We discuss the question of counting of stable equilibria in  Section~\ref{subsec:stable}. 

In Section~\ref{sec:NbForFixedLevel}, we also show how to extend the method to calculate the asymptotic number of equilibria $\mean{\mathcal{N}_\mathrm{tot}(H)}\sim\exp\{\rate(m^2;H/L)\,L\}$ for which the line has a given energy $H$. 
This allows us to deduce the large deviation rate function $\Psi(h)=\rate(m^2)-\rate(m^2;h)$ controlling the (annealed) distribution of the energy density over the set of equilibria
\begin{equation}
  \frac{ \mean{\mathcal{N}_\mathrm{tot}(H)} }{ \mean{\mathcal{N}_\mathrm{tot}} }
  \sim\exp\{-L\,\Psi(H/L)\}
  \hspace{0.5cm}\mbox{for } L\to\infty
  \:.
\end{equation}
We obtain the precise limiting behaviours of the large deviation function~:
%We show that the large deviation function is quadratic for large negative energy, $\Psi(h)\sim h^2$ as $h\to-\infty$, and linear for large positive energy, $\Psi(h)\sim h$ as $h\to+\infty$. Precisely~:
\begin{equation}
  \Psi(h) 
  \simeq
  \begin{cases}
    \displaystyle
    \frac{h^2}{2R(0)}  
      & \mbox{for } h\to-\infty
    \\[0.25cm]
    \displaystyle
      \frac{1}{2\sigma_h^2}\left( h - h^* \right)^2
      & \mbox{for } h\sim h^*
    \\[0.25cm]
    \displaystyle
    \frac{m^2}{|R''(0)|}\,h 
      & \mbox{for } h\to+\infty
  \end{cases}
\end{equation} 
where $h^*$ is a typical value of the energy density described for various regimes below.
We also study the annealed distribution, over the set of equilibria, of the elastic and disorder energy densities, respectively. 
Moreover, for the elastic energy, we can go beyond the large deviation analysis and study the distribution for arbitrary length. % (assuming periodic boundary conditions).
For weak confinement ($L_m\gg L_c$), the elastic energy dominates in typical configurations of equilibria and the typical energy density is $h^*\simeq|R''(0)|/(4m\sqrt{\kappa})>0$.
We also show that the typical disorder energy is negative in general and, for weak confinement, takes the form 
$h_d^*  = - a_1\,|R''(0)|/(R''''(0)\kappa)^{1/3}$, with $a_1\simeq0.47$.
The variance of the annealed distribution is also dominated by fluctuations of the elastic energy $\mathrm{Var}_a(h)=\sigma_h^2/L\simeq[R''(0)^2/(8m^2\kappa)](L_m/L)\propto m^{-3}$.
For strong confinement ($L_m\ll L_c$), the typical energy density is dominated by the disorder energy, which is negative and twice the elastic energy density in absolute value, thus $h^*\simeq-|R''(0)|/(4m\sqrt{\kappa})<0$.
The variance is dominated by the disorder energy $\mathrm{Var}_a(h)=\sigma_h^2/L\simeq R(0)/L$.
Some connections with the simplified Larkin model (see definition in \ref{app:Larkin})
are made explicit: we show that 
(\textit{i})~the (annealed) roughness of the elastic line, averaged over the set of all equilibria
(\textit{ii})~the annealed distribution of the elastic energy, both coincide exactly 
with the predictions of the Larkin model.

Finally, we further extend the upper bound for the depinning threshold $f_c$, to elastic
manifolds of internal dimensionality $d$, relating it in Eq. \eqref{fcd} to the problem of evaluating the mean 
modulus of the spectral determinant of the Laplace operators with a random potential,
as arises in Anderson localization problems. 
%We also show how to extend the method to calculate the asymptotic number of equilibria at fixed energy in Section~\ref{sec:NbForFixedLevel} , and discuss the question of counting of stable equilibria in  Section \ref{subsec:stable}. 

\subsection{Outline}

In Section~\ref{sec:CountingEquilibria}, starting from the Kac-Rice formula, we provide a representation of the mean number of equilibria as the disorder-averaged ratio of determinants (determinants of matrices for the discrete model and functional determinants of differential operators for the continuous model). In Section~\ref{sec:Depinning}, we show that the rate $\rate$ controlling the number of equilibria can be related to the depinning threshold.
Section~\ref{sec:NumericsAlberto} briefly exposes the numerical simulations of the pinning of the elastic line.
In Section~\ref{sec:Configurations} we determine the roughness of the elastic line over the annealed
set of equilibria.
In Section~\ref{sec:GLE}, we establish the relation between the counting of equilibria of the elastic line and the Anderson localization problem for a 1D Schr\"odinger equation with a random white-noise potential.
Precisely, the rate $\rate$ is given in terms of the generalized Lyapunov exponent (GLE), 
which is a moment-generating function characterizing the large-deviation type fluctuations 
of the solution $y(L)$ to the initial value problem associated with the random Schr\"odinger operator, 
$\smean{|y(L)|^q}\sim\EXP{L\,\Lambda(q)}$.
Namely, the rate $\rate$ is directly related to GLE function $\Lambda(q)$ at a special value of the argument equal to unity, $q=1$.
This allows us to perform a first-principle analysis of the GLE and to identify that $\rate$ depends non-perturbatively on the disorder strength for small disorder.
Starting from the interpretation of the GLE as the eigenvalue of a spectral problem, we perform in Section~\ref{sec:GLEspectral} accurate numerical determination of the GLE.
In Section~\ref{sec:WKBE1}, by using an improved WKB method, we obtain the value for the GLE, i.e. for the rate $\rate$.
Section~\ref{sec:LambdaEvenQ} shows that a simple determination of the GLE for even integer argument is possible, however it does not lead to the non-perturbative contribution which fully controls the rate $\rate$.
This indicates that the attempts to calculate the mean number of equilibria by a naive replica-type
analytic continuation may be problematic, %are doomed to fail
reflecting that non-analyticity of the modulus of the determinant
 plays an important role in this problem.
Section~\ref{sec:NbForFixedLevel} presents the calculation of the large deviation rate functions for the elastic, disorder and total energy densities.

\subsection{Guideline}

This brief description of the content of the article leads us to propose two possible ways to read the article~:
\begin{itemize}
\item
  Elastic line in disordered medium~:
  Sections~\ref{sec:CountingEquilibria}, \ref{sec:Depinning}, \ref{sec:Configurations} and \ref{sec:NbForFixedLevel}. 

\item
  1D Schr\"odinger equation with a random potential and generalized Lyapunov exponent~:
  Sections~\ref{sec:GLE}, \ref{sec:GLEspectral}, \ref{sec:WKBE1} and \ref{sec:LambdaEvenQ}.
\end{itemize}

%%%%%%%%%%%%%%%%%%%%%%%%%%%%%%%%%%%%%%%%%%%%%%%%%%%%%%%%%%%%%%%%%%%%%%%%%%%%%%%%%%%%%%%%%%
%%%%%%%%%%%%%%%%%%%%%%%%%%%%%%%%%%%%%%%%%%%%%%%%%%%%%%%%%%%%%%%%%%%%%%%%%%%%%%%%%%%%%%%%%%

\section{Counting of equilibria}
\label{sec:CountingEquilibria}

As we shall see, the problem of  counting equilibria for the energy function describing
 the lattice version of our model can be treated in a well-established mathematical framework \footnote{Developing the corresponding formalism directly for the continuum model may represent an interesting mathematical problem.}.
%with a lattice version of the problem:
An equilibrium configuration is found as a solution of the system of $K$ stationarity conditions
which can be conveniently written as
\begin{equation}
  \label{equil}
  \partial_i\mathcal{H}( {\bf u} )
  = \left[(m^2 \identity_K-\kappa \Delta){\bf u}\right]_i+V_i'(u_i)=0
  \:, \quad i=1,\ldots, K
\end{equation}
where $\partial_i\equiv\partial/\partial u_i$ is the partial derivative, $\identity_K$ is the identity matrix of size $K$ and $\Delta$ is the {\it discrete Laplacian matrix} for the underlying one-dimensional lattice, with the only non-zero entries for our choice of boundary conditions being $\Delta_{i,i}=-2$ $, i=1,\ldots, K$ and $\Delta_{i,i-1}= \Delta_{i-1,i}=+1$ (see~\ref{app:bc}).
Note that in the continuum limit such a matrix approximates the standard one-dimensional Laplacian operator $\D ^2/\D \tau^2$ with Dirichlet boundary conditions.

The total number, $\mathcal{N}_\mathrm{tot}$, of solutions of such equations is then known to be given by the famous {\it Kac-Rice} formula, see Ref.~\cite{AzaWsc09} for a rigorous exposition, and Ref.~\cite{Fyo15} for a review. It can be written as follows
\begin{align}
  \label{Ntot1}
  \mathcal{N}_\mathrm{tot} &= \int_{\mathbb{R}^K}  \D{\bf u} \, \rho({\bf u})
  \\
  \label{Ntot2}
  \rho({\bf u}) &=  \left|\det{\left(\partial_i\partial_j \mathcal{H}\right)}\right|
  \:
   \prod_{i=1}^K \delta\left(\partial_i \mathcal{H}\right)
  \:,
\end{align}
where the Hessian is a $K \times K$ matrix given explicitly by
\begin{equation}
  \label{eq:Hessian}
  \partial_i\partial_j\mathcal{H}
  = \left[ m^2 + V''_i(u_i)\right] \delta_{i,j}
  - \kappa\,\Delta_{i,j}
  \:.
\end{equation}
The notation
(\ref{Ntot1}) allows to express more general quantities such as the total number of solutions
$ \mathcal{N}_{A}= \int_{A} \D{\bf u} \, \rho({\bf u})$ where ${\bf u}$ belongs to a subset $A$ of $\mathbb{R}^K$.

Eqs. (\ref{Ntot1},\ref{Ntot2}) are valid in every disorder realization such that all equilibria are
distinct (which happens almost surely). However for technical reasons we restrict our attention
to the simplest, yet informative characteristics,
the disorder average $\left\langle \mathcal{N}_\mathrm{tot}\right\rangle$.
To perform the disorder average we will use that (a) the potentials $V_i(u)$ and $V_j(u)$ are statistically independent for $i\ne j$ and (b) the variables $V_i'(u)$ are independent of $V_i''(u)$ for any $i$: indeed, from (\ref{corr2}), one has
\begin{equation}
  \label{eq:Rsymmetric}
  \mean{ V_i'(u)V_i''(u) } = R'''(0) =0
\end{equation}
since we have assumed differentiability ($R(u)$ being obviously an even function).
More generally, the property (b) is
an important consequence of translational invariance and the Gaussian character of the random function $V_i(u)$ \cite{AzaWsc09}. Moreover, after taking the average the mod-Hessian factor is obviously independent of ${\bf u}$, and
the average of each of the $K$ $\delta-$ factors can be done independently over the distribution of
the Gaussian variable  $V_i'(u)$ with the variance $\langle [V_i'(u)]^2\rangle=-R''(0)$, which gives
\begin{equation}
  \mean{ \mathcal{N}_\mathrm{tot} }
  =\mean{  |\det{\big(\partial_i\partial_j\mathcal{H}\big)}| } \, J(m^2)
\end{equation}
where
\begin{equation}
  \label{eq:IntegralI}
  J(m^2)
  =\int  \frac{\D {\bf u}}{(2\pi |R''(0)|)^{K/2}}
   \exp\left\{- \frac{\left[(m^2\, \identity_K -\kappa\,\Delta){\bf u}\right]^2}{2  |R''(0)|} \right\}
    %\left[\mu\, u_i+\kappa\,(2u_i-u_{i-1}-u_{i+1})\right]^2}
    \:.
\end{equation}
The Gaussian integral yields the constant Jacobian factor 
$J(m^2)=|\det(m^2\, \identity_K -\kappa\,\Delta)|^{-1}$ finally implying that
\begin{equation}
\label{eq:DetRepres}
  \mean{ \mathcal{N}_\mathrm{tot} }
  = \frac{
          \mean{
           \left|
              \det \left( (m^2/\kappa)\, \delta_{ij}-\Delta_{ij}+U_i\,\delta_{ij} \right)
           \right|
            }
          }
          {
            \left| \det \left( (m^2/\kappa)\, \delta_{ij}-\Delta_{ij}\right) \right|
          }
          \:,
\end{equation}
where the averaging goes over the set of independent and identically distributed (i.i.d.) mean-zero Gaussian random variables
\begin{equation}
  U_j \equiv V_j''(u_j)/\kappa
\end{equation}
with the covariance structure $\mean{ U_iU_j } = 2D\,\delta_{ij}$, where the parameter
\begin{equation}
  D = \frac{ R''''(0)}{2\kappa^2} = \frac{1}{2L_c^3}
  \:,
\end{equation}
measures the strength of the disorder in the problem, and is directly related to the Larkin length at $m=0$ as defined in Eq.~\eqref{LL}.

It is easy to see that Formula~\eqref{eq:DetRepres} obeys the exact bound $\mean{ \mathcal{N}_\mathrm{tot} } \geq 1$. 
Indeed, inside the average in the numerator one can use $|\det| \geq \det$, together with the
following property 
\begin{equation}
  \label{eq:MeanDetWithoutAbsoluteValue}
  \mean{\det \left( (m^2/\kappa)\, \delta_{ij}-\Delta_{ij}+U_i\,\delta_{ij} \right)} 
  = 
  \det \left( (m^2/\kappa)\, \delta_{ij}-\Delta_{ij} \right)
\end{equation}
which holds from the fact that the $U_i$ are centered i.i.d random variables, and
the determinant is a polynomial in the $U_i$ of degree at most one in each
variable. 
The disorder realizations with a single minimum $\mathcal{N}_\mathrm{tot}=1$
satisfy $|\det|=\det$. Note that replacing $|\det| \to \det$ in our calculations would amount to
neglect possibility of multiple minima and lead to the (simpler) Larkin model, described in \ref{app:Larkin},
which leads to a roughness exponent $\zeta=\zeta_L(d)=(4-d)/2 = \zeta_{\rm th}(d-2)$
where $\zeta_{\rm th}(d)=(2-d)/2$ is the thermal roughness of the disorder-free model.
This is an example of the dimensional reduction property which holds for
a broader class of models see \cite{EfeLar77,ParSou79,GiaLeD94,Fis85b,Fel00,Fel01} and short review in \cite{WieLeD07},
by which zero temperature observables in a disordered model -- upon the replacement
$|\det|=\det$ -- formally identify with those of a pure model at finite temperature in two dimension less.
\\

%Since this replacement 
%is invisible in perturbation theory in the disorder, all correlation functions 
%are given by the Larkin theory

The continuous version of the problem is related to the pair of Schr\"odinger operators
\begin{equation}
  \label{eq:DisorderedHamiltonian}
  H = -\deriv{^2}{\tau^2} + U(\tau)
  \hspace{0.25cm}
  \mbox{and}
  \hspace{0.25cm}
  H_\mathrm{free} = -\deriv{^2}{\tau^2}
  \:,
\end{equation}
where $U(\tau)$ is the Gaussian white-noise potential with mean zero and covariance
\begin{equation}
  \label{eq:WhiteNoisePot}
  \mean{ U(\tau)U(\tau') }=2D\,\delta(\tau-\tau')
  \:.
\end{equation}
Taking the appropriate continuous limit makes it rather apparent that  the mean number of equilibria should be given in this case by a similar mean modulus of the ratio of two functional determinants for the operators \eqref{eq:DisorderedHamiltonian} acting on functions vanishing at the two boundaries (Dirichlet boundary conditions) so that \eqref{eq:DetRepres} takes the form
\begin{equation}
\label{eq:DetRepres2}
  \mean{ \mathcal{N}_\mathrm{tot} }
  = \frac{
          \mean{
           \left|
              \det \left( m^2/\kappa -\partial_\tau^2 + U(\tau) \right)
           \right|
            }
          }
          {
            \left| \det \left( m^2/\kappa -\partial_\tau^2 \right) \right|
          }
          \:.
\end{equation}
The calculation of these determinants will be performed in Section~\ref{sec:GLE}.
In addition we will generalize the method to count equilibria at a given value of the
energy in Section~\ref{sec:NbForFixedLevel}.

%%%%%%%%%%%%%%%%%%%%%%%%%%%%%%%%%%%%%%%%%%%%%%%%%%%%%%%%%%%%%%%%%%%%%%%%%%%%%%%%%%%%%%%%%%
%%%%%%%%%%%%%%%%%%%%%%%%%%%%%%%%%%%%%%%%%%%%%%%%%%%%%%%%%%%%%%%%%%%%%%%%%%%%%%%%%%%%%%%%%%

\section{Counting of equilibria in the presence of a force -- Depinning}
\label{sec:Depinning}

Before evaluating the main objects of our interest, namely the ratios \eqref{eq:DetRepres} or \eqref{eq:DetRepres2}, which we postpone to Section~\ref{sec:GLE}, we discuss the effect of a uniform force field and show that the calculation of the rate $\rate$ is modified in a simple way.
This will allow us to establish the connection to depinning threshold.

%\subsection{Growth of the number of {\color{blue} total and stable} equlibria.}
\subsection{Growth of the total number of equilibria and of the number of stable equilibria}

%To connect to depinning,
Let us define, in a given disorder realization and with a uniform force $f$ applied to the polymer, the total number equilibria $\mathcal{N}_\mathrm{tot}(f)$ and the number of stable equilibria $\mathcal{N}_\mathrm{st}(f)$. A stable equilibrium is characterized
by a strictly positive Hessian matrix~; to connect to the notations used earlier we note that $\mathcal{N}_\mathrm{tot}(f=0)\equiv\mathcal{N}_\mathrm{tot}$.
We expect that before the depinning threshold is reached both $\mathcal{N}_\mathrm{st}(f)$ and $\mathcal{N}_\mathrm{tot}(f)$
grow exponentially with $L$, and define the rate functions as the following limits~\footnote{  The limits presumably also exist in a given sample, i.e. these quantities are likely to be self-averaging.} $\rho_\mathrm{st}(f) := \lim_{L \to \infty} (1/L) \mean{ \ln \mathcal{N}_\mathrm{st}(f) }$, and similarly $\rho_\mathrm{tot}(f) := \lim_{L \to \infty} (1/L) \mean{ \ln \mathcal{N}_\mathrm{tot}(f) }$. These two rates are expected to vanish at the \textit{same} value of the applied force $f$, which
defines
the depinning threshold $f_c$. This is because
\begin{itemize}
\item[(\textit{i})]
 the no-crossing rule (or Middleton theorems) \cite{Mid92,BaeMac98,Hir85}, known to hold for interface depinning, implies that in any given sample the last equilibrium which disappears  upon increasing $f$ is a stable equilibrium,
\item[(\textit{ii})]
 the sample-dependent threshold force at which this happens has fluctuations decaying to zero at large $L$ (see e.g. Ref.~\cite{KolBusFerRos13}).
\end{itemize}
In this article, instead, we consider the "annealed" rates
\begin{equation}
r_\mathrm{tot}(f) := \lim_{L \to \infty} \frac{1}{L}  \ln \mean{ \mathcal{N}_\mathrm{tot}(f) }
\hspace{0.25cm}\mbox{ and }\hspace{0.25cm}
r_\mathrm{st}(f) := \lim_{L \to \infty} \frac{1}{L}  \ln \mean{ \mathcal{N}_\mathrm{st}(f) }
\end{equation}
and have defined the corresponding thresholds $f_c^\mathrm{tot}$ and $f_c^\mathrm{st}$ as the forces for which these rates vanish (i.e. when $\mean{ \mathcal{N}_\mathrm{tot,st}(f) } \sim 1$ for $f=f_c^\mathrm{tot,st}$).
From $\mathcal{N}_\mathrm{st}\leq\mathcal{N}_\mathrm{tot}$ and the convexity of the logarithm we obtain
\begin{equation}
  f_c  \leq f_c^\mathrm{st} \leq f_c^\mathrm{tot}
  \:.
\end{equation}
In the next subsections, we determine
%The calculations of
the exact $f$-dependence of $\rate_\mathrm{tot}(f)$ and $\rate_\mathrm{st}(f)$.
% and of the exact value
Together with the determination of the rate $\rate_\mathrm{tot}\equiv\rate_\mathrm{tot}(0)$ in Section~\ref{sec:GLE}, this will provide the value for $f_c^\mathrm{tot}$ displayed in \eqref{fclast0}.
Obtaining the rates $\rho_\mathrm{tot,st}(f)$ and the threshold $f_c$ (and even $f_c^\mathrm{st}$ as we argue below) looks to us quite a challenging problem and goes beyond the scope of the present paper.

%%%%%%%%%%%%%%%%%%%%%%%%%%%%%%%%%%%%%%%%%%%%%%%%%%%%%%%%%%%%%%%%%%%%%%%%%%%%%%%%%%%%%%%%%%%%%%%%%%%%

\subsection{The rate $\rate_\mathrm{tot}(f)$ in the presence of an external force field}

\subsubsection{Toy model ($d=0$)}

For simplicity let us start with a toy model of a single particle, of energy
$\mathcal{H}(u) = m^2 \,{u^2}/{2} + V(u) - f u$ (i.e. our discrete model  \eqref{eq:Hdiscrete} with $K=1$ and $\kappa=0$), in the presence of an external force $f$. Let us calculate the number of
equilibria, denoted here $\mathcal{N}_\mathrm{tot}^w(f)$, weighted by a function $\phi(u)$ of width $w$:
\begin{align}
 \mathcal{N}_\mathrm{tot}^w(f)
 &:= \int \D u\, \rho(u) \, \phi(u)
 \\
 \rho(u)
 &= |m^2 + V''(u)|\: \delta(m^2 u - f + V'(u))
\end{align}
where $\rho(u)$ is the mean density of equilibria at point $u$.
For convenience, we will choose below the function $\phi(u)$ to be a Gaussian of width $w$. 
We will see that its introduction is necessary in order to connect the calculation of the number of equilibria to the threshold $f_c^\mathrm{tot}$.
From the text, its mean value is given by
\begin{equation}
   \mean{ \mathcal{N}_\mathrm{tot}^w(f) }
 = G(m^2)\,   J_{w}(m^2)
\end{equation}
where
\begin{align}
 G(m^2)
 := \left\langle |m^2 + V''(u)| \right\rangle
 = \sqrt{\frac{2 R''''(0)}{\pi}} \EXP{-m^4/(2R''''(0))}
 + m^2\, \mathrm{erf}\left(\frac{m^2}{\sqrt{2 R''''(0)}}\right)
\end{align}
where $\mathrm{erf}(x)$ is the error function~\cite{gragra}.
We have $G(0)=\sqrt{{2 R''''(0)}/{\pi}}$ and
\begin{align}
\label{Imupart}
  J_{w}(m^2)
  := \left\langle \delta(m^2 u - f + V'(u)) \right\rangle
   =  \int\frac{\D u\, \phi(u) }{\sqrt{2\pi |R''(0)|}} 
   \, \exp\left\{-\frac{1}{2 |R''(0)|}\left(m^2 u - f \right)^2\right\}
\end{align}
Now note that if we choose $\phi(u)=1$, i.e. $w\to\infty$, then $J_\phi(m^2)=1/m^2$ independently of $f$, and we
obtain the mean total number of equilibria as $\mean{ \mathcal{N}_\mathrm{tot}^\infty(f) } = G(m^2)/m^2$.
For $m^2>0$ it makes sense that it is independent of $f$, since the position of the center of the harmonic confining well
is just shifted by $f/m^2$. In the case most relevant for depinning, $m^2 \to 0^+$, this number diverges (as the particle can explore the
whole real axis) and we cannot extract any information about the depinning threshold.

To have a better defined problem in the limit $m^2 \to 0^+$ we will now count the mean number of equilibria in a
finite region of space $u$, of width $w$. In practice this is how a depinning force is defined, e.g. by restricting the particle to move on a cylinder of perimeter $w$. Here we could restrict $u \in [-w/2,w/2]$, but for calculational simplicity let us choose instead $\phi(u)=\EXP{-u^2/(2 w^2)}$.
%, a choice indicated by the subscript $w$.
This gives
\begin{align}
  \label{eq:JToyModel}
 J_{w}(m^2)
 = \int \frac{\D u}{\sqrt{2\pi |R''(0)|}}\, \EXP{ - u^2/(2 w^2) - \frac{1}{2 |R''(0)|}\left(m^2 u - f \right)^2}
 = \frac{\EXP{- f^2/[2(m^4 w^2 + |R''(0)|)]}}{\sqrt{ m^4 + |R''(0)|/  w^2}}
\end{align}
In the limit $m^2 \to 0^+$ we find
\begin{equation}
  \left\langle \mathcal{N}_\mathrm{tot}^w(f) \right\rangle = w \sqrt{\frac{2 R''''(0)}{\pi |R''(0)|}} \EXP{- f^2/(2|R''(0)|)}
\end{equation}
Hence we find that the mean number drops below unity when $f > f_c^\mathrm{tot}$ with
\begin{equation}
  f_c^\mathrm{tot} = \sqrt{2 |R''(0)| \left( \ln w + \frac{1}{2} \ln \frac{2 R''''(0)}{\pi |R''(0)|} \right) }
\end{equation}
This is the force such that the mean total number of equilibria drops below unity in the interval
of width $\sim w$, in the $u$-space. Note that the effect of the mass $m$ can be neglected
as long as $w \ll \sqrt{|R''(0)|}/m^2$, and that in $d=0$ the number of equilibria is at most
of order twice the number of metastable states.
The square-root logarithmic dependence in the width $w$ is thus not surprising for the model of a particle as it originates from rare large barriers in the tail of the Gaussian{-distributed potential}, and is compatible with
calculations of the depinning threshold in Ref.~\cite{LeDWie09a} on related models.
Due to this effect there is no true finite depinning threshold force for the particle
for $w \to +\infty$ (except for a bounded disorder).
We now turn to the elastic line, which does admit a well-defined depinning threshold force
in the thermodynamic limit, even for the Gaussian disorder.

\subsubsection{Elastic line ($d=1$) -- discrete model}
\label{subsubsec:discrete}

Let us now generalise the previous calculation to the DP model. Introducing the restriction
\begin{equation}
\phi({\bf u})= \EXP{- {\bf u}^2/(2 w^2)}
\end{equation}
in the integrals in Eqs.~\eqref{Ntot1} and \eqref{eq:IntegralI},
we need to calculate
\begin{align}
\label{j6}
  J_w(m^2)=\int \frac{\D{\bf u}}{\left(2\pi |R''(0)|\right)^{K/2}}
  \,
   \exp
  \left\{
  - \frac{{\bf u}^2}{2 w^2}
    - \frac{
       \left[(m^2 \identity_K-\kappa \Delta){\bf u}- {\bf f}\right]^2
           }{2 |R''(0)|}
           \right\}
\end{align}
where we have added a term $-\sum_{i=1}^K f_i\, u_i=- {\bf f}^\mathrm{T}{\bf u}$ to the energy (\ref{eq:Hdiscrete}).
This equation obviously generalizes \eqref{eq:JToyModel}.
Performing the integrals, this leads to a modification of (\ref{eq:DetRepres})
\begin{align} 
\label{n7}
 \mean{  \mathcal{N}_\mathrm{tot}^w(f) }
  =&
     \frac{
       \mean{ \left| \det \left(m^2\,\identity_K - \kappa\,\Delta +  \kappa\,U \right) \right| } }
          {\sqrt{  \det  \left((m^2\,\identity_K - \kappa\,\Delta)^2 + \frac{|R''(0)|}{w^2}\,\identity_K \right)  } }
         \\
&
\times
  \exp\left\{
    -
    \frac{1}{2w^2}
        {\bf f}^\mathrm{T}
        \left[ (m^{2}\,\identity_K - \kappa\,\Delta )^2 + \frac{|R''(0)|}{w^2}\,\identity_K \right]^{-1}
        {\bf f}
  \right\}
  \nonumber
\end{align}
where $U$ denotes the diagonal matrix with elements $U_i\,\delta_{ij}$.
Using
$
\det\big[( a\,\identity_K - \Delta )^2 + b^2\,\identity_K \big]
=
\det\big[ a\,\identity_K - \Delta  +\I\,b \,\identity_K ]
\det\big[ a\,\identity_K - \Delta  -\I\,b \,\identity_K ]
$,
the expression can be further simplified as
\begin{align}
\label{eq:NiceFormula}
 \mean{  \mathcal{N}_\mathrm{tot}^w(f) }
  =&
     \frac{ \mean{ \left| \det\left(L_m^{-2}\,\identity_K - \Delta + U \right) \right| } }
          {  \left| \det  \left((L_m^{-2}+\I\,L_w^{-2})\,\identity_K - \Delta\right) \right|  }
         \\
&
\times
  \exp\left\{
    -
    \frac{1}{2(w\kappa)^2}
        {\bf f}^\mathrm{T}
        \left[ (L_m^{-2}\,\identity_K - \Delta )^2 + L_w^{-4}\,\identity_K \right]^{-1}
        {\bf f}
  \right\}
  \nonumber
\end{align}
We have introduced the two length scales \eqref{eq:DefLm} and 
%$L_m$ and $L_w$ defined by
\begin{equation}
  \label{Lw}
  L_w 
  := \frac{\sqrt{w\kappa}}{|R''(0)|^{1/4}}
   = L_c \, \sqrt{\frac{w}{\rf L_c^{1/2}}} 
  \:.
\end{equation}
%up to boundary terms at $i=1$, $i=K$, unimportant in the large $K$ limit.
The formula \eqref{eq:NiceFormula} is valid for an arbitrary number of monomers $K$ and
general boundary conditions, i.e. for all three
types studied in~\ref{app:bc}. Note that for $w=+\infty$ the result
\eqref{n7} is independent of ${\bf f}$~: this is because one can shift the 
Gaussian integration measure on ${\bf u}$ in \eqref{j6} by ${\bf u}_0$, using that there exist (for $m>0$) a unique
solution ${\bf u}_0$ to the equation $(m^2 \identity_K-\kappa \Delta){\bf u}_0= {\bf f}$,
which represents the new equilibrium position in the absence of disorder, displaced by the force. 
\\

The criterion $\mean{ \mathcal{N}_\mathrm{tot}^w(f) } \sim 1$ then
allows to obtain a bound on the depinning threshold in the discrete setting (see below
for an application).

\subsubsection{Elastic line ($d=1$) -- continuous model}
\label{subsec:ElasticLineContinuous}

Taking the continuum limit is now straightforward.
The presence of an external force is accounted for by adding the term $- \int_0^L\D\tau\, f(\tau) u(\tau)$ to the energy functional \eqref{eq:H}.
To restrict to a finite region in $u$-space we similarly
introduce a functional
\begin{equation}
\phi[u] = \exp\left\{ -\frac{1}{2w^2} \int_0^L \D\tau\, u(\tau)^2\right\}
\:.
\end{equation}
Note that the dimension of $w^2$ is
now $[u]^2 [L]$.
The extension of \eqref{eq:NiceFormula} is
\begin{align}
\label{eq:NiceFormula2}
 \mean{  \mathcal{N}_\mathrm{tot}^w(f) }
  =&
     \frac{ \mean{ \left| \det\left(L_m^{-2} - \partial_\tau^2 + U(\tau) \right) \right| } }
          {  \left| \det  \left( L_m^{-2}+\I\,L_w^{-2} - \partial_\tau^2 \right) \right|  }
          \\\nonumber
  &\times\exp\bigg\{
    -    \frac{1}{2(w\kappa)^2}
      \int_0^L
      \D\tau\D\tau'\,
         f(\tau)
        \bra{\tau} \frac{1}{(L_m^{-2} - \partial_\tau^2 )^2 + L_w^{-4} } \ket{\tau'}
         f(\tau')
  \bigg\}
\end{align}
where we have used a quantum mechanical notation for the propagator.
For $f=0$ we recover \eqref{eq:DetRepres2}.
Expanding the force over the eigenmodes $\psi_n(\tau)$ of the Laplace operator (given in~\ref{app:bc} for several boundary conditions) with components
$\tilde{f}_n=\int_0^L\D\tau\,\psi_n^*(\tau)f(\tau)$ we rewrite more explicitly
\begin{align}
\label{eq:NiceFormula3}
 \mean{  \mathcal{N}_\mathrm{tot}^w(f) }
  =
     \frac{ \mean{ \left| \det\left(L_m^{-2} - \partial_\tau^2 + U(\tau) \right) \right| } }
          {  \left| \det  \left( L_m^{-2}+\I\,L_w^{-2} - \partial_\tau^2 \right) \right|  }
  \exp\bigg\{
    -    \frac{1}{2(w\kappa)^2}
     \sum_n \frac{|\tilde{f}_n|^2}{(L_m^{-2} + q_n^2 )^2 + L_w^{-4} }
  \bigg\}
\end{align}
where $- \partial_\tau^2\psi_n(\tau)=q_n^2\psi_n(\tau)$. % (see Appendix~\ref{app:bc}).
The determinant in the denominator is easily obtained for various boundary conditions, Dirichlet, Neumann or periodic (cf.~\ref{app:bc})~\cite{Tex10}~:
\begin{equation}
  \label{eq:FreeDetVariousBC}
  \det( \gamma - \partial_\tau^2  ) =
  \begin{cases}
    \frac{2}{\sqrt{\gamma}}\sinh\sqrt{\gamma}L
    & \mbox{ (Dir/Dir)}
    \\[0.1cm]
    2\cosh\sqrt{\gamma}L
    & \mbox{ (Dir/Neu)}
    \\[0.1cm]
    2\sqrt{\gamma}\sinh\sqrt{\gamma}L
    & \mbox{ (Neu/Neu)}
    \\[0.1cm]
    2(\cosh\sqrt{\gamma}L-1)
    & \mbox{ (periodic)}
  \end{cases}
\end{equation}
where the $\gamma$-independent prefactor is fixed by zeta-regularisation.
We stress that the leading behaviour of the determinant is independent of the boundary conditions in the large $L$ limit
\begin{equation}
  \det( \gamma - \partial_\tau^2  ) \sim\exp(\sqrt{\gamma}L)
\:.
\end{equation}

We consider a constant force $f(\tau)=f$ and the case of free (Neumann) or periodic boundary conditions (the case of fixed (Dirichlet) boundary conditions is discussed in~\ref{app:PinningDirichlet}).
The  double integral in the exponential of Eq.~\eqref{eq:NiceFormula2} is most easily computed for Neumann and periodic boundary conditions, as the Laplacian possesses a zero mode $\psi_0(\tau)=1/\sqrt{L}$ in this case.
Using \eqref{eq:NiceFormula3} with $\tilde{f}_n= \delta_{n,0}\,f\,\sqrt{L}$ shows that (minus) the argument of the exponential
in \eqref{eq:NiceFormula3} is
\begin{equation}
  \frac{|\tilde{f}_0|^2}{2(w\kappa)^2(L_m^{-4}  + L_w^{-4})}
 %  \frac{f^2L^2|\psi_0|^2}{2(w\kappa)^2(L_m^{-4}  + L_w^{-4})}
  =
  \frac{f^2L}{2( m^4w^2 + |R''(0)|)}
\end{equation}

The modulus of the determinant in the denominator of (\ref{eq:NiceFormula2},\ref{eq:NiceFormula3}) is
$$
\left| \det  \left( L_m^{-2}+\I\,L_w^{-2} - \partial_\tau^2 \right) \right|
\sim\exp\big[ L\,\re\sqrt{{1}/{L_m^2}+{\I}/{L_w^2}}\big]
$$
when $L \gg L_m, L_w$.
Finally we deduce
\begin{align}
  \label{eq:Form4.19}
  \rate_{w,m}(f):=\lim_{L\to\infty} \frac{\ln \mean{  \mathcal{N}_\mathrm{tot}^w(f) }}{L}
  = \Lambda(1) - \re\sqrt{\frac{1}{L_m^2}+\frac{\I}{L_w^2}}
    - \frac{f^2}{2(m^4w^2 + |R''(0)|)}
\end{align}
where $\Lambda(1)=\lim_{L\to\infty}(1/L)\ln\mean{ \left| \det\left(L_m^{-2} - \partial_\tau^2 + U(\tau) \right) \right| }$ will be determined in Sections~\ref{sec:GLE} and~\ref{sec:GLEspectral}.
In this section we only need to known that taking $L_m\to\infty$ yields the value $\Lambda(1)=C/L_c$, Eq.~\eqref{eq:RateAtZeroMass}, where $C$ is a dimensionless number of order unity and $L_c$ the Larkin length.

The form \eqref{eq:Form4.19} is now appropriate to consider first the limit $m^2\to0$ at fixed $w$, and second
the limit $w\to\infty$ (which clearly do not commute). The first limit leads to
$\rate_w(f):=\lim_{m^2\to0} \rate_{w,m}(f)$, with
\begin{equation}
\label{rwf}
  \rate_w(f)= \frac{C}{L_c} - \frac{1}{\sqrt{2} L_w} - \frac{f^2}{2|R''(0)|}
\end{equation}
which is valid for $L_m,\, L \gg L_w,\, L_c$. 
Taking now the limit $w \to \infty$, i.e. $L_w \gg L_c$, we obtain $\rate_\mathrm{tot}(f):=\lim_{w\to\infty} \rate_w(f)$, with
\begin{equation}
\label{rr}
  \rate_\mathrm{tot}(f) = \frac{C}{L_c} - \frac{f^2}{2|R''(0)|}
  \:.
\end{equation}
This is one of the central results of the paper.
We define the threshold $f_c^\mathrm{tot}=\sqrt{2r|R''(0)|}$ as the value of $f$ for which this rate vanishes (i.e. $\mean{  \mathcal{N}_\mathrm{tot} }\sim1$). We get
$f_c^\mathrm{tot} = \sqrt{{2 C} |R''(0)|/L_c }$, i.e.
%\color{black}
\begin{equation}
  \label{fclast}
  f_c^\mathrm{tot} %= \sqrt{\frac{{2 C} |R''(0)|}{L_c} }
  = \sqrt{2C}\frac{\rf\kappa}{L_c^2}
  =  \frac{\sqrt{2 C|R''(0)| R''''(0)^{1/3}}}{\kappa^{1/3} }
%  =\sqrt{2 C}\frac{ |R''(0)|^{2/3}}{ (\rf\kappa)^{1/3} }
\end{equation}
as displayed in the text. It is interesting to note that it
has  a similar order of magnitude as the Larkin-Ovchinnikov (LO)
formula
\cite{BlaFeiGesLarVin94,LeD11}
\begin{equation}
  f_c^\mathrm{LO} = \frac{ |R''(0)|^{2/3}}{ (\rf\kappa)^{1/3} }
\end{equation}
with $R''''(0) = |R''(0)|/\rf^2$, which, however is only an order of magnitude estimate.
As discussed in the text, our result (\ref{fclast}) is
an exact upper bound for the true $f_c$
in the continuous model, or the discrete one in the limit $L_c \gg a$.

\paragraph{Concluding remarks ---}

Let us conclude by some remarks on the validity of the present calculation.
Our starting formula \eqref{eq:NiceFormula3} for the mean number of equilibria
$\left\langle \mathcal{N}_\mathrm{tot}^w(f) \right\rangle$ is exact for arbitrary $m,w,L$. Our strategy to obtain a
robust value for $f_c^\mathrm{tot}$ (independent of details of the procedure) was to consider $m \to 0$ first,
then $L \to +\infty$ and only at the end $w \to +\infty$,
so that the condition $m^4 w^2 \ll |R''(0)|$ is always satisfied.
With that procedure we found that the criterion
$\left\langle \mathcal{N}_\mathrm{tot}^w(f) \right\rangle \sim 1$
identifies $f_c^\mathrm{tot}$ unambiguously, independently of any further details of the boundary conditions.
Furthermore it is independent of the type and range of correlations of the disorder, whether random field, random bond, or random periodic, i.e. independent of the precise shape of the function $R(u)$.
Hence it is an upper bound on $f_c$ in {\it any} type of disorder.

It is useful to recall that a similar robustness of the depinning threshold
$f_c$ with respect to boundary conditions was observed in Ref.~\cite{KolBusFerRos13}
(and previous works cited there). There $f_c$ was studied for an elastic line
on a cylinder of width 
$W = \alpha \, \rf (L/L_c)^\zeta$, 
where $\zeta$ is the roughness exponent
at depinning (and $m=0$) 
and $\alpha$ a dimensionless constant. 
The latter is measured from the roughness of the
last metastable configuration encountered as $f$ is increased towards $f_c$.
The value of $f_c$ was found independent of the aspect ratio $\alpha$ of the cylinder when both $L$ and $W$ become large. 
Only finite size corrections, which are subdominant, depend on the aspect ratio and other details.
These subdominant sample to sample fluctuations of the depinning threshold force
were also studied in Ref.~\cite{FedLeDWie06,LeDWie09a}.

Let us now comment on the respective order of limits of
large $L$ and large $w$. In the present calculation, a finite value of $w$ means that any equilibrium
configuration which extends beyond a width $W_L = w/\sqrt{L}$ is not counted in $\mathcal{N}_\mathrm{tot}^w(f)$.
If we want the upper bound argument to hold, we only need that this width be larger
(or of the same order) than the typical width at depinning, i.e. $W_L \gg \rf (L/L_c)^\zeta$, equivalent to
$w \gg \rf L_c^{1/2} (L/L_c)^{\zeta + 1/2}$. On the other hand, our intermediate result \eqref{rwf} for the
finite $w$ rate, $r_w(f)$, requires $L/L_w \gg 1$. From the definition
\eqref{Lw} of $L_w$, that is equivalent to $w \ll \rf L_c^{1/2} (L/L_c)^2$.
Hence whenever $\zeta < 3/2$, which is the case for most classes of
disorder, both conditions can be met simultaneously: one can safely use formula \eqref{rwf}
and the term $1/L_w$ is then negligible in the thermodynamic limit $W \sim L^\zeta \to \infty$,
leading to \eqref{rr} and \eqref{fclast}. 
For the case of periodic depinning studied numerically
in the main text, $\zeta=3/2$ (the roughness of the Larkin model, see~\ref{app:Larkin}).
In that case $W \sim L^\zeta$ is equivalent to $L \sim L_w$ and one cannot
use the second term in the asymptotic rate formula \eqref{rwf}. Instead one
must replace it by $- \frac{1}{L} \ln \left| \det  \left(L_w^{-2} - \partial_\tau^2 \right) \right|$.
An estimate of this quantity however, shows that it is again negligible in the
thermodynamic limit.

%Finally one can ask whether our finite $L$ expressions could
%give some insight on the finite $L$ corrections to $f_c$ and
%its sample to sample distribution. It may prove subtle however.
%Indeed for a cylinder of width $u = k L^\zeta$, these
%finite size corrections are known to scale \cite{KolBusFerRos13} as $L^{-2+\zeta}$
%where $\zeta \simeq 1.25$ is the non-trivial roughness
%exponent at the depinning transition, whose
%calculation requires more powerful methods, such
%as the Functional RG \cite{LeDWieCha02}. A similar
%behavior was also observed \cite{RosLeDWie09} for the finite mass corrections
%$f_c(m)$. Finally, these corrections also depend on
%the class, random periodic versus non-periodic disorder
%correlator $R(u)$. The question a better understanding
%of sample to sample fluctuations of $\mathcal{N}_w$, neglected here,
%may also become of importance.

\subsubsection{Interface model on arbitrary graph and dimension $d$}

As mentioned earlier the present work establishes a connection between the pinning and localization theories.
In the 1D continuum elastic line model it is
best illustrated by rewriting the threshold force $f_c^\mathrm{tot}$ \eqref{fclast} as
\begin{equation}
  \frac{(f^\mathrm{tot}_c)^2}{2 |R''(0)|}
  =
  \lim_{L\to\infty}
  \frac{\ln\mean{ \left| \det\left(- \partial_\tau^2 + U(\tau) \right) \right| }}{L}
\end{equation}

From \eqref{eq:NiceFormula} and following the same steps,
a similar formula can be written for the discrete model
\eqref{eq:Hdiscrete}. It can in fact be generalized further, to an interface $u_i$, $i \in \mathbb{Z}^d$
of internal dimension $d$, with an arbitrary elastic matrix $- \Delta_{ij} \to K_{ij}$, for instance
$\widehat{K}_q \sim |q|^a$ in Fourier representation, with $a=2$ ($a<2$) for short-range (long-range)
elasticity - and an on-site Gaussian random potential correlated as in \eqref{eq:Hdiscrete}.
The same method leads to a general formula for the threshold force $f_c^\mathrm{tot}$, {\it which is
again an upper bound for the depinning threshold force}, $f_c \leq f_c^\mathrm{tot}$ with
\begin{equation} 
  \label{fcd} 
  \frac{(f_c^\mathrm{tot})^2}{2 |R''(0)|} =  \lim_{ \Omega \to \infty }
  \frac{  \ln\mean{  \left|\det\big( K_{ij}  + U_i\,\delta_{ij} \big)\right|} }{\Omega}
\end{equation}
where $\Omega$ is the volume of the system and $U_i$ a Gaussian random potential
with correlator $\mean{ U_iU_j }=\big(R''''(0)/\kappa^2\big)\,\delta_{ij}$. This formula
further generalizes to an arbitrary graph.

\subsection{Rate for the number of stable equilibria~: $\rate_\mathrm{st}(f)$}
\label{subsec:stable} 

We show that the analysis for $\rate_\mathrm{tot}(f)$ can be extended for the computation of the number of \textit{stable} equilibria in the presence of the external force.
In order to consider only stable equilibria, (\ref{Ntot1},\ref{Ntot2}) must be modified as follows
$$
  \mathcal{N}_\mathrm{st}(f) =
  \int_{\mathbb{R}^K} \hspace{-0.15cm} \D{\bf u} \,
  \det \left(\partial_i\partial_j \mathcal{H}({\bf u})\right)
  \,
  \Theta\left( \partial_i\partial_j \mathcal{H}({\bf u}) \right)
  \,
  \prod_{i=1}^K
  \left(
  \EXP{-u_i^2/(2w^2)}
  \delta\left(\partial_i \mathcal{H}({\bf u}) - f_i\right)
  \right)
$$
where $\Theta(M)=1$ if all eigenvalues of the matrix $M$ are positive, and $0$ otherwise.
Because $V'_i(u_i)$ and $V''_i(u_i)$ are independent, Eq.~\eqref{eq:Rsymmetric}, we have the same simplification as for the calculation of the total number of equilibria~:
\begin{align}
  \mean{ \mathcal{N}_\mathrm{st}(f) }
  =
  &\mean{
    \det \left(\partial_i\partial_j \mathcal{H} \right)
    \,
    \Theta\left( \partial_i\partial_j \mathcal{H} \right)
  }
  %\\\nonumber
  %&\times
  \int_{\mathbb{R}^K} \hspace{-0.15cm} \D{\bf u} \,
  \mean{
  \prod_{i=1}^K
  \left(
  \EXP{-u_i^2/(2w^2)}
  \:
  \delta\left(\partial_i \mathcal{H}({\bf u}) - f_i\right)
  \right)
  }
\end{align}
where we have also made use of the fact that $\partial_i\partial_j \mathcal{H}$ does not depend explicitly on ${\bf u}$, but only through $V''_i(u_i)$, which are i.i.d. Gaussian random variables.
Finally we can perform the same sequence of manipulations for a constant force
\begin{align}
  \mean{ \mathcal{N}_\mathrm{st}(f) }
  =
  &
  \overbrace{
  \mean{
    \det \left(\partial_i\partial_j \mathcal{H} \right)
    \,
    \Theta\left( \partial_i\partial_j \mathcal{H} \right)
  }
  }^{\sim \exp[ L\,r_\mathrm{st} ]}
%  \\\nonumber
%  &\times
  \underbrace{
    \frac{
    \exp\left\{
    -
    \frac{1}{2(w\kappa)^2}
        {\bf f}^\mathrm{T}
        \left[ (L_m^{-2}\,\identity_K - \Delta )^2 + L_w^{-4}\,\identity_K \right]^{-1}
        {\bf f}
  \right\}
     }
          {  \left| \det  \left((L_m^{-2}+\I\,L_w^{-2})\,\identity_K - \Delta\right) \right|  }
  }_{
    \sim \exp\left\{ -L \left[ \re\sqrt{\frac{1}{L_m^2}+\frac{\I}{L_w^2}}
    + \frac{f^2}{2(m^4w^2 + |R''(0)|)} \right] \right\}
  }
\end{align}
where we have assumed the exponential behaviour in the absence of the external force $\mean{\mathcal{N}_\mathrm{st}(0)}\big|_{w=\infty} \sim \EXP{L \rate_\mathrm{st}}$ .
This allows us to define a new rate controlling the number of stable equilibria $\mean{\mathcal{N}_\mathrm{st}(f)} \sim \EXP{L r_\mathrm{st}(f)}$ and take the limits
$\lim_{w\to\infty}\lim_{m^2\to0}$.  Finally we obtain
\begin{equation}
   \rate_\mathrm{st}(f) = \rate_\mathrm{st} - \frac{f^2}{2|R''(0)|}
   \:,
\end{equation}
i.e. the same $f$-dependence as $\rate_\mathrm{tot}(f)$, Eq.~\eqref{rr}.

We have not been able to compute analytically the rate $\rate_\mathrm{st}$ so far, however, by dimensionless analysis, we can write that 
$\rate_\mathrm{st}=C_\mathrm{st}/L_c$ is another dimensionless constant, still unkown.
The knowledge of $C_\mathrm{st}$ would provide
%This result provides 
another, more stringent ($C_\mathrm{st}\leq C$), upper bound for the depinning threshold as the value $f_c^\mathrm{st}$ of the force such that $\rate_\mathrm{st}(f)=0$ which gives $f_c^\mathrm{st}=\sqrt{2\rate_\mathrm{st}|R''(0)|}\leq f_c^\mathrm{tot}$ as we have obviously $\mathcal{N}_\mathrm{st}\leq\mathcal{N}_\mathrm{tot}$.

%%%%%%%%%%%%%%%%%%%%%%%%%%%%%%%%%%%%%%%%%%%%%%%%%%%%%%%%%%%%%%%%%%%%%%%%%%%%%%%%%%%%%%%%%%
%%%%%%%%%%%%%%%%%%%%%%%%%%%%%%%%%%%%%%%%%%%%%%%%%%%%%%%%%%%%%%%%%%%%%%%%%%%%%%%%%%%%%%%%%%

\section{Numerical calculation of the depinning threshold}
\label{sec:NumericsAlberto}

We have computed $f_c$ numerically for a discrete elastic chain of $K$ monomers using the algorithm developed in Ref.~\cite{RosKra02}.
The energy of the chain is given by Eq.~\eqref{eq:Hdiscrete} with $m^2=0$ and $\kappa=1$.
The DP lives on a torus $(\tau,u)\in[0,L]\times[0,2\pi/\rf]$ (i.e. we choose periodic boundary conditions in the two directions). 
A convenient model for the disordered force is the harmonic model
\begin{equation}
  \label{eq:CosModel}
  V_i(u) = \xi_{1,i}\, \cos(u/\rf) + \xi_{2,i}\, \sin(u/\rf)
  \:,
\end{equation}
where $\xi_{\alpha,i}$ are independent real Gaussian numbers of zero mean,
$\mean{\xi_{\alpha,i}\xi_{\beta,j}}=\sigma^2\,\delta_{\alpha,\beta}\delta_{i,j}$. 
This leads to the correlations \eqref{corr2} for the periodic correlation function
\begin{equation}
  R(u)=\sigma^2\,\cos(u/\rf)
  \:.
\end{equation}
This model provides a simple practical way to ensure that the disordered strength is translational invariant.
Setting $\rf=1$ for convenience, the variance of the disorder is $-R''(0)=R''''(0)=\sigma^2$. 
The presence of a stationary state is detected for a given force $f$. 
The equations of motion are (for $\kappa=1$)
\begin{equation}
  \label{eq:EqMotionMonomer}
  \dot{u}_i(\tau) = -\derivp{\mathcal{H}({\bf u})}{u_i} = u_{i+1} -2 u_i +u_{i-1}  - V'_i(u_i) + f
  \hspace{0.5cm}\mbox{for }
  i=1,\cdots,K
\end{equation}
%for $i=1,\cdots,K$ 
(these are the Langevin equations in the zero temperature limit of vanishing Langevin noise).
For a given force $f$ we target the first metastable state by using an efficient algorithm \cite{RosKra02}.
Monomers are considered one by one and moved from their positions to the first equilibrium point (for the model \eqref{eq:CosModel}, the equilibrium point of each monomer can be easily found by a bisection method).
The iteration stops when all monomers have a velocity smaller than a small threshold $v_\mathrm{min}$ (here we set $v_\mathrm{min}=10^{-6}$).
The procedure is repeated by increasing the force by steps $\delta f=0.01$. 
The search is ended when no stationary state are found and all monomers have moved forward of at least two periods of the disorder, giving the value of the critical force $f_c$.
Our results are shown in Fig.~\ref{fig:DataAlbe} for DP of $K=512$ monomers, varying the disorder strength $\sigma^2$.
When compared to the upper bound found in the paper, Eq.~\eqref{fclast0}, the critical force is
\begin{equation}
 f_c^\mathrm{tot}/f_c=1.64\pm0.02
  \:.
\end{equation}
The range of disordered strength considered in the simulation corresponds to $L_c/a=\sigma^{-2/3}$ between $1/\sqrt{2}$ to $2$. 
Thus it is quite remarkable that we do not observe any significant deviation from the behaviour $f_c\propto\sigma^{4/3}$ (Fig.~\ref{fig:DataAlbe}), which is expected to hold in the continuum limit ($L_c\gg a$).

%%%%%%%%%%%%%%%%%%%%%%%%%%%%%%%%%%%%%%%%%%%%%%%%%%%%%%%%%%%%%%%%%%%%%%%%%%%%%%%%%%%%%%%%%%
%%%%%%%%%%%%%%%%%%%%%%%%%%%%%%%%%%%%%%%%%%%%%%%%%%%%%%%%%%%%%%%%%%%%%%%%%%%%%%%%%%%%%%%%%%

\section{Equilibria configurations: spatial correlations in the annealed measure}
\label{sec:Configurations}

It is interesting to ask what is the typical spatial configuration ${\bf u}$ of an elastic line at a force-free stationary point (equilibrium) chosen at random. 
In a fixed environment, it involves a quenched measure, which is difficult to study. 
It is simpler, but still quite instructive, to ask the same question over the set of all stationary points in all environments, i.e. to define the annealed measure
\begin{equation} 
  \label{defAnn} 
  \mathcal{P}_a({\bf u}) := \frac{\rho({\bf u})}{\mean{{\cal N}_{\rm tot}}} 
\end{equation}
where $\rho({\bf u})$ was defined in \eqref{Ntot2}. 
We also introduce the annealed averaging of any function of the monomers positions~:
\begin{equation}
  \label{eq:DefAnnealedAveraging}
  \mean{ F({\bf u}) }_a 
  := \mean{\int_{\mathbb{R}^K}  \D{\bf u}\, \mathcal{P}_a({\bf u})\,F({\bf u}) }
  \:,
\end{equation}
where $K$ is the number of monomers and $\mean{\cdots}$ denotes averaging over the disorder.
The normalization in \eqref{defAnn} obviously ensures that $\mean{1}_a=1$. 
%It is such that $\mean{\int_{\mathbb{R}^K}  \D{\bf u}\, \mathcal{P}_a({\bf u})} =1$. 
We thus calculate the generating function in presence of a source ${\bf j}$ as
\begin{equation}
  Z ({\bf j}) 
  := 
  \mean{ \EXP{ {\bf j} \cdot {\bf u}} }_a 
 % \mean{\int_{\mathbb{R}^K}  \D{\bf u}\, \mathcal{P}_a({\bf u})  \, \EXP{ {\bf j} \cdot {\bf u}} }
  \:.
\end{equation}
Its calculation is very similar, but slightly different from the one of Section~\ref{sec:Depinning} for an external force,
see in particular the Subsection~\ref{subsubsec:discrete}.
We can examine Eq.~\eqref{j6} where we can set for now $w=+\infty$ and ${\bf f}=0$
but we need to add the factor $\EXP{{\bf j} \cdot {\bf u}}$. Performing the integration
over ${\bf u}$ we now obtain
\begin{equation}
  Z ({\bf j}) 
  = \exp\left( \frac{|R''(0)|}{2} \, \,{\bf j} \cdot (m^2 \identity_K-\kappa \Delta)^2 \cdot {\bf j} \right)
\end{equation}
Hence the annealed measure $\mathcal{P}_a({\bf u})$ over ${\bf u}$ is centered Gaussian 
and its two point correlation function is
\begin{equation} 
  \label{larkin1} 
  \mean{ u_i u_j }_a = |R''(0)|\, ([m^2 \identity_K-\kappa \Delta]^{-2})_{i j}
\end{equation}
which is the same result as for the so-called Larkin model recalled in~\ref{app:Larkin}. In the limit of zero mass
it yields a roughness exponent~\footnote{
  The roughness exponent is given by considering 
  $\mean{ u(\tau) u(\tau') }\sim \bra{\tau}\Delta^{-2}\ket{\tau'}\sim|\tau-\tau'|^{2\zeta_L}$, 
  as clearly the above calculation generalizes to arbitrary internal dimension and elasticity.
  Simple dimensional analysis gives $\zeta_L=2-d/2$ for an interface in (spatial) dimension $d$ (i.e. $\tau\in\mathbb{R}^d$).
} 
%$\langle (u_i - u_j)^2 \rangle_{\mathcal{P}_a} \sim |i-j|^{2 \zeta_L}$ 
%with 
$\zeta_L=3/2$, see also~\ref{app:Larkin}. 
Hence we have established here by
a precise calculation, the fact (anticipated from the physics of the dimensional reduction) 
that the annealed measure over all equilibria (stable and
unstable) and environments yields configurations with the Larkin roughness. 

It is then immediate to provide a few extensions. For instance
\begin{equation}
 \mean{ u_i u_j }_{a,w}
  = |R''(0)| \,
  \left(
  \left[ (m^2\,\identity_K - \kappa\,\Delta)^2 + \frac{|R''(0)|}{w^2}\,\identity_K \right]^{-1}
  \right)_{i j}
\end{equation}
when $\mathcal{P}_{a,w}$ is the measure where displacements are restricted, see subsection \ref{subsubsec:discrete},
defined as in \eqref{defAnn} with $\rho({\bf u}) \to \rho({\bf u}) \phi({\bf u})$
and $\mean{{\cal N}_{\rm tot}} \to  \mean{ \mathcal{N}_\mathrm{tot}^w } $. 
Finally in the case $w=+\infty$ and ${\bf f} \neq 0$ it is immediate (in view of the remark
at the end of Subsection~\ref{subsubsec:discrete}) that the result
\eqref{larkin1} still holds but for the shifted deviation from disorder-free equilibrium, i.e. for 
${\bf u} - {\bf u}_0$.

%%%%%%%%%%%%%%%%%%%%%%%%%%%%%%%%%%%%%%%%%%%%%%%%%%%%%%%%%%%%%%%%%%%%%%%%%%%%%%%%%%%%%%%%%%
%%%%%%%%%%%%%%%%%%%%%%%%%%%%%%%%%%%%%%%%%%%%%%%%%%%%%%%%%%%%%%%%%%%%%%%%%%%%%%%%%%%%%%%%%%
\section{Relation with one-dimensional Anderson localization}
\label{sec:GLE}

The main outcome of Section~\ref{sec:CountingEquilibria} was the representation of the number of equilibria of the elastic line in terms of the determinant of a discrete random Schr\"odinger operator,
%random matrix, 
Eq.~\eqref{eq:DetRepres}, or a random differential operator \eqref{eq:DetRepres2}.
Such determinants are common in the problem of one-dimensional localization of a wave by a random medium.
A famous example is the Herbert-Jones-Thouless relation \cite{HerJon71,Tho72,Luc92} (see also the appendix of Ref.~\cite{GraTexTou14} for a discussion of the continuous case).
This remark will allow us to make a precise connection with the study of the one-dimensional Schr\"odinger equation for a random (time-independent) potential.

\subsection{From the number of equilibria to the generalized Lyapunov exponent}
\label{sec:IntroGLE}

Although it is not essential, we find more transparent to formulate the calculation of the mean value
%of the determinant
$\mean{ \mathcal{N}_\mathrm{tot} }$ in the continuum limit, i.e. our starting point is the representation \eqref{eq:DetRepres2}.
As is well-known \cite{GelYag60,LevSmi77}, such ratios of functional determinants for the Schr\"odinger operators \eqref{eq:DisorderedHamiltonian} can be evaluated by so-called Gelfand-Yaglom method as
\begin{equation}
 \frac{\det \left( m^2/\kappa + H \right) }
          { \det \left( m^2/\kappa + H_\mathrm{free} \right) }
%  \frac{ \det{\mathcal{M}} }{ \det{\mathcal{M}_\mathrm{free}} }
  = \frac{ y(L) }{ y_\mathrm{free}(L) }
\end{equation}
in terms of the solutions $y(\tau)$ of the {\it initial value} (Cauchy) problem for the same operators~:
$(m^2/\kappa + H)\, y(\tau)=0 $, i.e.
\begin{equation}
  \label{eq:yDiffEq}
  -\deriv{^2y(\tau)}{\tau^2}+U(\tau)\,y(\tau) = -\frac{m^2}{\kappa}\, y(\tau)
\end{equation}
 for
\begin{equation}
   \label{eq:CauchyIntialValue}
  \begin{cases}
    y(0)=0 \\
    y'(0)=1
  \end{cases}
\end{equation}
(see~\ref{app:bc}).
We have obviously $y_\mathrm{free}(\tau)=L_m \sinh{(\tau/L_m)}$ where $L_m=\sqrt{\kappa}/m$.
We deduce%thus conclude that in the continuous case
\begin{equation}
  \label{eq:DetRepresCont}
  \mean{ \mathcal{N}_\mathrm{tot} }
  =\frac{ \mean{ \left|y(L)\right| } }{L_m \sinh{(L/L_m)}}
  \:.
\end{equation}
Cauchy problems related to products of random $2\times 2$ matrices of various types as well as their continuum limits were under active consideration recently, with many important analytical insights, see Refs.~\cite{Kol10,ComLucTexTou13,ComTexTou13} and references therein.

At this point it is appropriate to note that the spectral problem defined by \eqref{eq:yDiffEq} with
$y(0)=y(L)=0$ is the classical problem of one-dimensional localization for
the Schr\"odinger equation with a Gaussian white-noise potential, studied extensively since the seminal work \cite{Hal65}, see Refs.~\cite{LifGrePas88,Luc92} for further details.
To make contact to notations used for that problem in the literature we introduce
\begin{equation}
  E = -\frac{m^2}{\kappa} = -\frac{1}{L_m^2}
  \:,
\end{equation}
%$E=-m^2/\kappa=1/L_m^2$,
which plays the role of %an effective energy.
the energy in the Schr\"odinger equation \eqref{eq:yDiffEq}.
As is well-known the main qualitative feature of the Cauchy problem (\ref{eq:yDiffEq},\ref{eq:CauchyIntialValue}) is the exponential growth of its solution from chosen initial conditions in every realization of the disorder \cite{LifGrePas88}.
It is conventional to define the localization length as the inverse of the Lyapunov exponent
\begin{equation}
  \gamma_1 := \lim_{L \to \infty} \frac{1}{L}\smean{\ln|y(L)|}  \geq 0
%  \:,
\end{equation}
(in principle the average is not needed as $\ln|y(L)|$ is self-averaging in the limit and it was only introduced for comparison with the definition of the GLE below).
%which is self-averaging in the limit.
However the average $\mean{|y(L)|}$ depends crucially on the fluctuations,
and is not given by $\exp\{\gamma_1 L\}$. To discuss the role of the fluctuations,
as was done in the description of turbulence one uses the
multifractal formalism developed by Paladin and Vulpiani~\cite{PalVul87,PalVul87b,CriPalVul93}.
In this approach one introduces the moments $\mean{|y(L)|^q}$, with arbitrary positive parameter $q$. These are known to grow exponentially at large $L$ as $\mean{ |y(L)|^q } \sim \EXP{L\,\Lambda(q)}$ where
\begin{equation}
  \label{eq:DefGLE}
  \Lambda(q) := \lim_{L \to \infty} \frac{1}{L}\ln\mean{ |y(L)|^q }
\end{equation}
is the large deviation rate function, also known as the {\it generalized} Lyapunov exponent (GLE) \cite{PalVul87,PalVul87b,CriPalVul93}.~\footnote{
  The GLE has another physical interpretation~:
  if the Schr\"odinger equation is solved in a scattering configuration, when the disordered potential has support $[0,L]$ and vanishes outside this interval, the transmission probability through the disordered region is related to the solution of the Cauchy problem as $T\simeq|y(L)|^{-2}$ (if the length $L$ is larger than the localization length, the precise matching of the wave function between the disordered region and the regions free of disorder is not important).
  The distribution of the reflection probability $1-T$ has been determined in Ref.~\cite{AntPasSly81} in the limit $E\gg D^{2/3}$.
  The Landauer formula relates this probability to the electric resistance (in unit $2e^2/h$) as $\rho=1/T-1$. We conclude that the GLE also controls the moments of the resistance
  $\mean{\rho^\theta}\sim\EXP{L\,\Lambda(2\theta)}$.
}
It can be written as a series
\begin{equation}
  \label{eq:LambdaExpansion}
  \Lambda(q)
  = 
  \sum_{n=1}^{\infty} \frac{\gamma_n}{n!}\,q^n
\end{equation}
where $\gamma_n L$ is the cumulant of order $n$ of $\ln|y(L)|$ at large $L$.

\begin{figure}[!ht]
\centering
\includegraphics[width=8cm]{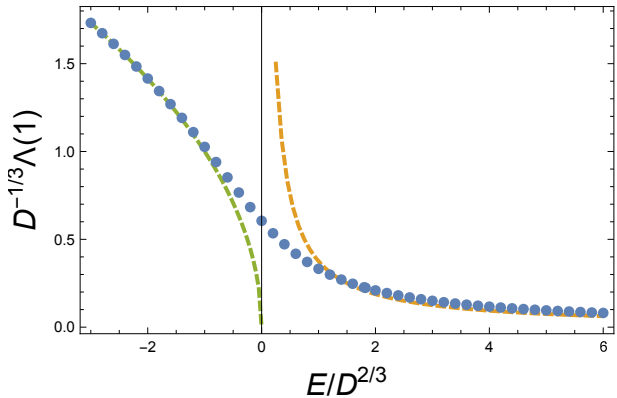}
\caption{
\it Generalized Lyapunov exponent $\Lambda(1)$ as a function of the energy $E$ (blue dots). 
The GLE is computed %obtained by diagonalisation of the discretised operator $\mathscr{O}_q$, as 
thanks to the method presented in Subsection~\ref{subsec:FPDiscrete}. 
Dashed lines are $\sqrt{-E}$ (green), Eq.~\eqref{eq:LambdaPert} or~\eqref{eq:HeuristicEstimation}, and $3/(8E)$ (orange), Eq.~\eqref{eq:GLEforLargePositiveEnergy}. 
%The linear fit at the origin is $\Lambda(1)\simeq0.60\,D^{1/3}-0.37\,D^{-1/3}\,E$ (dotted line).
  The dependence in $q$ will be discussed in Sections~\ref{sec:GLE} and~\ref{sec:LambdaEvenQ}.
}
\label{fig:GLEforPosNegEnergy}
\end{figure}

From Eq.~\eqref{eq:DetRepresCont}, it is related to the rate, 
governing the exponential growth $\mean{ \mathcal{N}_\mathrm{tot} }\sim\exp[\rate L]$ for $L\to\infty$,
as
\begin{equation}
  \label{eq:Rate}
  \rate = \Lambda(1) - \sqrt{-E} = \Lambda(1) - 1/L_m
  \:,
\end{equation}
defined for $E<0$.
For the study of the localization properties it is natural to consider the so-called {\it semiclassical regime} of large positive energy, $E \gg D^{2/3}$.
In that regime the fluctuations can be considered as Gaussian with subdominant higher cumulants \cite{SchTit02},
$\Lambda(q) \simeq \gamma_1\, (q + q^2/2)$, where the Lyapunov exponent can be computed perturbatively \cite{AntPasSly81}, $\gamma_1\simeq D/(4E)$, thus 
\begin{equation}
  \label{eq:GLEforLargePositiveEnergy}
    \Lambda(q) \simeq \frac{D}{4E} \, \left( q + \frac12\,q^2 \right)
    \hspace{1cm}
    \mbox{for }
    E\gg D^{2/3}
    \:.
\end{equation}
The property $\gamma_2 \simeq \gamma_1$ is known as ``single parameter scaling'' \cite{CohRotSha88} (see also the discussion in Ref.~\cite{RamTex14}). 
The GLE is plotted in Fig.~\ref{fig:GLEforPosNegEnergy} for $q=1$.
This regime would correspond to %large negative
$m^2 <0$ with $|L_m|\ll L_c$ while here, in the elastic line counting problem, we are interested in the opposite case $m^2 \geq 0$ of negative effective energies $E<0$.
The explicit evaluation of \eqref{eq:DetRepresCont} remains an outstanding and non-perturbative problem where non-Gaussian fluctuations dominate (for early works see e.g. Refs.~\cite{BouGeoHanLeDMai86,BouGeoLeD86,Pen94} where it was analysed for integer value of $q$~; see also Section~\ref{sec:LambdaEvenQ}).

\subsection{Stochastic Riccati equation}

A recursive method was developed in Ref.~\cite{SchTit02} allowing to obtain integral representations for the $\gamma_n$'s in terms of multiple integrals, but they are quite complicated and not convenient to obtain limiting behaviours.
An alternative way of calculating these quantities, suitable for the $E\to-\infty$ limit (large mass limit
for the DP problem), was proposed in Ref.~\cite{RamTex14} for a different model. We adjust this approach here:
the transformation
\begin{equation}
  |y(L)|=\exp\left\{\int_0^L\D\tau\,z(\tau)\right\}
\end{equation}
 relates \eqref{eq:yDiffEq} to the stochastic Riccati equation \footnote{Note that this equation also describes the thermally activated
  motion at temperature $T$ of a particle near depinning with $-E=f-f_c$ and $D=T$}
\begin{equation}
  \label{Riccati}
  \deriv{z(\tau)}{\tau} = -E - z(\tau)^2 + U(\tau) = - \potRic'(z(\tau)) + U(\tau)
  \:,
\end{equation}
where
\begin{equation}
  \label{eq:RiccatiPotential}
  \potRic(z)=Ez+\frac13z^3
  \:.
\end{equation} 
For $E<0$ and  $U(\tau)=0$, $z=\sqrt{|E|}$ is a fixed point.
The ``noise'' $U(\tau)$ thus generates fluctuations around this point.

\subsubsection{A perturbative analysis of the stochastic differential equation~\eqref{Riccati}}
\label{subsec:Peturbative}

We present first a perturbative approach to the determination of the GLE, valid for small disorder $D$ and large negative energy $E$.
This method provides the main $q$-dependence of $\Lambda(q)$ in this regime, however it will turn insufficient in order to obtain the GLE for the specific value $q=1$, which will be shown to involve non perturbative contributions.
We follow the method introduced in Ref.~\cite{RamTex14} (section 6 of this reference) in a different situation~:
since in the limit $E\to-\infty$ the process $z(x)$ is most of the time trapped near $z=\sqrt{-E}$, this suggests  to linearize the ``force'', $- \potRic'(z)=-E-z^2\simeq-2\sqrt{|E|}(z-\sqrt{|E|})$, leading to the Ornstein-Uhlenbeck process.
The method developed below is a systematic perturbative expansion around the Ornstein-Uhlenbeck process.
This can be most conveniently achieved at the level of the stochastic differential equation (SDE) \eqref{Riccati} (this is more straightforward that from the Fokker-Planck equation (FPE)).
For convenience, we write 
\begin{equation}
  E=-k^2
\end{equation}
 and rescale the coordinate and the process as
\begin{equation}
  z(x) = k\, \left( 1+\epsilon\, \rr(u) \right)
  \quad\mbox{with}\quad
  u = kx
  \:.
\end{equation}
Eq.~\eqref{Riccati} leads to the SDE in terms of dimensionless variables
\begin{equation}
    \rr'(u) = -2\, \rr(u) + \eta(u) - \epsilon \, \rr(u)^2
    \:,
\end{equation}
where $\eta(u)$ is a normalized Gaussian white noise, $\mean{ \eta(u) \eta(u')}=\delta(u-u')$, and
\begin{equation}
  \epsilon := \sqrt{\frac{2D}{k^3}}
\end{equation}
is the small perturbative parameter.
We now expand the process in powers of $\epsilon$ as $\rr=\rr_0+\rr_1+\rr_2+\cdots$.
The different terms can be found recursively
\begin{align}
  \label{eq:Zeta0}
  \rr_0(u) &= \int_0^u\D t\, \EXP{-2(u-t)} \,  \eta(t)
  \\
  \label{eq:Zeta1}
  \rr_1(u) &= - \epsilon \int_0^u\D t \, \EXP{-2(u-t)} \,  \rr_0(t)^2
  \\
  \label{eq:Zeta2}
  \rr_2(u) &= 2\epsilon^2 \int_0^u\D t\,   \EXP{-2(u-t)} \,  \rr_0(t)\int_0^t\D t'  \EXP{-2(t-t')}\,  \rr_0(t')^2
    \\
  \label{eq:Zeta3}
  \rr_3(u) &= -\epsilon  \int_0^u\D t\,   \EXP{-2(u-t)} \, \left[ \rr_1(t)^2 + 2\rr_0(t)\rr_2(t) \right]
\end{align}
etc.
Making use of the Gaussian nature of the Ornstein-Uhlenbeck process and of
\begin{equation}
  \mean{ \rr_0(u)\,\rr_0(u') }_\mathrm{stat} = \frac{1}{4} \EXP{-2|u-u'|}
\end{equation}
we can compute any correlations functions.
$\mean{\cdots}_\mathrm{stat}$ denotes averaging in the stationary regime (i.e. for $u,\,u'\gg1$).

We can check the method on the Lyapunov exponent: it leads to the expansion
\begin{align}
  \frac{\gamma_1^\mathrm{(pert)}}{k} &=  1+\epsilon\, \mean{\rr(u)}_\mathrm{stat}
   = 1 - \frac{\epsilon^2}{8}  -\frac{5\epsilon^4}{128} + \mathcal{O}(\epsilon^6)
\end{align}
which is in perfect correspondence with the analytic part of the expansion obtained from the exact result.
The complex Lyapunov exponent for $D=1$ is~\cite{LifGrePas88} (see also Ref.~\cite{GraTexTou14})
\begin{equation}
  \Omega = \gamma_1 - \I\pi \, N =
  \frac{\mathrm{Ai}'(-E)-\I\,\mathrm{Bi}'(-E)}{\mathrm{Ai}(-E)-\I\,\mathrm{Bi}(-E)}
\end{equation}
leading to
\begin{align}
  \label{eq:Gamma1}
  \frac{\gamma_1}{k} = 1 - \frac{\epsilon^2}{8}  -\frac{5\epsilon^4}{128}
  + \mathcal{O}(\epsilon^6)
  - \frac{\EXP{-{16}/{(3\epsilon^2)}} }{2}
  \left[ 1 - \frac{\epsilon^2}{12} + \mathcal{O}(\epsilon^4) \right]
\end{align}
The Lyapunov exponent exhibits non-analytic contributions, $\sim\exp\big[-8|E|^{3/2}/(3D)\big]$, which are associated to the possibility of rare excursions of the process $z(x)$ to $\pm\infty$ related to the exponentially small probability current (this problem was not present in the case studied in Ref.~\cite{RamTex14} by the same method).

The variance is given by
\begin{equation}
  \gamma_2^\mathrm{(pert)} = 2k\epsilon^2\,\lim_{u\to\infty}
  \smean{ \rr(u)\int_0^u\D v\, \rr(v) }_c
\end{equation}
where $\smean{XY}_c=\smean{XY}-\smean{X}\smean{Y}$. The limit $u\to\infty$ ensures that the correlator is computed in the stationary regime.
Using the expressions (\ref{eq:Zeta0},\ref{eq:Zeta1},\ref{eq:Zeta2}), some lengthy algebra gives
\begin{equation}
  \label{eq:Gamma2}
  \frac{\gamma_2^\mathrm{(pert)}}{k} = \frac{\epsilon^2}{4}  +\frac{9\epsilon^4}{64}
  + \mathcal{O}(\epsilon^6)
\end{equation}

\begin{figure}[!ht]
\centering
\includegraphics[width=8cm]{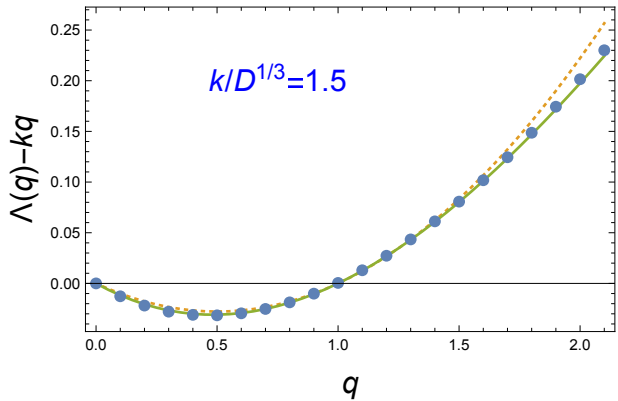}
\caption{\it Generalized Lyapunov exponent obtained numerically (dots) according to the method of Subsection~\ref{subsec:FPDiscrete}.
  Dotted line is $-q(1-q)\,D/(4k^2)$ and continuous line corresponds to~\eqref{eq:LambdaPert}. We have set $D=1$.}
  \label{fig:LambdaDeQ}
\end{figure}
% Prg LambdaDeQ.nb

The leading term to the third cumulant is more easy to compute
\begin{align}
  \gamma_3^\mathrm{(pert)} &= k\,\epsilon^3\,
  \lim_{u\to\infty}\frac{1}{u}
  \mean{ \left(\int^u (\rr-\mean{\rr})\right)^3 }
  \\
  \nonumber
  &
  = 3 k\,\epsilon^3\,
  \lim_{u\to\infty}\frac{1}{u}
  \mean{ \left(\int^u \rr_0\right)^2 \int^u (\rr_1-\mean{\rr_1})  }
  +\cdots
\end{align}
hence
\begin{equation}
  \label{eq:Gamma3}
  \frac{\gamma_3^\mathrm{(pert)}}{k} = - \frac{3\epsilon^4}{16}  + \mathcal{O}(\epsilon^6)
  \:.
\end{equation}
One can check that $\gamma_4=\mathcal{O}(\epsilon^6)$, hence gathering (\ref{eq:Gamma1},\ref{eq:Gamma2},\ref{eq:Gamma3}) leads the perturbative result $\Lambda(q)\simeq\Lambda_\mathrm{pert}(q)$ with
\begin{equation}
  \label{eq:LambdaPert}
%  \frac{\Lambda_\mathrm{pert}(q)}{\sqrt{|E|}}
%  = q-\frac{s}{4}q(1-q)-\frac{s^2}{32}q(1-q)(5-4q)+\mathcal{O}(s^3)
  \Lambda_\mathrm{pert}(q) 
  =  k\,q - \frac{D}{4k^2}\, q(1-q) - \frac{D^2}{32k^5}\, q(1-q)(5-4q)
  +\mathcal{O}(D^3/k^8)
  \:.
\end{equation} 
This approximate expression of the GLE is compared with numerics of Section~\ref{sec:GLEspectral} in Fig.~\ref{fig:LambdaDeQ}~: the agreement is excellent.
Quite remarkably, Eq.~\eqref{eq:LambdaPert} shows that the rate~$r$ vanishes up to third order
in $s\ll1$ (large mass limit).
We conjecture that this remains true at all orders in $D$, i.e.
\begin{equation}
  \Lambda_\mathrm{pert}(1) - k =0
\end{equation}
which is confirmed by numerics (see Figs.~\ref{fig:LambdaDeQ} and~\ref{fig:rate}, and discussion below).

\subsubsection{The need of a non perturbative analysis and a first estimate}
\label{subsec:EstimateRate}

For $E<0$,
the potential \eqref{eq:RiccatiPotential} is not confining, thus
the noise is not only responsible for small fluctuations around $+\sqrt{|E|}$, characterized by \eqref{eq:LambdaPert}, but can also produce large excursions of the process at $\pm\infty$:
if the process overcomes the potential barrier at $-\sqrt{|E|}$, it is rapidly driven towards $-\infty$, reinjected at $+\infty$, from which it eventually goes back to $+\sqrt{|E|}$.
The rare jumps are separated by time intervals exponentially distributed~\cite{Tex00}.
The probability rate for a jump is exponentially small~:
\begin{equation}
  \IDoS 
  \simeq
  \frac{\sqrt{|E|}}{\pi}\,\exp\big[-4|E|^{3/2}/(3D)\big]
\end{equation}
 (this is the integrated density of states of the disordered model \eqref{eq:DisorderedHamiltonian}, cf. Ref.~\cite{Hal65}).
This picture suggests to split the process into two parts, describing the small fluctuations around $z=\sqrt{|E|}$ and the rare jumps: $z(\tau)=z_\mathrm{trapped}(\tau)+z_\mathrm{jump}(\tau)$.
Because the two processes involve different time scales, they can be assumed independent and the GLE splits in two parts: $\Lambda(q)\simeq\Lambda_\mathrm{pert}(q)+\Lambda_\mathrm{jump}(q)$.
The contribution of the jumps is estimated by writing $z_\mathrm{jump}(\tau)=\sum_nh_n(\tau-\tau_n)$ where $h_n(\tau)$ is a narrow function describing the jump at time $\tau_n$.
Over large scale, $\int_0^L\D\tau\,z_\mathrm{jump}(\tau)$ is equivalent to a Compound Poisson process with L\'evy exponent
$\Lambda_\mathrm{jump}(q)\simeq\IDoS\,\big(\smean{\EXP{qv_n}}-1\big)$ where $v_n=\int\D\tau\,h_n(\tau)$
(see Ref.~\cite{GraTexTou14} and references therein).
This simple argument shows that the strongest  dependence of the rate in $E$ is
mostly controled by $N(E)$, hence presents the non-analytic behaviour
\begin{equation}
 \label{eq:HeuristicEstimation}
  \rate
  =\Lambda(1)-\sqrt{-E}
  \simeq\Lambda_\mathrm{jump}(1)
  \sim N(E)\sim\exp\big[-4|E|^{3/2}/(3D)\big]
  \:.
\end{equation}
We stress an important point~:
as noticed above, the Lyapunov exponent $\gamma_1$ provides a non-analytic contribution to the GLE $\Lambda(q)=\gamma_1\,q+\gamma_2\,q^2/2+\cdots$. This contribution, $\sim\exp\big[-8|E|^{3/2}/(3D)\big]$, cf.~Eq.~\eqref{eq:Gamma1}, is however much smaller than \eqref{eq:HeuristicEstimation}, which is therefore fully controlled by \textit{fluctuations}.
In the next section, we confirm the behaviour \eqref{eq:HeuristicEstimation} by a more detailed analysis and provide the pre-exponential function.

%%%%%%%%%%%%%%%%%%%%%%%%%%%%%%%%%%%%%%%%%%%%%%%%%%%%%%%%%%%%%%%%%%%%%%%%%%%%%%%%%%%%%%%%%%%%%%%%%%%%%%%%%%%%%

\section{Generalized Lyapunov exponent and spectral analysis}
\label{sec:GLEspectral}

The study of the non-analytic contribution to the GLE requires powerful methods.
In this section we use the relation with a spectral problem to develop some accurate numerical methods.
%We now go beyond the previous analysis, which focused only on the large mass limit.
The diffusion
%stochastic equation
\eqref{Riccati} is characterized by the (``backward'') generator $\mathscr{G}$ and its adjoint $\mathscr{G}^{\dagger}$ (``forward generator'')
\begin{equation}
  \mathscr{G}=D\deriv{^2}{z^2} - (E+z^2)\deriv{}{z}
  \:,
  \hspace{0.25cm}\mbox{ and }\hspace{0.25cm}
  \mathscr{G}^{\dagger}=D\deriv{^2}{z^2}+\deriv{}{z}(E+z^2)
\end{equation}
governing the conditional probability density $\mathcal{P}_{\tau}(z|z_0)=\bra{z}\exp\big(\tau\mathscr{G}^{\dagger}\big)\ket{z_0}$, solution of the Fokker-Planck equation (we recall that $(\D/\D z)^\dagger=-\D/\D z$). 
%It will be convenient to write the forward generator as
%\begin{equation}
%  \mathscr{G}^\dagger = D\deriv{^2}{z^2} + \deriv{}{z}\potRic'(z)
%  = D\deriv{}{z}\EXP{-\potRic(z)/D}\deriv{}{z}\EXP{\potRic(z)/D}
%  \:.
%\end{equation}
The moments
\begin{equation}
  \label{eq:MomentsAndGenerator}
  \mean{|y(L)|^q}
  =\mean{ \EXP{q\int_0^L\D\tau\,z(\tau)} }
  =\int\D z\, \bra{z}\EXP{L\,\mathscr{O}_q}\ket{z_0}
\end{equation}
are controlled by the operator
\begin{equation}
\label{eq:DefOq}
 \mathscr{O}_q=\mathscr{G}^{\dagger}+qz
 \:,
\end{equation}
where the initial condition is $z_0=\infty$ for Dirichlet boundary condition $y(0)=0$. 
%, what provides a spectral interpretation of the GLE: $\Lambda(q)$ is the largest eigenvalue of the non Hermitian operator $\mathscr{O}_q$.
%Based on the spectral analysis of this non-Hermitian operator, we now propose two different and complementary numerical methods allowing to determine the generalized Lyapunov exponent (GLE) $\Lambda(q)$.
%We propose two different and complementary numerical methods allowing to determine the generalized Lyapunov exponent (GLE) $\Lambda(q)$. Both are based on the spectral analysis of the non-Hermitian operator $\mathscr{O}_q=\mathscr{G}^\dagger+qz$ controlling the evolution of the moments $\mean{|y(L)|^q}$. The spectral analysis leads to
We introduce the biorthogonal set of right and left eigenvectors
\begin{align}
  \label{eq:EqForPhiR}
  \left( \mathscr{G}^\dagger+qz \right)\Phi_n^\mathrm{R}(z;q) &= - \mathscr{E}_n(q)\, \Phi_n^\mathrm{R}(z;q)
  \\
  \label{eq:EqForPhiL}
  \left( \mathscr{G} +qz \right)\Phi_n^\mathrm{L}(z;q) &= - \mathscr{E}_n(q)\, \Phi_n^\mathrm{L}(z;q)
\end{align}
where $n\in\mathbb{N}$ and $\mathscr{E}_0<\mathscr{E}_1<\cdots$, with normalization condition
\begin{equation}
\int\D z\,\Phi_n^\mathrm{L}(z;q)\,\Phi_m^\mathrm{R}(z;q)=\delta_{n,m}
\end{equation}
(in~\ref{app:SpectrumGdagger}, we show that the spectrum of $\mathscr{O}_q$ is discrete).
The representation \eqref{eq:MomentsAndGenerator} shows that the moments present the exponential behaviour with the length $\mean{|y(L)|^q}\sim\EXP{-L\mathscr{E}_0(q)}$, and thus the GLE is the \textit{largest} eigenvalue 
\begin{equation}
  \Lambda(q)= - \mathscr{E}_0(q) 
\end{equation}
of the operator $\mathscr{O}_q$ \cite{PalVul87}.
To make connection with the study of elastic line, we can rewrite the number of equilibria as
\begin{equation}
  \label{eq:NtotAndSpectrum}
  \mean{\mathcal{N}_\mathrm{tot}}
  =\frac{\sqrt{-E}}{\sinh(\sqrt{-E}L)}
  \sum_{n=0}^\infty
  \EXP{-L\mathscr{E}_n(1)}\int\D z\,\Phi_n^\mathrm{R}(z;1)\,\Phi_n^\mathrm{L}(z_0;1)\big|_{z_0\to\infty}
  \:.
\end{equation}

We now study the spectral problem with two complementary methods to determined $\Lambda(q)$~:
In Subsection~\ref{subsec:FPDiscrete}, we discretize the operator $\mathscr{O}_q=\mathscr{G}^{\dagger}+qz$ and perform a direct diagonalisation.
In Subsection~\ref{subsec:DiffEq}, we use a more precise method based on a neat analysis of the differential equation for $\Phi_0^\mathrm{R}(z;q)$.

\subsection{Continuous time random walk with drift}
\label{subsec:FPDiscrete}

The FPE $\partial_\tau\mathcal{P}_\tau(z)=\mathscr{G}^\dagger\mathcal{P}_\tau(z)$ can be discretized in space as follows~:
we write $z=n\,b$ and introduce the transition rate from site $m$ to site $n=m\pm1$
\begin{equation}
  t_{n,m} = \frac{1}{b^2} \, \EXP{(\potRic_m-\potRic_{n})/2}
\end{equation}
where $\potRic_n=\potRic(n\,b)$.
The continuous time random walk is thus described by the master equation
\begin{align}
  \label{eq:DiscretePFE}
   \partial_\tau P_\tau(n)
   = t_{n,n+1}\, P_\tau(n+1)
   + t_{n,n-1}\, P_\tau(n-1)
   - (t_{n+1,n}+t_{n-1,n})\, P_\tau(n)
\end{align}
(in the limit $b\to0$ we recover the continuous diffusion for $D=1$~; cf. Ref.~\cite{HagTex08} for example).
The boundaries must be discussed in detail.
In order to mimic the absorption at $z=-\infty$ with reinjection at $z=+\infty$, we consider a finite lattice $n\in\{-\LatNb ,-\LatNb+1,\cdots,+\LatNb\}$ and choose the following transition rates connecting the two boundaries
\begin{equation}
  t_{\LatNb,-\LatNb} = \frac{1}{b^2} \,p_\mathrm{inj}
  \quad\mbox{and}\quad
  t_{-\LatNb,\LatNb} = 0
\end{equation}
where $p_\mathrm{inj}$ is a positive parameter.
Eq.~\eqref{eq:DiscretePFE} must be supplemented by
\begin{align}
   \partial_\tau P_\tau(-\LatNb)
   &= t_{-\LatNb,-\LatNb+1}\, P_\tau(-\LatNb+1)
   - (t_{-\LatNb+1,-\LatNb}+t_{\LatNb,-\LatNb})\, P_\tau(-\LatNb)
   \\
   \partial_\tau P_\tau(\LatNb)
   &= t_{\LatNb,\LatNb-1}\, P_\tau(\LatNb-1) + t_{\LatNb,-\LatNb}\, P_\tau(-\LatNb)
   - t_{\LatNb-1,\LatNb}\, P_\tau(\LatNb)
   \:.
\end{align}
These equations define the $(2\LatNb+1)\times(2\LatNb+1)$ matrix $\big(\mathscr{G}^\dagger\big)_{n,m}$.
Adding $(q\,n\,b)\,\delta_{n,m}$ we obtain the matrix $\big(\mathscr{O}_q\big)_{n,m}$ and perform exact diagonalization.

\begin{figure}[!ht]
\centering
\includegraphics[width=7cm]{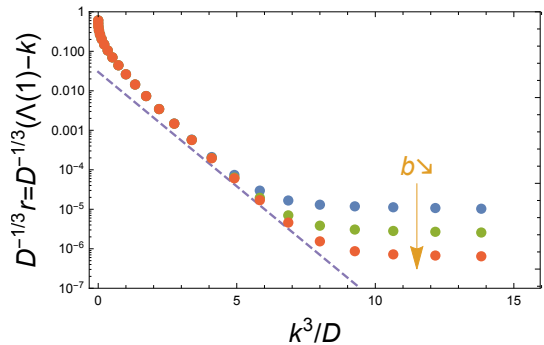}
\caption{\it Rate $\rate=- \mathscr{E}_0(1)-k=\Lambda(1)-k$, where $E=-k^2$, obtained by diagonalization of $(2\LatNb+1)\times(2\LatNb+1)$ matrix $\big(\mathscr{O}_q\big)_{n,m}$.
  Lattice spacing is $b=0.02,\, 0.01$ and $0.005$, with $\LatNb\times b=5$ fixed ($0\leq k\leq2.4$).
  The line corresponds to $\propto\exp[-4|E|^{3/2}/(3D)]$.
  }
  \label{fig:rateForDiffA}
\end{figure}
% prg LambdaDeQ-diag-scaling.nb

For $E=-k^2$, we first check the method in the small $D/k^3$ (perturbative) limit for $q\neq1$~:
Fig.~\ref{fig:LambdaDeQ} shows that the agreement with the perturbative result is excellent already for $D/k^3=2/3$ (for lattice spacing $b = 0.005$ and $2\LatNb+1=2001$ sites).
The study of the case $q=1$ is more tricky as it requires to identify an exponentially small non-analytic correction.
First we have checked that the finite size effect is negligible (provided $\LatNb\times b\gg k$), as well as the role of the parameter $p_\mathrm{inj}$ (it can be changed by several orders of magnitude without modifying significantly the result).
The most important parameter is the lattice spacing $b$.
Fig.~\ref{fig:rateForDiffA} shows the different data obtained by diminishing $b$ while keeping $\LatNb\times b$ constant.
This confirms~:
\begin{itemize}
\item[(\textit{i})] 
  that $\rate=\Lambda(1)-k$ is non-analytic in the small parameter $D/k^3$,
\item[(\textit{ii})] 
  its main exponential behaviour is $\sim\exp[-4k^3/(3D)]$, corresponding to \eqref{eq:HeuristicEstimation}.
\item[(\textit{iii})] 
  We find the value $\rate=\Lambda(1)\simeq0.59\,D^{1/3}$ for $E=0$ (i.e. $m^2=0$).
\end{itemize}

\subsection{Analysis of the differential equation $\mathscr{O}_q\Phi_0^\mathrm{R}=\Lambda(q)\Phi_0^\mathrm{R}$}
\label{subsec:DiffEq}

The numerical study of the rate \eqref{eq:Rate} at large $k=\sqrt{-E}$ requires to determine the eigenvalue $\mathscr{E}_0(q)=-\Lambda(q)$ with extreme precision.
For Hermitian one-dimensional operators, such as Schr\"odinger operators, a well-known algorithm, based on the Sturm-Liouville theorem stating that the $n$-th excited state has $n$ nodes, provides the spectrum with high accuracy.
The method cannot be used for the problem of interest here because the two first eigenfunctions of the operator $\mathscr{O}_q$, denoted $\Phi_0^\mathrm{R}(z;q)$ and $\Phi_1^\mathrm{R}(z;q)$ above, \textit{are both strictly positive}.
% (and $\Phi_n^\mathrm{R}$ with $n\geq1$ have $n-1$ nodes).
The nature of these eigenvectors is discussed in~\ref{app:SpectrumGdagger}.

\begin{figure}[!ht]
\centering
\includegraphics[width=0.425\textwidth]{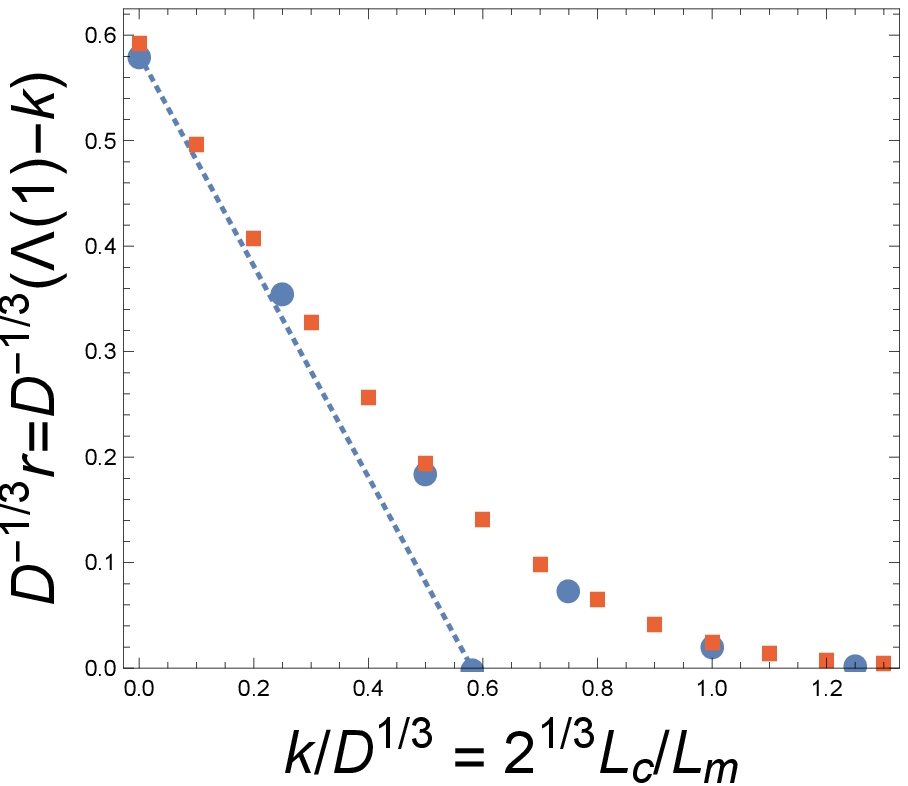}
\hspace{0.25cm}
\includegraphics[width=0.525\textwidth]{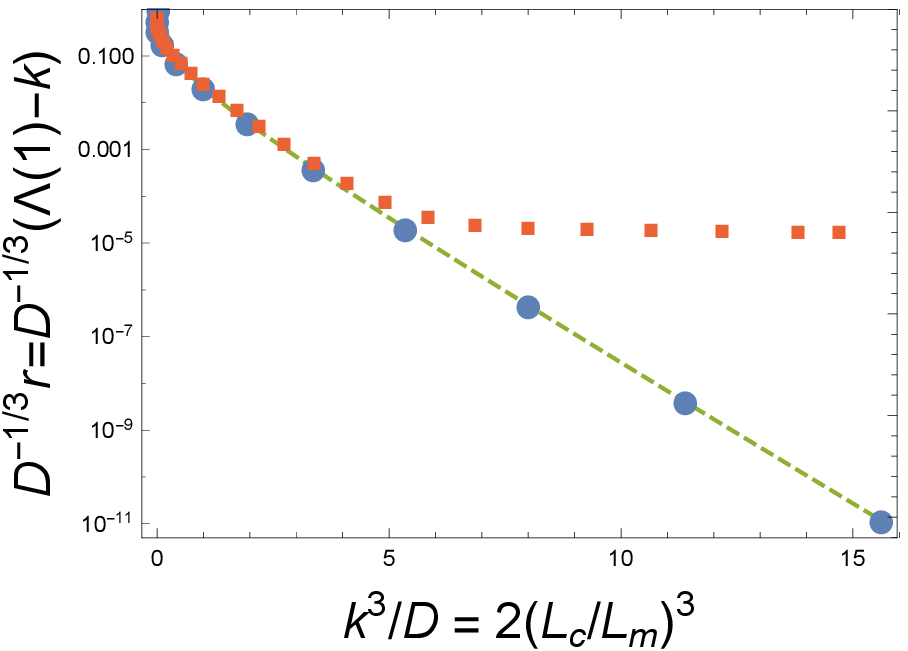}
\caption{\it
  Generalized Lyapunov exponent $\Lambda(1)$.---
  Left~: the rate $\rate=\Lambda(1)-k$ is plotted as a function of $k$ (with $E=-k^2$) in linear scale (dotted blue line is $C\,\,(2D)^{1/3}-k$).
  Right~: the rate $\rate=\Lambda(1)-k$ is plotted as a function of $k^3$ in log-linear scale. 	
  The GLE is obtained numerically by two methods~:
  diagonalization of the discretised operator $\mathscr{O}_q$, Subsection~\ref{subsec:FPDiscrete}, (red squares) and analysis of the differential equation (blue dots), Subsection~\ref{subsec:DiffEq}~; in the first case (red squares), we recall that the saturation of the data at large $k$ is a lattice effect.
  The dashed green line is $0.08D|E|^{-1}\exp[-4|E|^{3/2}/(3D)]$.
  }
  \label{fig:rate}
\end{figure}

We have found a criterion allowing to determine $\Lambda(q)=-\mathscr{E}_0(q)$ by a careful study of the function $\Phi_0^\mathrm{R}(z;q)$ obtained numerically (Fig.~\ref{fig:Phi0R}).
The starting point is the differential equation
\begin{equation}
  \label{eq:EDforPhi0R}
     \varphi''(z) + (E+z^2)\varphi'(z) + \left[(2+q)\,z - \lambda \right]\varphi(z) = 0
\end{equation}
corresponding to
$\mathscr{O}_q\,\Phi_0^\mathrm{R}(z;q) = \Lambda(q)\,\Phi_0^\mathrm{R}(z;q)$ with $\Phi_0^\mathrm{R}(z;q)\to\varphi(z)$ and for arbitrary value of the spectral parameter $\Lambda(q)\to\lambda$ (and for $D=1$).
A spectral problem is defined by specifying boundary conditions (or asymptotic behaviours).
Here, it is easy to see that the two linearly independent asymptotic behaviours of the solution of \eqref{eq:EDforPhi0R} are
$$
|z|^{-2-q}
\hspace{1cm}
\mbox{and}
\hspace{1cm}
|z|^q\,\EXP{-\potRic(z)}
\:.
$$
As argued in~\ref{app:SpectrumGdagger}, the first eigenvector $\Phi_0^\mathrm{R}(z;q)$ is the only one which behaves algebraically at infinity on both sides, so that we must select the solution with asymptotic behaviours~:
\begin{equation}
   \varphi(z) \simeq A_\pm(\lambda)\,  |z|^{-2-q}
  \quad\mbox{for}\quad z\to\pm\infty
  \:.
\end{equation}
For $q=0$, when $\Lambda(0)=0$, we have shown that the two coefficients are equal, $A_+(0)=A_-(0)$, cf. Eq.~\eqref{eq:AsympPhi0Rq0}. We have verified numerically that this property remains true for $q>0$, see an example of solution in Fig.~\ref{fig:Phi0R}.
This key observation has led us to propose the following method for the determination of the lowest eigenvalue $\mathscr{E}_0(q)=-\Lambda(q)$~:
we solve \eqref{eq:EDforPhi0R} and choose $\lambda$ large enough to get the power law behaviours $\varphi(z)\sim|z|^{-2-q}$. 
Then, decreasing progressively $\lambda$, the eigenvalue is given when the two coefficients match exactly~:
\begin{equation}
  \label{eq:QuantifForE0}
  A_-(\lambda)=A_+(\lambda)
  \quad\mbox{for}\quad
  \lambda=\Lambda(q) = -\mathscr{E}_0(q)
  \:.
\end{equation}
We have compared the numerical values obtained in this way with the one deduced by direct diagonalization of the matrix $\big(\mathscr{O}_q\big)_{n,m}$ (previous Subsection): we have obtained a perfect agreement for the smallest values of the parameter $k$ (Fig.~\ref{fig:rate}), when diagonalization is reliable (cf Fig.~\ref{fig:rateForDiffA}).
The method is sufficiently accurate to make accessible relatively large $k$ (up to $k=2.5$) leading to the precision on the eigenvalue, and thus the rate $\rate=\Lambda(1)-k$, up to~$\sim10^{-11}$.

% ctse size [-5,+5]
% lattice spacing a = 0.005 (L=1000)
% prg LambdaDeQ-diffeq2.nb

%The method allows to extract unambiguously the behavior of the pre-exponential function of the rate: in Fig. \ref{fig:preexp} we fit $r \times \exp\big[4|E|^{3/2}/(3D)\big]$
%by $\simeq 0.08 (D/|E|)$.
%
%\begin{figure}[!ht]
%\centering
%\includegraphics[width=0.6cm]{pre-exp}
%\caption{\it Pre-exponential function of the rate~$r$, extracted from the same
%numerical data as in Fig.~\ref{fig:rate} in the text. The dashed line corresponds to
%$0.08 D/|E|$.}
%  \label{fig:preexp}
%\end{figure}

We obtain the value of the GLE in the limit $E \to 0^{-}$ (i.e. $m^2 \to0^{+}$)~:
\begin{equation}
%  \rate = \Lambda(1) \simeq0.581\,D^{1/3} \simeq 0.461/L_c
  \rate = \Lambda(1) \simeq 0.58\,D^{1/3} \simeq 0.46/L_c
  \:,
\end{equation}
i.e. a value close to the one obtained by the diagonalisation (\S~\ref{subsec:FPDiscrete}).
This value corresponds to the limiting behaviour \eqref{eq:MainResultG} for $x\to\infty$.
The result also agrees with the one deduced from the first 8 cumulants for $E=0$, given in Table~1 of Ref.~\cite{SchTit02}.
Fig.~\ref{fig:rate} shows the agreement of the two numerical methods in the small $k=\sqrt{|E|}$ regime.
The $E\to0$ behaviour will also be needed later.
As the parameter $E$ enters the spectral problem $\mathscr{O}_q\Phi_0^\mathrm{R}=\Lambda(q)\Phi_0^\mathrm{R}$ as an additive constant in the drift of the diffusion operator, we expect that $\Lambda(q)$ is an analytic function of $E$ at $E=0$~; moreover,  in the localization problem, the value of  the energy $E=0$ does not play any special role in the presence of the disorder.
This is supported by the study of Ref.~\cite{RamTex14} (cf. Appendix~3 of this article).
As a consequence, the GLE presents the behaviour
\begin{equation}
  \label{eq:Lambda1SmallE}
  \Lambda(1) = C\,(2D)^{1/3} - a_1\,(2D)^{-1/3}\,E + \mathcal{O}(E^2)
  \hspace{0.5cm}\mbox{as } E\to 0
\end{equation}
where we have obtained numerically $a_1\simeq0.47$ (cf. Fig.~\ref{fig:GLEforPosNegEnergy}).
Correspondingly, we get
\begin{equation}
  \label{eq:RateSmallE}
  \rate = C\,(2D)^{1/3} - \sqrt{-E} - a_1\,(2D)^{-1/3}\,E + \mathcal{O}(E^2)
  \hspace{0.5cm}\mbox{as } E\to 0^-
\end{equation}
(cf. Fig.~\ref{fig:rate}).

In the limit $E \to -\infty$ (i.e. $m^2 \to+\infty$), we confirm once again the main exponential behaviour \eqref{eq:HeuristicEstimation}.
Moreover, the method allows to extract unambiguously the behaviour of the pre-exponential function~: cf. Fig.~\ref{fig:prefE0andE1}.
The results of the numerical calculation are compatible with
\begin{equation}
  \label{eq:ConclusionSDE1}
     \Lambda(1)= - \mathscr{E}_0(1) \simeq \sqrt{|E|} + \frac{0.08\,D}{|E|}\EXP{-4|E|^{3/2}/(3D)}
  \hspace{0.5cm}\mbox{for }
  |E|^{3/2}\gg D
  \:.
\end{equation}
%for $|E|^{3/2}\gg D$.
This asymptotic behaviour corresponds to the limiting behaviour \eqref{eq:MainResultG} for $x\to0$.
In~\ref{app:Saykin}, written by D. Saykin, it is demonstrated that the above
coefficient $0.08$ is actually $1/(4 \pi)$.

%\begin{figure}[!ht]
%\centering
%\includegraphics[width=0.35\textwidth]{pre-exp}
%\caption{\it Pre-exponential function of the rate~$r$, extracted from the same
%numerical data as in Fig.~\ref{fig:rate} in the text. The dashed line corresponds to
%$0.08 D/|E|$ (recall that $D=\sigma/2$).
%Figure from \cite{FyoLeDRosTex17}.
%}
%  \label{fig:preexp}
%\end{figure}

\subsection{Higher eigenvalues }
\label{subsec:E1}

\begin{figure}[!ht]
\centering
\includegraphics[width=8cm]{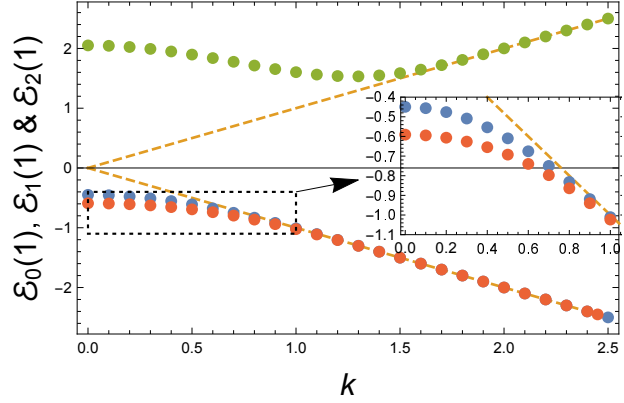}
\caption{\it The three first energies obtained by diagonalization for $q=1$.
  Straight dashed lines are $\pm k$. We have set $D=1$.}
  \label{fig:E0E1E2}
\end{figure}

The other eigenvalues of the operator $\mathscr{O}_q$, $\mathscr{E}_n(q)$ for $n>0$, are also of interest as they control the relaxation towards equilibrium.
We have obtained it by a direct diagonalization of the discretized operator $\big(\mathscr{O}_q\big)_{n,m}$~:
we have plotted the three first energies in Fig.~\ref{fig:E0E1E2}
(we have also checked that the diagonalization of the discretized operator $\big(\mathscr{H}_q\big)_{n,m}$ gives the same result).
Interestingly, we see that the two lowest energy levels become exponentially close in the large $k/D^{1/3}$ limit (cf. Fig.~\ref{fig:E0E1E2}). The higher excited eigenvalues are well separated.

%prg diag_Hq.nb

\begin{figure}[!ht]
\centering
\includegraphics[height=4cm]{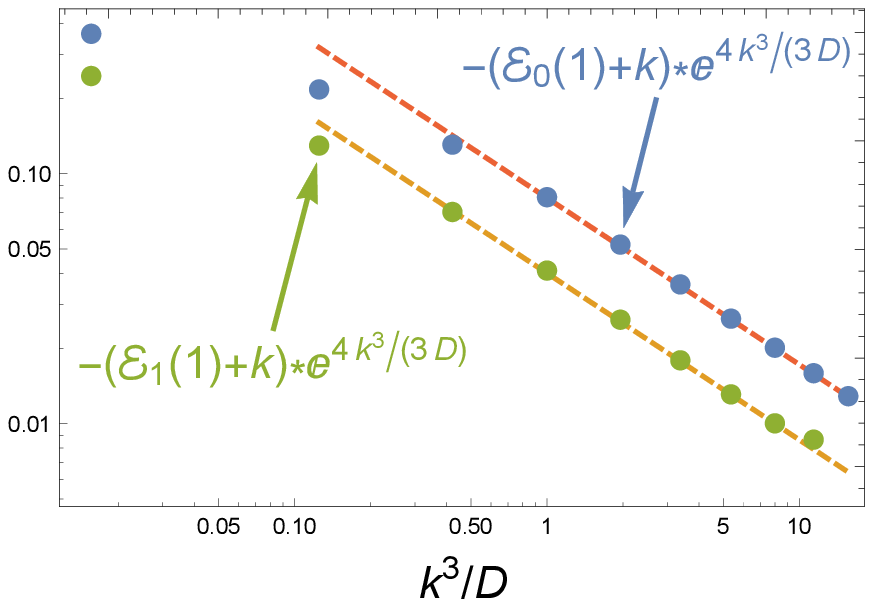}
\hspace{1cm}
\includegraphics[height=4cm]{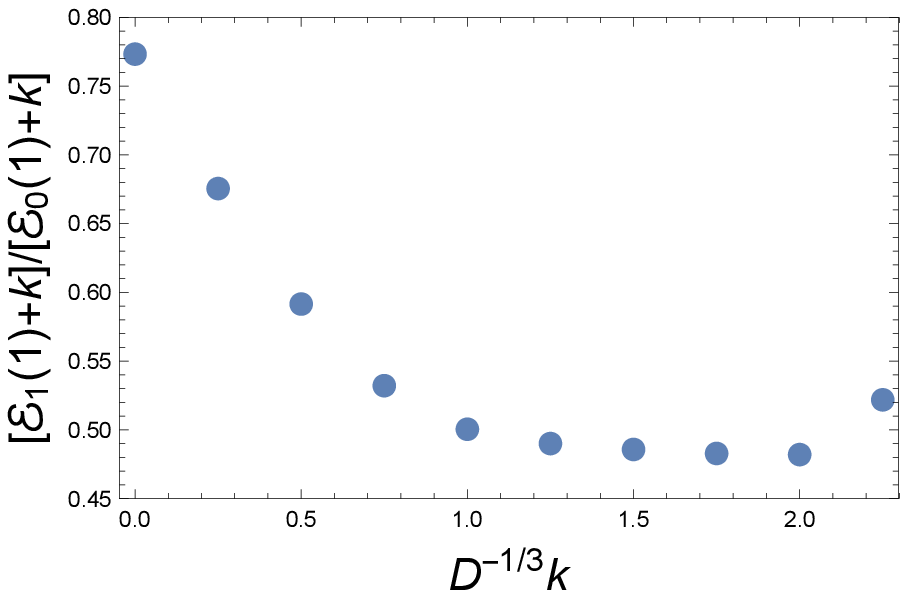}
\caption{  \textit{The two non-analytic corrections to $\mathscr{E}_0(1)$ and $\mathscr{E}_1(1)$.
The two dashed lines correspond to
$0.08 D/|E|$ and $0.04 D/|E|$ (where $E=-k^2$). }
  Right~: \textit{ratio of the two corrections.}}
\label{fig:prefE0andE1}
\end{figure}
% prg LambdaDeQ-diffeq2.nb

A more precise method is to study neatly the differential equation \eqref{eq:EDforPhi0R}~:
the first node in the solution $\varphi(z)$ appears when $\lambda\leq-\mathscr{E}_1(q)$. This allows to determine the non-analytic contribution to  $\mathscr{E}_1(q)$ for relatively large $|E|^{3/2}/D$.
Such a precise analysis of $\mathscr{E}_1(1)$ shows that the non-analytic contribution to $\mathscr{E}_1(1)$ is \textit{half} of the correction to $\mathscr{E}_0(1)$~:
\begin{equation}
  \label{eq:E1fromNumerics}
   \mathscr{E}_1(1) \simeq -\sqrt{|E|} - \frac{0.04\,D}{|E|}\EXP{-4|E|^{3/2}/(3D)}
  \hspace{0.5cm}\mbox{for }
  |E|^{3/2}\gg D
  \:,
\end{equation}
compare with Eq.~\eqref{eq:ConclusionSDE1}.
See Fig.~\ref{fig:prefE0andE1}~: 
the left plot shows the $1/|E|$ behaviour of the pre-exponential factor and the right plot exhibits the ratio $1/2$ in the large $|E|/D^{2/3}$ limit (we attribute the deviation of the last point to some numerical error).
In Section~\ref{sec:WKBE1}, it will be shown that such non-analytic contribution to $\mathscr{E}_1(1)$ can indeed be obtained by a judicious extension of the classical WKB analysis.
Furthermore, we will demonstrate that the dimensional factor is $1/(8\pi)\simeq0.04$.

We have also checked numerically that, not only for $q=1$ but for arbitrary values of $q$, the two eigenvalues become exponentially close in the large $|E|/D^{2/3}$ limit.
From the analysis of Sections~\ref{sec:WKBE1}, we conclude that the gap is 
$\mathscr{E}_1(q)-\mathscr{E}_0(q)=\mathcal{O}\big(\EXP{-4|E|^{3/2}/(3D)}\big)$ where only the pre-exponential function carries the $q$-dependence.

We can connect this spectral analysis with the study of equilibria for the elastic line~:
Fig.~\ref{fig:E0E1E2} shows that only the two lowest energies are positive and lead to exponentially growing contributions in Eq.~\eqref{eq:NtotAndSpectrum}~:
\begin{equation}
  \label{eq:FiniteSizeEffect}
  \mean{ \mathcal{N}_\mathrm{tot}} 
  \simeq
  A_0\, \EXP{\rate\,L} + A_1\, \EXP{\rate_1\,L}
  + \mathcal{O}\left(\EXP{-c_2\,L}\right)
\end{equation}
where $\rate_1=-\mathscr{E}_1(1)-\sqrt{|E|}<\rate$ and $c_2=\mathscr{E}_2(1)+\sqrt{|E|}>0$.
Moreover, in the limit $|E|^{3/2}\gg D$ ($L_m\ll L_c$), Eqs.~(\ref{eq:ConclusionSDE1},\ref{eq:E1fromNumerics}) show that $\rate_1\simeq\rate/2$.

%We have first checked the perturbative result \eqref{eq:LambdaPert}:
%we perform a numerical calculation for $D |E|^{-3/2}=8/27$ and see an excellent agreement (Fig.~\ref{fig:LambdaDeQ}).
%The rate is plotted in Fig.~\ref{fig:rate} as a function of $|E|^{3/2}/D$, which confirms the non-analytic behaviour  for small disorder $D\to0$:
%the numerical precision allows to determine the pre-exponential factor, yielding \Cred{\bf[*** Ref to SM ***]}
%$\rate=\Lambda(1) -\sqrt{|E|} \simeq 0.08 (D/|E|) \exp\big[-4|E|^{3/2}/(3D)\big]$
%(the next eigenvalue of the operator $\mathscr{O}_q$
%can be found by an elaborate WKB analysis
%and presents the same behaviour with a dimensionless factor half \Cred{\bf[*** Ref to SM ***]}, which thus provides a lower bound for the rate~$r$).
%The value of the rate in the opposite $E \to 0^{-}$ (i.e. $m^2 \to0^{+}$) limit is also of interest:
%we have found $\rate=\Lambda(1)\simeq0.581\,D^{1/3} \simeq0.461/L_c$ (Fig.~\ref{fig:rate}), a value agreeing with the one deduced from the first 8 cumulants given in Table~1 of \cite{SchTit02}.

%%%%%%%%%%%%%%%%%%%%%%%%%%%%%%%%%%%%%%%%%%%%%%%%%%%%%%%%%%%%%%%%%%%%%%%%%%%%%%%%%%%%%%%%%%
%%%%%%%%%%%%%%%%%%%%%%%%%%%%%%%%%%%%%%%%%%%%%%%%%%%%%%%%%%%%%%%%%%%%%%%%%%%%%%%%%%%%%%%%%%

\section{WKB analysis of $\mathscr{E}_1(q)$}
\label{sec:WKBE1}

%The generalized Lyapunov exponent $\Lambda(q)=-\mathscr{E}_0(q)$ studied in the article is related to the largest eigenvalue of the operator $\mathscr{O}_q=\mathscr{G}^\dagger+qz$.
In this section we study the eigenvalue $\mathscr{E}_1(q)$, which controls finite size effect (elastic line of finite length $L$), 
Eq.~\eqref{eq:FiniteSizeEffect}, and also provides a lower bound for the rate $\rate$.
The calculation confirms the $D\to0$ behaviour obtained numerically in the previous section, Eq.~\eqref{eq:E1fromNumerics}.

\subsection{The double well effective potential }

\begin{figure}[!ht]
\centering
\includegraphics[width=5cm]{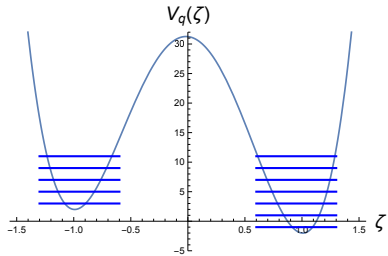}
\caption{\it The effective potential for $q=1$.
   The two spectra associated with the two harmonic wells are in correspondence for integer $q$ (apart for the $q+1$ lowest levels~; $q+1=2$ on the plot).}
\label{fig:SusyPotential}
\end{figure}

In the previous section, we have shown that the largest eigenvalue of the operator $\mathscr{O}_q$ defined by \eqref{eq:DefOq} coincides with the GLE.
It is convenient to perform a non-unitary transformation in order to relate $\mathscr{O}_q$ to a Hermitian operator $\mathscr{H}_q$, which can be interpreted as the quantum Hamiltonian for a particle in a one-dimensional double well potential, as we will see below (cf. details in \ref{app:SpectrumGdagger} and more precisely Eq.~\eqref{eq:Hq}).
Here, we analyse the ground state energy $\mathscr{E}_1(q)$ of the Hamiltonian $\mathscr{H}_q$, which coincides with the second eigenvalue of the operator $\mathscr{O}_q$. 
We find convenient to rescale the coordinate as $\zeta=|E|^{-1/2}\,z$ and the energy as $\mathcal{H}_q=|E|^{-1/2}\mathscr{H}_q$, hence we denote $\mathcal{E}_n(q)=|E|^{-1/2}\mathscr{E}_n(q)=L_m\mathscr{E}_n(q)$ ($n\in\mathbb{N}^*$) the spectrum of the new Hamiltonian
\begin{equation}
  \label{Hq}
  \mathcal{H}_q = -s \deriv{^2}{\zeta^2} + V_q(\zeta)
\end{equation}
where
\begin{equation}
   V_q(\zeta) =  \frac{1}{4s}(\zeta^2-1)^2-(q+1)\zeta
   \hspace{1cm}\mbox{and}\hspace{1cm}
   s=\frac{D}{|E|^{3/2}}
\end{equation}
%and $s=D/|E|^{3/2}$
(see Fig.~\ref{fig:SusyPotential}).
One can show that the eigenvalues of $\mathcal{H}_q$ are $\mathcal{E}_n(q)$ with $n\in\mathbb{N}^*$.
Note that the spectra of $-\mathscr{O}_q$ and $\mathscr{H}_q$ coincide apart from the eigenvalue $\mathcal{E}_0(q)$, which is in the spectrum of $\mathscr{O}_q$ but not of $\mathcal{H}_q$, as discussed in the previous Section.
Note that for $q=0$, the Hamiltonian (\ref{Hq}) has the correct expected exact non-normalizable eigenstate at zero energy.

% prg SusyPotential.nb

As we have seen in the article,  the results seem to give a strong evidence in favour of very fast, hence non-perturbative vanishing of the $\Lambda(1)=-|E|^{1/2}\mathcal{E}_0(1)$ in this regime.
Although $\Lambda(1)$ is given by an eigenvalue which does not belong to the spectrum of $\mathcal{H}_q$, we still can analyse semiclassically the ground state energy $\mathcal{E}_1(q)$ of the latter Hamiltonian,
which provides a lower bound for $\Lambda(q)=-\mathscr{E}_0(q)>-\mathscr{E}_1(q)$
(moreover the numerics has showed that the non-analytic corrections to
$\mathscr{E}_0(1)$ an $\mathscr{E}_1(1)$ only differ by a factor $1/2$ in the $s\to0$ limit~; cf. Section D).
The only way the energy $\mathscr{E}_1(q)$ of the ground state (located semiclassically
in the right well around $\zeta\approx 1$) can get such a non-perturbative shift %$\delta \mathcal{E}(1)$
is by a tunnelling admixture from the lowest-level eigenstate located semiclassically in the higher (left) potential well.

We now develop an improved WKB procedure in order to analyse the ground state
of the Schr\"odinger equation
\begin{equation}
  \label{eq:SchrodEq}
  \mathcal{H}_q \Psi(\zeta) = \mathcal{E}\, \Psi(\zeta)
\end{equation}
as $s\to 0$.
The potential $V_q(\zeta)$ has two deep minima around $\zeta=\pm 1$ separated by a big barrier around $\zeta=0$ (Fig.~\ref{fig:SusyPotential}).
The potential is well approximated by two quadratic wells near these two points
\begin{align}
  \label{Leftharmonicwell}
  V_q(\zeta)
  &\simeq  %\underset{\zeta\sim-1}{\simeq}
  \phantom{-(}q+1\phantom{)} + \frac{1}{s}(\zeta+1)^2
  \quad \mbox{for }\zeta\sim-1
\\
  \label{Rightharmonicwell}
 %V_q(\zeta)
  &\simeq   %\underset{\zeta\sim+1}{\simeq}
  -(q+1) + \frac{1}{s}(\zeta-1)^2
  \quad \mbox{for }\zeta\sim+1
  \:.
\end{align}
Before starting the presentation of the WKB method, it is useful to have in mind the correspondence with the standard formulae in quantum mechanics. This is done by identifying $s=\hbar^2/2m$~:
we can set $\hbar=1$ and $m=1/(2s)$.
We see that the classical oscillator frequency in each well in our problem is given by $\omega=2$.
%$\omega=(2/s)^{1/2}$ so that $\hbar \omega=2$.
The spectra related to the two harmonic wells (\ref{Leftharmonicwell},\ref{Rightharmonicwell}) are~:
\begin{equation}
  \begin{cases}
  \mathcal{E}_n^\mathrm{R} = 2n - q \\
  \mathcal{E}_n^\mathrm{L} = 2n+2 +q
  \end{cases}
  \hspace{0.5cm}
  \mbox{with }
  n\in\mathbb{N}
\end{equation}
which shows that the two spectra are in correspondence for integer $q$, apart for the $q+1$ lowest levels (Fig.~\ref{fig:SusyPotential}).

\subsection{Weakly asymmetric double well ($|q+1|\ll1$)}

For $q=-1$ the double well potential is symmetric and it is possible to use known results to express the energy of the ground state \cite{LanLif66c,Gar00,Son08}.
%For $q >-1$ the right well is deeper, and we eventually consider the range $q\geq 0$.
The analysis of the bottom of the spectrum can be mapped onto a simple two level problem
\begin{equation}
  \begin{pmatrix}
    \mathcal{E}_0^\mathrm{R} & -\Delta/2 \\
    -\Delta/2                & \mathcal{E}_0^\mathrm{L}
  \end{pmatrix}
\end{equation}
For $q=-1$, we have $\mathcal{E}_0^\mathrm{R}=\mathcal{E}_0^\mathrm{L}$ and the two eigenvalues are $\pm\Delta/2$.
Going back to standard notations, the semiclassical calculation of the splitting for the Hamiltonian $H=-\big[\hbar^2/(2m)\big]\partial_x^2+V(x)$, with $V(x)=(1/8)ma^2\omega^2\big[(x/a)^2-1\big]^2$, is \cite{Gar00}
$\Delta=4\sqrt{3}\hbar\omega\,\sqrt{S/(2\pi\hbar)}\,\exp[-S/\hbar]$ where $S=\int_{-a}^{a}\D x\,\sqrt{2mV(x)}$ (the formula given in \S~50 of Ref.~\cite{LanLif66c} must be corrected by a factor $\sqrt{\pi/e}$ \cite{Gar00}), which gives the ground state energy
\begin{equation}
  \label{eq:CorrectionAtResonance}
  \mathcal{E}_1(-1) \simeq 1  - \frac{\Delta}{2} \simeq 1 - \frac{4}{\sqrt{\pi s}} \, \EXP{-2/(3s)}
  \:.
\end{equation}

If $q$ is still close to $-1$ but slightly deviates such that $|\mathcal{E}_0^\mathrm{L}-\mathcal{E}_0^\mathrm{R}|=2|q+1|\gg\Delta\sim\exp[-2/(3s)]$, we can assume that the coupling is still described by the same formula and use second order perturbation formula \cite{Tex15book}.
As a consequence the shift of energy levels is now much smaller~:
whereas it was proportional to the tunnelling amplitude $\sim\exp[-S/\hbar]$ in the degenerate case, it is now proportional to the tunnelling probability $\sim\exp[-2S/\hbar]\sim\exp[-4/(3s)]$~:
\begin{equation}
  \label{eq:CorrectionAlmostResonant}
  \mathcal{E}_1(q) \simeq
  \mathcal{E}_0^\mathrm{R} - \frac{|\Delta/2|^2}{\mathcal{E}_0^\mathrm{L}-\mathcal{E}_0^\mathrm{R}}
  \underset{q\simeq-1}{\simeq} -q - \frac{2}{\pi(q+1)s} \, \EXP{-4/(3s)}
  \:.
\end{equation}

Additionally to the non-analytic contributions \eqref{eq:CorrectionAtResonance} and \eqref{eq:CorrectionAlmostResonant}, there exist also some analytic contributions due to the anharmonicity of the potential, which are not studied here (such corrections have the same origin as the contribution $-\Lambda_\mathrm{pert}(q)$ to $\mathscr{E}_0(q)$ obtained above in \S~\ref{subsec:Peturbative}).~\footnote{
  We have also been able to obtain the $\mathcal{O}(s)$ correction to $\mathcal{E}_0^\mathrm{R}$ by a perturbative treatment of the non-anharmonic terms describing the deviation from the quadratic potential at $\zeta=+1$ in the Hamiltonian $\mathcal{H}_q$.
}

%\cite{Gar00,Ras12}

\subsection{Strongly asymmetric double well ($|q+1|\gtrsim1$)}

In the ``strongly'' asymmetric potential limit, i.e. for $|q+1|\gtrsim1$, it is not anymore possible to restrict the problem to a two level problem.
We have to develop a different strategy involving two different approximation schemes~:
in the neighbourhood of each potential well ($\zeta\sim\pm1$), the potential is replaced by parabolas, which allows to express the exact solution of the approximated Schr\"odinger equation locally.
In between, inside the potential barrier, we write the approximate WKB solution of the (exact) Schr\"odinger equation and match the three expressions.
This strategy is borrowed from Ref.~\cite{Son08}.

\subsubsection{Step 1: wave function for $\zeta\sim+1$}

In the neighbourhood of the right well, the Schr\"odinger equation \eqref{eq:SchrodEq} takes the form
\begin{equation}
  \label{eq:SERW}
  \left[
      -s \deriv{^2}{\zeta^2}    -q-1 + \frac{1}{s}(\zeta-1)^2
  \right]\Psi(\zeta) \simeq \mathcal{E} \,  \Psi(\zeta)
  \:.
\end{equation}
It will be convenient to introduce the variable $\xi=(\zeta-1)/\sqrt{s}$ and parametrize the energy as
$$
  \mathcal{E} = \mathcal{E}_0^\mathrm{R}+2\epsilon = -q+2\epsilon
\:,
$$
 where $\epsilon$ is a small (negative) shift to the ground state energy of the right well.
Extracting the Gaussian function $\Psi(\zeta)=y(\xi)\,\exp[-(1/2)\xi^2]$ we obtain that the function $y(\xi)$ obeys the Hermite equation $y''(\xi)-2\xi\,y'(\xi)+2\epsilon\,y(\xi)=0$, whose solution can be expressed in terms of the Hermite function \cite{NikOuv83} with integral representation (for $\epsilon<0$)
\begin{equation}
  \label{eq:Hermite}
  H_\epsilon(\xi)
  = \frac{1}{\Gamma(-\epsilon)} \int_0^\infty \D t \, t^{-1-\epsilon} \, \EXP{-t^2 - 2\xi t}
  \:.
\end{equation}
This integral representation is suitable to extract the asymptotic behaviour
\begin{equation}
  H_\epsilon(\xi) \simeq (2\xi)^\epsilon
  \quad \mbox{ for } \xi\to+\infty
\end{equation}
and, splitting the integral in \eqref{eq:Hermite} as $\int_0^\infty =\int_{-\infty}^{+\infty} -\int_{-\infty}^0$~:
\begin{equation}
  H_\epsilon(\xi) \simeq -(-2\xi)^\epsilon
  +\frac{\sqrt{\pi}}{\Gamma(-\epsilon)} (-\xi)^{-\epsilon-1}\EXP{+\xi^2}
  \quad \mbox{ for } \xi\to-\infty
  \:.
\end{equation}
This shows that the solution of \eqref{eq:SERW},
\begin{equation}
  \label{eq:PsiR}
  \Psi(\zeta) \simeq C_R \, H_\epsilon\left( \frac{\zeta-1}{\sqrt{s}} \right)
  \,\EXP{-(\zeta-1)^2/(2s)}
  \:,
\end{equation}
decays exponentially for $\xi=(\zeta-1)/\sqrt{s}\to+\infty$ and grows exponentially for $\xi=(\zeta-1)/\sqrt{s}\to-\infty$.
Keeping the variable $\xi$, we write~:
\begin{align}
  \label{eq:AsymptoticsRWell}
  \Psi(\zeta) \simeq C_R
    \times
  \begin{cases}
    -(-2\xi)^\epsilon \, \EXP{-\xi^2/2}
     + \frac{\sqrt{\pi}}{\Gamma(-\epsilon)} (-\xi)^{-\epsilon-1}\EXP{+\xi^2/2}
      & \mbox{as } \xi\to-\infty
      \\
    (2\xi)^\epsilon \, \EXP{-\xi^2/2} & \mbox{as } \xi\to+\infty
  \end{cases}
\end{align}

\subsubsection{Step 2: wave function for $\zeta\sim-1$}

In the neighbourhood of the left harmonic well, the Schr\"odinger equation \eqref{eq:SchrodEq} takes the form
\begin{equation}
  \label{eq:SELW}
  \left[
      -s \deriv{^2}{\zeta^2}    +q+1 + \frac{1}{s}(\zeta+1)^2
  \right]\Psi(\zeta) \simeq \mathcal{E} \,  \Psi(\zeta)
  \:.
\end{equation}
It is now convenient to write the energy as
$\mathcal{E}=\mathcal{E}_0^\mathrm{L}+2\tilde\epsilon=q+2+2\tilde\epsilon$ so that $\tilde\epsilon=-(q+1)+\epsilon\simeq-(q+1)$, as we expect that $\epsilon$ is exponentially small.
We choose the solution
\begin{equation}
  \label{eq:PsiL}
  \Psi(\zeta) \simeq C_L \, H_{\tilde\epsilon}\left( -\frac{\zeta+1}{\sqrt{s}} \right)
  \,\EXP{-(\zeta+1)^2/(2s)}
  \:,
\end{equation}
which now decays in the negative direction and grows in the positive direction~:
\begin{align}
  \label{eq:AsymptoticsLWell}
  \Psi(\zeta) \simeq C_L  \times
  \begin{cases}
    (-2\xi)^{\tilde\epsilon} \, \EXP{-\xi^2/2} & \mbox{as } \xi\to-\infty
    \\
    -(2\xi)^{\tilde\epsilon} \, \EXP{-\xi^2/2}
     + \frac{\sqrt{\pi}}{\Gamma(-{\tilde\epsilon})} \xi^{-{\tilde\epsilon}-1}\EXP{+\xi^2/2}
      & \mbox{as } \xi\to+\infty
  \end{cases}
\end{align}
where now $\xi=(\zeta+1)/\sqrt{s}$.

\subsubsection{Step 3: WKB solution inside the barrier}

Inside the potential barrier, the solution of \eqref{eq:SchrodEq} is well approximated by the WKB wave function
\begin{align}
  \label{eq:PsiWKB}
  \Psi(\zeta)&\simeq
  \Psi^\mathrm{WKB}(\zeta)
  =
  \left(\frac{\sqrt{s(2q+1)}}{2s\,p(\zeta)}\right)^{1/2}
  \left[
     A\,\EXP{\int_{-1}^{\zeta}\D\zeta'\, p(\zeta')}
     +
     B\, \EXP{-\int_{-1}^{\zeta}\D\zeta'\,  p(\zeta') }
   \right]
\end{align}
where
\begin{equation}
  p(\zeta) = \sqrt{\frac{1}{s}(V_q(\zeta)-\mathcal{E})}
  \:.
\end{equation}
The validity of the WKB expression \eqref{eq:PsiWKB} extends to the domain at the left of the turning point $\zeta_0$ where $\mathcal{E}=V_q(\zeta_0)$.
Using the approximate form \eqref{Rightharmonicwell} we find
$\zeta_0 = 1 - \sqrt{s}[1+\epsilon + \mathcal{O}(\epsilon^2)] $. 
We expect that $\epsilon$ is exponentially small $\sim\EXP{-4/(3s)}$, as is indeed confirmed below, thus 
\begin{equation}
  \label{eq:turningPoint}
  \zeta_0 = 1 - \sqrt{s} +  \mathcal{O}\big(\EXP{-4/(3s)}\big)
  \:.
\end{equation}

\subsubsection{Step 4: matching}

In order to match the WKB solution \eqref{eq:PsiWKB} with \eqref{eq:PsiR} and \eqref{eq:PsiL}, it is convenient to introduce a specific notation for the action~: we define
\begin{equation}
  \label{eq:PhiLPhiR}
  \Phi_L(\zeta)=\int_{-1}^{\zeta} p(\zeta')\,\D\zeta'
  \quad\mbox{ and }\quad
  \Phi_R(\zeta)=\int_{\zeta}^{\zeta_0} p(\zeta')\,\D\zeta'
  \:.
\end{equation}
The total action associated to the trajectory going from $\zeta=-1$ to the turning point is denoted
\begin{equation}
  \label{eq:DefSq0}
  S_q = \int_{-1}^{\zeta_0} p(\zeta')\,\D\zeta'
\end{equation}
thus $S_q=\Phi_L(\zeta)+\Phi_R(\zeta)$, obviously.

\paragraph{Matching of \eqref{eq:PsiR} with \eqref{eq:PsiWKB} ---}

In the vicinity of the right harmonic well at $\zeta=+1$, the asymptotic form \eqref{eq:AsymptoticsRWell} for $\xi\to-\infty$ should match with the WKB solution \eqref{eq:PsiWKB}, which provides a first constraint on the two coefficients $A$ and $B$.
We must therefore consider $\zeta$ sufficiently far from the turning point, i.e. $(\zeta_0-\zeta)\gg\sqrt{s}$, so that the asymptotic behaviour \eqref{eq:AsymptoticsRWell} holds.
$\zeta$ must however be sufficiently close to the turning point so that the parabolic approximation (for the potential) is still justified. This last observation allows us to simplify the action $\Phi_R(\zeta)$~:
the parameter $t_*=(1-\zeta)/(1-\zeta_0)\gg 1$ will be treated as a large parameter.
Using $V_q(\zeta_0)=\mathcal{E}$ we have $p(\zeta)=(1/s)\sqrt{(1-\zeta)^2-(1-\zeta_0)^2}$.
Introducing the variable $t=(1-\zeta')/(1-\zeta_0)$ we find in this approximation:
\begin{align}
  \Phi_R(\zeta)
  &=\frac{1}{s}\int_{\zeta}^{\zeta_0}\sqrt{(\zeta'-1)^2-(\zeta_0-1)^2}  \,\D\zeta'
  \\
  \nonumber
  &=\frac{(1-\zeta_0)^2}{s}\int_{1}^{t_*}\sqrt{t^2-1}\,\D t
  =\frac{(1-\zeta_0)^2}{s}\frac{1}{2}\left(t_*\sqrt{t_*^2-1}-\mbox{arccosh}(t_*)\right)
\end{align}
and finally using $t_*\sqrt{t_*^2-1}\approx t_*^2-1/2$ and $\mbox{arccosh}(t_*)\approx \ln{(2t_*)}$, we get
\begin{align*}
  \Phi_R(\zeta) \simeq
  \frac{(1-\zeta_0)^2}{2s}
  \left[ t_*^2 - \ln(2t_*) -\frac12 \right]
  \:.
\end{align*}
We now find convenient to express $\Phi_R(\zeta)$ in terms of the variable
$\xi=(\zeta-1)/\sqrt{s}=-\big[(1-\zeta_0)/\sqrt{s}\big]t_*$ introduced above~:
\begin{equation}
  \Phi_R(\zeta) \simeq \frac12\xi^2
  - \frac{1+2\epsilon}{2}\ln\left( \frac{-2\xi}{\sqrt{1+2\epsilon}} \right)
  - \frac{1+2\epsilon}{4}
\end{equation}
from which we write the WKB wave function as
\begin{align}
  \Psi^\mathrm{WKB}(\zeta)
  &\simeq
  A \,
  (2q+1)^{1/4}\, \EXP{S_q+1/4}\,(-2\xi)^{\epsilon} \,\EXP{-\xi^2/2}
  \nonumber\\
  &+B \,
  (2q+1)^{1/4}\, \EXP{-S_q-1/4}\,(-2\xi)^{-1-\epsilon} \,\EXP{+\xi^2/2}
  \:,
\end{align}
where we made use that $\epsilon\ll1$.
Matching with \eqref{eq:AsymptoticsRWell}  for $\xi\to-\infty$ gives
\begin{align}
  A &= - \EXP{-S_q-1/4} \, C_R                   \\
  B &= -2 \epsilon\sqrt{\pi}\,\EXP{S_q+1/4}\, C_R
\end{align}
where we have used $\Gamma(-\epsilon)\simeq-1/\epsilon$.
We get the first condition
\begin{equation}
  \label{eq:MatchingCondition2}
  \frac{A}{B} =  \frac{\EXP{-1/2}}{2\epsilon\sqrt{\pi}} \, \EXP{-2S_q}
  \:.
\end{equation}

\paragraph{Matching of \eqref{eq:PsiL} with \eqref{eq:PsiWKB} ---}

We proceed the same way in the neighbourhood of the left harmonic well and match the asymptotic behaviour  \eqref{eq:AsymptoticsLWell} for $\xi\to+\infty$ with the WKB approximation \eqref{eq:PsiWKB}.
In the region where the two expressions of the wave function match, we can use the parabolic approximation for the potential~:
\begin{equation}
  \label{PhiLA}
  \Phi_L(\zeta)=\frac{1}{s}\int_{-1}^{\zeta}\sqrt{(2q+1)s+(\zeta'+1)^2}  \,\D\zeta'
\end{equation}
 (we recall that we can neglect $\epsilon$ there).
Introducing the variable $t=(\zeta'+1)/\sqrt{s(2q+1)}$, this rewrites
\begin{align}
\Phi_L(\zeta)=(2q+1)\int_{0}^{t_*}\sqrt{1+t^2}\,\D t
=(2q+1)\frac{1}{2}\left(t_*\sqrt{1+t_*^2}+\mbox{arcsinh}(t_*)\right)
\end{align}
\eqref{eq:PsiWKB} and \eqref{eq:AsymptoticsLWell} match in the regime where $t_*=(\zeta+1)/\sqrt{s(2q+1)}\gg 1$, thus, using $t_*\sqrt{1+t_*^2}\approx t_*^2+1/2$ and $\mbox{arcsinh}(t_*)\approx \ln{(2t_*)}$, we find
\begin{align}
  \Phi_L(\zeta) \simeq
  \left(q+\frac{1}{2}\right)
  \left( t_*^2 + \ln(2t_*) + \frac{1}{2} \right)
  \:.
\end{align}
At this stage it is convenient to use the variable
$\xi=(\zeta+1)/\sqrt{s}=\sqrt{(2q+1)}t_*$
introduced above~:
\begin{align}
  \Psi^\mathrm{WKB}(\zeta)
  &\simeq
  A\, \EXP{(2q+1)/4}\left(\frac{2\xi}{\sqrt{(2q+1)}}\right)^q\EXP{+\xi^2/2}
%  \nonumber
%  \\
%  &
  +
  B\, \EXP{-(2q+1)/4}\left(\frac{2\xi}{\sqrt{(2q+1)}}\right)^{-q-1}\EXP{-\xi^2/2}
\end{align}
which matches with \eqref{eq:AsymptoticsLWell} for $\xi\to+\infty$ (we recall that $\tilde{\epsilon}\simeq-q-1$) if
\begin{align}
  A &= \frac{2^{-q}\sqrt{\pi}}{\Gamma(q+1)}(2q+1)^{q/2}\EXP{-(2q+1)/4} \, C_L \\
  B &= -(2q+1)^{-(q+1)/2}\EXP{(2q+1)/4} \, C_L
\end{align}
We deduce the second condition
\begin{equation}
  \label{eq:MatchingCondition1}
  \frac{A}{B} =  - \frac{2^{-q}\sqrt{\pi}}{\Gamma(q+1)}(2q+1)^{q+1/2}\EXP{-(2q+1)/2}
  \:.
\end{equation}

It is worth emphasizing the result $A\sim B$, obtained from the improved WKB method developed here. The naive (standard) WKB method would have given $B|_\mathrm{naive\:WKB}=0$, since the only term of \eqref{eq:PsiWKB} which is vanishing as $\zeta\to-\infty$, in the classically forbidden region, is the term with coefficient~$A$.

\subsubsection{Ground state energy}

Comparing \eqref{eq:MatchingCondition2} and \eqref{eq:MatchingCondition1} finally provides the expression of the shift of the ground state energy
 \begin{equation}
   \label{shift}
   \epsilon
   \simeq
   -  \frac{\Gamma(q+1)\,\EXP{q}}{2\pi\sqrt{2} (q+1/2)^{q+1/2}  } \, \EXP{-2S_q}
 \end{equation}
Making use of \eqref{eq:turningPoint}, we can write the action \eqref{eq:DefSq0} as 
\begin{equation}
  \label{eq:DefActionWKB}
  S_q = \frac{1}{2s}
  \int_{-1}^{1-\sqrt{s}}
 \D\zeta\,
 \sqrt{(1-\zeta^2)^2-4s(q+1)\zeta+4sq\,(1-\sqrt{s})-s^2}
 \:.
\end{equation}
In \ref{app:cq}, we show that it presents the limiting behaviour 
\begin{equation}
  \label{eq:ActionWKB}
  S_q
  \underset{s\to0}\simeq
  \frac{2}{3s} -  \frac{q}{2} \,\ln s + c_q
  +\mathcal{O}(s^{1/2}\ln s)
  \:,
\end{equation}
where 
\begin{equation}
  \label{eq:cq}
  2c_q=q+4q\ln2-\left(q+\frac{1}{2}\right)\ln(2q+1)
  \:.
\end{equation}
%where the constant $c_q$ is studied numerically ($c_1\simeq1.05$).
We conclude that the ground state energy of the Schr\"odinger operator $\mathcal{H}_q$ is given by
\begin{align}
  \label{final0}
  \mathcal{E}_1(q)= -q + 2\epsilon
  \simeq -q
  - \frac{\Gamma(q+1)\,\EXP{-2c_q+q}}{\pi\sqrt{2}\, (q+1/2)^{q+1/2}  }    \,
   s^{q}\, \EXP{-4/(3s)}
\end{align}
which presents a different pre-exponential dependence, compared to \eqref{eq:CorrectionAlmostResonant} obtained for weakly asymmetric double-well.
Going back to the initial notation
\begin{align}
  \label{final}
   \mathscr{E}_1(q) \simeq - q\sqrt{|E|}
   - 
     \frac{ \Gamma(q+1) \,D^q }{ 2^{3q}\pi\,|E|^{(3q-1)/2} } 
     \, \EXP{-4|E|^{3/2}/(3D)}
  \:.
\end{align}
It is however important to remember that (\ref{final0},\ref{final}) are \textit{not} the full result, but only the \textit{non-analytic} contribution to the ground state energy (which is not expected to be the dominant correction).
As discussed at length in the main text and in Section~\ref{sec:GLEspectral}, the ground state energy is also shifted by analytic contributions (in $s$) related to the non-harmonicity of the potential. 
As explained above, the analytic terms are also given by \eqref{eq:LambdaPert}~:
\begin{equation}
\mathscr{E}_1(q)\simeq-\Lambda_\mathrm{pert}(q)+\mathcal{O}(\EXP{-4|E|^{3/2}/(3D)})
\:.
\end{equation}
%(this is the physical content of the perturbative terms contributing to $\mathscr{E}_0(q)$, cf. subsection~\ref{subsec:Peturbative}).

For $q=1$, we have demonstrated that the first analytic contributions vanish (up to $\mathcal{O}(s^2)$) and observed numerically that this is true at all orders in $s$.
Hence the final result for $q=1$ is expected to be
\begin{equation}
  \mathscr{E}_1(1)+\sqrt{|E|}
  \simeq - 
 \frac{D}{8\pi|E|}
  \, \EXP{-4|E|^{3/2}/(3D)}   
  \:.
\end{equation}
Note that
not only the power law of the pre-exponential term perfectly agrees with the numerical result, cf. Eq.~\eqref{eq:E1fromNumerics}, but moreover the dimensionless factor $1/(8\pi)\simeq0.0397887$ coincides with the value extracted numerically in Subsection~\ref{subsec:E1}.
%Nonetheless, we recall that the generalized Lyapunov exponent $\Lambda(q)$ is not related to this eigenvalue, but is controlled by the largest eigenvalue $-\mathscr{E}_0(q)$ of the operator~$\mathscr{O}_q$.

%%%%%%%%%%%%%%%%%%%%%%%%%%%%%%%%%%%%%%%%%%%%%%%%%%%%%%%%%%%%%%%%%%%%%%%%%%%%%%%%%%%%%%%%%%
%%%%%%%%%%%%%%%%%%%%%%%%%%%%%%%%%%%%%%%%%%%%%%%%%%%%%%%%%%%%%%%%%%%%%%%%%%%%%%%%%%%%%%%%%%

\section{The generalized Lyapunov exponent for even integer argument}
\label{sec:LambdaEvenQ}

A central quantity of the paper is the generalized Lyapunov exponent $\Lambda(q)$ for a particular value of its argument, $q=1$.
In this section, we show that when its argument is an even integer, $q=2n$ with $n\in\mathbb{N}^*$, the analysis can be greatly simplified. 
The main idea is to consider $\mean{y(\tau)^q}$ instead of $\mean{|y(\tau)|^q}$,  
%where $y(\tau)$ is the solution of the Cauchy problem $-y''(\tau)+U(\tau)y(\tau)=Ey(\tau)$. Obviously the two quantities 
which obviously coincide for $q=2n$.
The method presented in this section is based on the observation that the correlation functions of the form $\smean{y(\tau)^ny'(\tau)^m}$ with $n+m>0$ can be analysed through a closed set of $n+m+1$ coupled linear stochastic differential equations. 
In relation with the functional determinants of stochastic operators, this means that, in the present Section, we consider $\mean{y(L)^q}=\mean{\det(-E-\partial_\tau^2+U(\tau))^q}$ instead of $\mean{|y(L)|^q}=\mean{|\det(-E-\partial_\tau^2+U(\tau))|^q}$.

To be concrete we introduce
\begin{equation}
  \label{eq:DefGLEtilde}
  \LamNoA(q) := \lim_{L \to \infty} \frac{1}{L}\ln\mean{ y(L)^q  }
\end{equation}
with $q\in\mathbb{N}^*$, to be compared with \eqref{eq:DefGLE}.
The case $q=1$ already emphasizes the difference between $\Lambda(q)$ and $\LamNoA(q)$~:
by using the technique exposed below, it is rather easy to verify that $\mean{y(L)}=\sinh(\sqrt{-E}L)/\sqrt{-E}$ is independent of the presence of the disorder. 
Note that a simple explanation of the fact that $\mean{y(L)}=\mean{\det(-E-\partial_\tau^2+U(\tau))}=\det(-E-\partial_\tau^2)$ is independent of the disorder was given above for the discrete model, cf. Eq.~\eqref{eq:MeanDetWithoutAbsoluteValue}.
We deduce
\begin{equation}
 \LamNoA(1) = \sqrt{-E}
 \hspace{0.5cm}\mbox{for }
 E<0
 \:,
\end{equation}
whereas $\Lambda(1) = \sqrt{-E}+\mathcal{O}(\EXP{-4|E|^{3/2}/(3D)})$, Eq.~\eqref{eq:ConclusionSDE1}.
For $E=0$, the difference is more stricking since $\LamNoA(1) = 0$ whereas the GLE is finite $\Lambda(1)\simeq0.58\,D^{1/3}$.
The non-analytic part in $\Lambda(1)$ is therefore related to the presence of the absolute value in the average $\mean{|y(\tau)|}$.

\subsection{Warming up~: exact calculation of $\Lambda(2)=\LamNoA(2)$}

\begin{figure}[!ht]
\centering
\includegraphics[width=7.5cm]{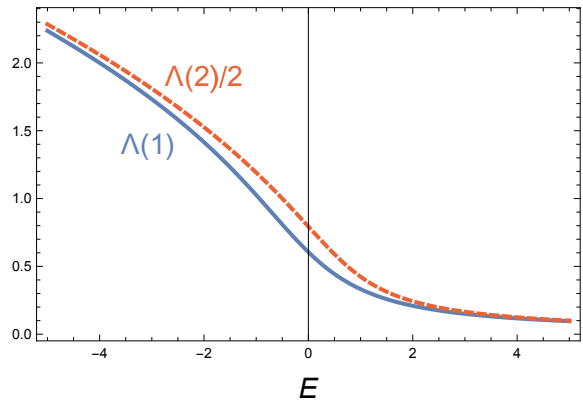}
\caption{\it Comparison between $\Lambda(1)$, obtained numerically (Subsection~\ref{subsec:FPDiscrete}), and $\Lambda(2)/2$, given by Eq.~\eqref{eq:Lambda2bis}~; we set $D=1$.}
\label{fig:Lambda2}
\end{figure}

We recall that the starting point is the Schr\"odinger equation 
\begin{equation}
  - y''(\tau) + U(\tau)\,y(\tau) = E\,   y(\tau)
  \:.
\end{equation}
Let us start from the simplest non trivial case $q=2$ which requires to analyse $\mean{y(\tau)^2}$.
Differentiating $y^2$ leads to consider the set of three coupled SDEs
\begin{equation}
\begin{cases}
  \deriv{}{\tau}y(\tau)^2&= 2y(\tau)\,y'(\tau)
  \\
  \deriv{}{\tau}y(\tau)\,y'(\tau)&= (U(\tau)-E)\, y(\tau)^2 + y'(\tau)^2
  \\
  \deriv{}{\tau}y'(\tau)^2&= 2(U(\tau)-E)\, y(\tau) \,y'(\tau)
\end{cases}
\hspace{0.5cm}\mbox{(Stratonovich)}
\end{equation}
which are interpreted in the Stratonovich convention, as usual in physics when Gaussian white noises arise  in order to model regular physical noises~\cite{Gar89}.
The SDEs can be rewritten in the It\^o convention as
\begin{equation}
\begin{cases}
  \deriv{}{\tau}y(\tau)^2&= 2y(\tau)\,y'(\tau)
  \\
  \deriv{}{\tau}y(\tau)\,y'(\tau)&= (U(\tau)-E)\, y(\tau)^2 + y'(\tau)^2
  \\
  \deriv{}{\tau}y'(\tau)^2&= 2(U(\tau)-E)\, y(\tau) \,y'(\tau) +2D\,y(\tau)^2
\end{cases}
\hspace{0.5cm}\mbox{(It\^o)}
\end{equation}
leading to
\begin{equation}
  \deriv{}{\tau}
  \begin{pmatrix}
     \mean{y^2} \\ \mean{y\,y'} \\ \mean{y'^2}
  \end{pmatrix}
  =
  \begin{pmatrix}
    0 & 2 & 0 \\
    -E & 0 & 1 \\
    2D & -2 E & 0
  \end{pmatrix}
  \begin{pmatrix}
     \mean{y^2} \\ \mean{y\,y'} \\ \mean{y'^2}
  \end{pmatrix}
  =: M_2
  \begin{pmatrix}
     \mean{y^2} \\ \mean{y\,y'} \\ \mean{y'^2}
  \end{pmatrix}
\end{equation}
The largest eigenvalue of the matrix $M_2$ controls the exponential growth of $\mean{y(\tau)^2}$ and therefore coincides with $\LamNoA(2)=\Lambda(2)$.
The characteristic polynomial is
$
  P_2(\lambda) = \det(M_2 - \lambda\,\identity_3) = - \lambda^3 - 4E\,\lambda + 4D
$,
whose appropriate root is (for $E=-k^2<0$)
\begin{align}
  \label{eq:Lambda2}
  \Lambda(2) %= \LamNoA(2) 
  &= k
  \left[
    \left(2s+2\sqrt{s^2-16/27}\right)^{1/3}
    +
    \frac{4/3}{\left(2s+2\sqrt{s^2-16/27}\right)^{1/3}}
  \right]
  \mbox{ with } s=\frac{D}{k^3}\:,
  \\
  \label{eq:Lambda2bis}
  &
  =
  D^{1/3}
  \left[
    \left(2+2\sqrt{1+16E^3/(27D^2)}\right)^{1/3}
    -
    \frac{4E/(3D^{2/3})}{\left(2+2\sqrt{1+16E^3/(27D^2)}\right)^{1/3}}
  \right]
\end{align}
%where $s=D/k^3$.
We thus have recovered a result of~\cite{ZilPik03}.
For $D=0$ we have $\Lambda(2)=2k$ as it should, and for $E=0$, one gets $\Lambda(2) = (4D)^{1/3}$.
One can also check that the $s\to0$ expansion coincides with \eqref{eq:LambdaPert} for $q=2$,
which corresponds to the limit $E\to-\infty$.
For large positive energy $E\gg D^{2/3}$, we have $\Lambda(q)\simeq (q+q^2/2)\,\gamma_1\simeq (q+q^2/2)\,D/(4E)$ (cf. Section~\ref{sec:IntroGLE})~:
we can check that the expansion of \eqref{eq:Lambda2bis} for $E\to+\infty$ gives $\Lambda(2)\simeq D/E$, as it should.
The GLE is plotted in Fig.~\ref{fig:Lambda2} for $q=1$ and $q=2$~:
in order to match the asymptotic behaviours, $\Lambda(q)\simeq q\sqrt{-E}$ for $E\to-\infty$, we choose to plot $\Lambda(q)/q$.

\subsection{Analysis of $\LamNoA(q)$ with $q\in\mathbb{N}^*$}

We can generalise the above calculation to the case $q\in\mathbb{N}^*$. Clearly, this leads to consider a linear system with $q+1$ equations, controlled by a $(q+1)\times(q+1)$ matrix $M_{q}$, given below, whose largest eigenvalue coincides with $\LamNoA(q)$.
When $q=2n$ is an \textbf{even} integer, we have furthermore $\Lambda(2n)=\LamNoA(2n)$.
Because the eigenvalue $\LamNoA(q)$ is the root of a polynomial of degree $q+1$, it is necessary analytic in the disorder parameter $D$, which appears in the coefficients of the matrix.
Hence, throughout the paper, we have made the two remarkable observations~:
\begin{itemize}
\item
  for $q=1$, the analytic part of $\Lambda(q)-kq$ vanishes and $\Lambda(1)$ is a purely non-analytic function of $s=D/|E|^{3/2}$ (Section~\ref{sec:GLEspectral}).

\item
  For $q=2n$, the non-analytic contributions to $\Lambda(q)$ vanish and the GLE is analytic in $s=D/|E|^{3/2}$.
\end{itemize}
In particular, this shows that the replica trick must be used with caution as it is not possible to deduce 
$\smean{|y(\tau)|^{q}}$ for arbitrary real $q$ from $\smean{y(\tau)^{2n}}$. 
%\\
%\Cred{\textbf{Ch} : Pierre, in Ref.~\cite{BouGeoHanLeDMai86} you wrote that the analytical continuation $q\to0$, \og \textit{when possible} \fg{},  gives the Lyapunov exponent.... **** {\sc what are the conditions ??? }****}

\subsubsection{Linear system of SDEs}

Let us now discuss systematically the calculation of $\Lambda(2n)$.
For $q=2n$, one has to consider the $q+1$ equations
\begin{align}
  \deriv{}{\tau} y(\tau)^{q-m}\,y'(\tau)^{m}
  =&m\,(U(\tau)-E)\,y(\tau)^{q-m+1}\,y'(\tau)^{m-1}
  \\\nonumber
  &+ (q-m)\,y(\tau)^{q-m-1}\,y'(\tau)^{m+1}
  \hspace{0.5cm}\mbox{(Stratonovich)}
\end{align}
for $m=0,\,1,\cdots,\,q$.
The stochastic differential equation can be rewritten in the It\^o convention~:
\begin{align}
  &\deriv{}{\tau} 
  y(\tau)^{q-m}\,y'(\tau)^{m}
  =m\,(U(\tau)-E)\,y(\tau)^{q-m+1}\,y'(\tau)^{m-1} 
  \\\nonumber
  &  + (q-m)\,y(\tau)^{q-m-1}\,y'(\tau)^{m+1}
+m(m-1)\,D\,y(\tau)^{q-m+2}\,y'(\tau)^{m-2}
\hspace{0.5cm}\mbox{(It\^o)}
\end{align}

This leads to consider the linear system
\begin{equation}
  \deriv{}{\tau}
  \begin{pmatrix}
     \mean{y^{q}} \\
     \mean{y^{q-1}\,y'} \\
     \mean{y^{q-2}\,y'^2}  \\
     \vdots \\
     \mean{y'^{q}}
  \end{pmatrix}
  =
  M_{q}
  \begin{pmatrix}
     \mean{y^{q}} \\
     \mean{y^{q-1}\,y'} \\
     \mean{y^{q-2}\,y'^2}  \\
     \vdots \\
     \mean{y'^{q}}
  \end{pmatrix}
\end{equation}
where the $(q+1)\times(q+1)$ matrix is
\begin{equation}
  M_{q}=
  \begin{pmatrix}
    0    & q   &   0   & 0   & 0   & \cdots & 0
    \\
    -E   & 0    & q-1  & 0   & 0   & \cdots & 0
    \\
    2D& -2E  & 0     & q-2 & 0  & \cdots  & 0
    \\
    0 &6D   & -3E  & 0     & \ddots  & \cdots &\vdots
    \\
    0 & 0 &12D   & -4E     & \ddots  &   &
    \\
    \vdots &  \vdots & \ddots &\ddots &  \ddots &\ddots &
    \\
    0 & 0 & \cdots & & & 0& 1
    \\
    0 & 0 & \cdots  & &q(q-1)\,D & -q\,E&0
  \end{pmatrix}
  \:.
\end{equation}
The idea that the case of integer $q$ leads to consider a closed system of equations for correlators of $y$ and $y'$ has been used earlier \cite{MalMar02,SchTit02,ZilPik03} (the matrix $M_q$ was obtained in Refs.~\cite{SchTit02,ZilPik03}).

\subsubsection{Perturbative analysis ($D/k^3\ll1)$}

As a check, let us recover the perturbative result \eqref{eq:LambdaPert}.
For $E=-k^2$ and in the absence of disorder, the matrix $M_q$ has the spectrum
$$
\mathrm{Spec}(M_q) = \{ -qk,\, (-q+2)k,\, \cdots, -2k,\, 0,\, 2k,\, \cdots , +qk \}
\hspace{0.5cm}\mbox{for }
D=0
\:.
$$
The largest eigenvalue $\LamNoA(q)=+qk$ is associated with the right and left eigenvectors
\begin{equation}
  R = 2^{-q}
  \begin{pmatrix}
  k^{-q} \\ k^{-q+1} \\ \vdots \\ k^{-1} \\ 1
  \end{pmatrix}
  \hspace{1cm}\mbox{and}\hspace{1cm}
  L = \begin{pmatrix}
  C_q^0\,k^{q} \\ C_q^1\,k^{q-1} \\ \vdots \\ C_q^{q-1}\,k \\ C_q^q
  \end{pmatrix}
\end{equation}
with obviously $L^\mathrm{T}\cdot R=1$.
Perturbation theory allows to get the $\mathcal{O}(D)$ correction to the eigenvalue~:
we write the perturbation to $M_q^{(0)}=M_q|_{D=0}$ as
$W_{n,m}=\delta_{n,m+2}\,(n-1)(n-2)\,D$.
Then
\begin{align}
  \LamNoA(q) &= qk + L^\mathrm{T}\cdot W\cdot R
  +\mathcal{O}(D^2)
%  =qk + 2^{-q}\sum_{n=3}^{q+1}C_{q}^{n-1}k^{q-n+1}\,\sigma\,\frac{(n-1)(n-2)}{2}\,k^{-q+n-3}  +\mathcal{O}(\sigma^2)
  \\\nonumber
  &=qk + \frac{D}{k^2}2^{-q}
%  \underbrace{
  \sum_{n=2}^{q}n(n-1)C_{q}^{n}
%  }_{   =\deriv{^2}{u^2}(1+u)^q\big|_{u=1} =q(q-1)2^{q-2}   }
  +\mathcal{O}(D^2)
  = qk + \frac{D}{4k^2}q(q-1) + \mathcal{O}(D^2)
  \:.
\end{align}
The result is in perfect agreement with \eqref{eq:LambdaPert}. 
We emphasize that only for even integer $q=2n$ does the systematic perturbative expansion provide the exact result, without any additional non-analytic contribution in $D/k^3$~: according to the notation of Section~\ref{sec:GLE} we can write
\begin{equation}
  \Lambda(2n) = \LamNoA(2n) = \Lambda_\mathrm{pert}(2n)
  \:.
\end{equation}

\subsubsection{Zero energy $E=-k^2=0$}

The determination of $\LamNoA(q)$ reduces to analyse the largest eigenvalue of a $(q+1)\times(q+1)$ matrix $M_{q}$, which is easy to implement.
For $E=0$, we can find the analytic expression of $\LamNoA(q)$ up to $q=6$ (see Table~\ref{tab:Lambda2n}).

\begin{table}[!ht]
\centering
\begin{tabular}{lcl}
$q$ & $\LamNoA(q)$ for $E=0$ &
\\
\hline
\\[-0.25cm]
$1$ & $0$ &
\\[0.125cm]
$2$ &  $(4D)^{1/3}$ & $=\Lambda(2)$
\\[0.125cm]
$3$ &  $(24D)^{1/3}$ & 
\\[0.125cm]
$4$ &  $(84D)^{1/3}$ & $=\Lambda(4)$
\\[0.125cm]
$5$ &  $(112+24\sqrt{19})^{1/3}\,D^{1/3}$ & 
\\[0.125cm]
$6$ & $(252+24\sqrt{79})^{1/3}\,D^{1/3}$ &  $=\Lambda(6)$
\\
$\vdots$&$\vdots$ \hspace{1cm} $\vdots$ &$\vdots$
\\
$\gg1$  &$\simeq (3/4)\,q^{4/3}\,D^{1/3} $
  & $\simeq\Lambda(q)$
\\[0.125cm]
\hline
\end{tabular}
\caption{\it Generalized Lyapunov exponent $\Lambda(q)$ for even integer argument.}
\label{tab:Lambda2n}
\end{table}

\begin{figure}[!ht]
\centering
\includegraphics[width=8cm]{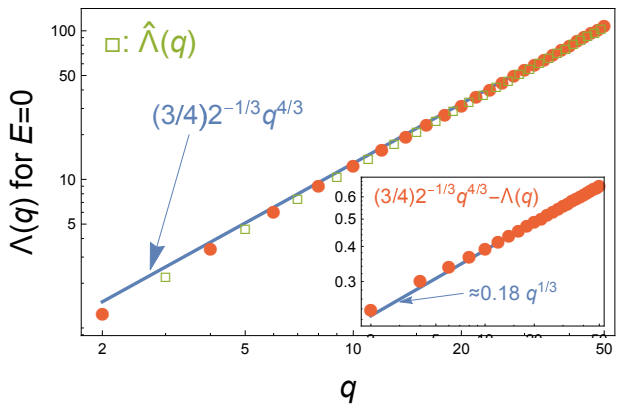}
%\includegraphics[width=0.475\textwidth]{gle-for-q-integer}
%\hfill
%\includegraphics[width=0.475\textwidth]{gle-for-q-integer-corr}
\caption{
\textit{Generalized Lyapunov exponent $\Lambda(q)$ for integer $q=2n$ (red dots) obtained numerically by finding the largest eigenvalue of the matrix $M_q$ (squares correspond to even integers). 
Green squares correspond to $\LamNoA(q)$ for positive odd integers.
Line is the asymptotic behaviour $\sim q^{4/3} $ given in the text.}
Inset~:
\textit{Subleading term (straight line is $0.18\,q^{1/3}$).
We set $2D=1$.}
}
\label{fig:gle-for-q-integer}
\end{figure}

This simple method also allows us to study the large $q$ behaviour.
We obtain numerically (Fig.~\ref{fig:gle-for-q-integer})
\begin{equation}
   \label{eq:GLEforLargeQ}
  \Lambda(q) \simeq \frac34\, q^{4/3} \,D^{1/3} - c_\mathrm{even}\,(2q\,D)^{1/3}
\end{equation}
with $c_\mathrm{even}\simeq0.18$ (cf. Fig.~\ref{fig:gle-for-q-integer}).
Although this behaviour was obtained here for even integer arguments, we expect a smooth dependence as a function of $q$ (as it was observed for large negative energy, Fig.~\ref{fig:LambdaDeQ}).
This is supported by comparing $\mathscr{E}_0(q)=-\Lambda(q)$ with the ground state energy $\mathscr{E}_1(q)$ of the Hamiltonian \eqref{eq:Hq} (or \eqref{Hq}, up to a rescaling), which provides a lower bound for the GLE.
A simple harmonic approximation around the minimum of the double well potential, at $z\simeq(Dq)^{1/3}$, gives
\begin{equation}
   \mathscr{E}_1(q)\simeq- \frac34 \,q^{4/3} \,D^{1/3}  +c_\mathscr{H} \,(2q\,D)^{1/3}
\end{equation}
with $c_\mathscr{H}=2^{-4/3}\sqrt{6}\simeq0.97$.
As was expected, only the subleading terms of the two eigenvalues differ.
We have $c_\mathscr{H}>c_\mathrm{even}$ as it should, since $\mathscr{E}_1(q)>\mathscr{E}_0(q)=-\Lambda(q)$.

Finally, we note that the main behaviour $\Lambda(q)\sim D^{1/3}q^{4/3}$ was obtained for integer $q$ in the conference's proceedings~\cite{BouGeoLeD86}.
It is also in agreement with the numerical calculations of $L(q)=\Lambda(q)/q\sim q^{\alpha-1}$ of Zillmer \& Pikovsky \cite{ZilPik03}, who obtained the exponents $\alpha\simeq1.28$ and $\alpha\simeq1.38$ for two different values of the energy.

% prg gle-for-q-integer.nb

\subsection{Large deviations for the wave function amplitude}

As it is clear from its definition \eqref{eq:DefGLE}, the generalized Lyapunov exponent is the generating function of the cumulants of the logarithm of the wave function $\ln|y(x)|$ [more precisely, $y(x)$ is the solution of the Cauchy initial value problem (\ref{eq:CauchyIntialValue},\ref{eq:yDiffEq})].
As such, the GLE can be related to the large deviation function $\Phi(u)$ controlling the distribution
\begin{equation}
  \mathcal{Q}_L(\Upsilon) = \mean{\delta(\Upsilon-\ln|y(L)|)}
  \underset{L\to\infty}{\sim}\exp\left\{ - L\,\Phi\left(\Upsilon/L\right)\right\}
  \:.
\end{equation}
$\Lambda(q)$ and $\Phi(u)$ are related by a Legendre transform
\begin{equation}
  \Phi(u) = q_*\,u - \Lambda(q_*)
  \hspace{0.5cm}\mbox{where }
  u = \Lambda'(q_*)
  \:.
\end{equation}
We can therefore relate the behaviour \eqref{eq:GLEforLargeQ} for $q\to+\infty$ to the asymptotic behaviour of the large deviation function
\begin{equation}
  \label{eq:TailLDF}
  \Phi(u)\simeq \frac{1}{4D}u^4  \hspace{0.5cm}\mbox{for } u\to+\infty
\end{equation}
characterizing the tail of the distribution of $\ln|y(L)|$, i.e. for $|y(L)|\to\infty$ (the tail for $\ln|y(L)|\to-\infty$, i.e. $|y(L)|\to0$, should be related to $\Lambda(q)$  for $q\to-\infty$, which we have not studied).
To conclude, we comment on the scaling of the fluctuations with the length $L$.
The typical fluctuations, encoded in the variance of $\ln|y(L)|$, display the standard behaviour 
$
\big[\ln|y(L)|-\gamma_1L\big]_\mathrm{typ}\sim(\gamma_2L)^{1/2}\sim(D^{1/3}L)^{1/2}
$,
characteristic of the central limit theorem (which can be applied to $\ln|y(L)|=\int_0^L\D t\,z(t)$ because $z(t)$ decorrelates over a finite length scale).
On the other hand, the behaviour \eqref{eq:TailLDF} shows that some rare events are responsible for atypically large fluctuations characterized by a different scaling
$
\big[\ln|y(L)|\big]_\mathrm{atyp}\sim (D^{1/3} L)^{3/4}
$.

\section{Mean number of equilibria at fixed value of the energy and annealed distribution of the energy of the line over the set of equilibria}
\label{sec:NbForFixedLevel}

\subsection{Definitions and determinant formulae for Laplace transforms}

Here we show how our method can be extended to study the  mean number of equilibria {\it at fixed values of the energy}, defined, for our discrete model of an elastic line, as 
\begin{align}
\label{eq:DefNtotDeH}
 \mean{ \mathcal{N}_\mathrm{tot}(H) }
  =
  \int_{\mathbb{R}^K}  \D{\bf u} \, 
  \mean{  
   \delta(H-\mathcal{H}({\bf u}))
     \,
   |\det \big(\partial_i\partial_j\mathcal{H}\big)|  \, 
   \prod_{i=1}^K \delta\left(\partial_i \mathcal{H}\right)
   }
\end{align} 
It is in fact convenient to split the total energy \eqref{eq:Hdiscrete} into an elastic part and
a disorder part as follows
\begin{equation}
  \mathcal{H}( {\bf u} )
  = \mathcal{H}_\mathrm{ela}({\bf u}) +  \mathcal{H}_\mathrm{dis}({\bf u})
  \hspace{0.5cm}
  \mbox{with }
  \begin{cases}
     \displaystyle
     \mathcal{H}_\mathrm{ela}({\bf u}) = \frac{1}{2} {\bf u}^\mathrm{T} \left( -\kappa\,\Delta+m^2\right) {\bf u}
     \\[0.25cm]
     \displaystyle
     \mathcal{H}_\mathrm{dis}({\bf u}) = \sum_{i=1}^K V_i(u_i)
  \end{cases}
\end{equation}
($\Delta$ is here the discrete Laplacian)
and define the mean number of equilibria
at fixed elastic energy ${H_e}$ and disorder energy ${H_d}$ as
\begin{align}
 &\mean{ \mathcal{N}_\mathrm{tot}(H_e,H_d) }
% \\\nonumber
%  &
  =
  \int_{\mathbb{R}^K}  \D{\bf u} \, 
  \mean{  
   \delta(H_e-\mathcal{H}_\mathrm{ela}({\bf u}))
   \,
   \delta(H_d-\mathcal{H}_\mathrm{dis}({\bf u}))
   \,
   |\det \big(\partial_i\partial_j\mathcal{H}\big)|  \, 
   \prod_{i=1}^K \delta\left(\partial_i \mathcal{H}\right)
   }
\end{align}
The easiest observable to study is the Laplace transform
\begin{align}
\label{Ldef} 
  \widetilde{N}(s_e,s_d) 
  &= \int_0^{+\infty} \D H_e\int_{-\infty}^{+\infty} \D H_d\: 
  \EXP{-s_e H_e -s_d H_d} \mean{ \mathcal{N}_\mathrm{tot}(H_e,H_d) }
\end{align}
from which we can deduce \eqref{eq:DefNtotDeH} from 
\begin{equation}
  \widetilde{N}(s,s) = \int_{-\infty}^{+\infty} \D H \EXP{-s H} \mean{ \mathcal{N}_\mathrm{tot}(H) }
\end{equation}
using that the total energy is $H=H_e+H_d$. Note that the elastic energy is always positive, while the disorder part can be of either sign (hence the $s_d$ dependence really involves a double-sided Laplace transform). 

We now use that $V_i(u_i)$ and $V_i''(u_i)$ are correlated Gaussian variables, described by the covariance matrix
\begin{equation}
  \begin{pmatrix}
    R(0)   & R''(0) \\
    R''(0) & R''''(0)
  \end{pmatrix}
  \:.
\end{equation}
and independent from $V'_i(u_i)$. We also use, for each monomer, the following property
of Gaussian variables (the "complete the square" trick)
\begin{equation} 
  \label{property}
  \mean{ \EXP{-s_d V_i(u_i)} \, F(V_i''(u_i)) }
  = 
  \EXP{\frac12 R(0)s_d^2}
  \mean{ F(V_i''(u_i) - s_d R''(0)) }  
  \:,
\end{equation}
for any function $F(x)$,
where the second average runs over the marginal distribution of $V_i''(u_i)$. Equivalently
one can write for any constant $V$
\begin{equation} 
  \label{property1}
  \mean{ \delta(V_i(u_i)-V)  \, F(V_i''(u_i)) }
  =  \mean{ F(V_i''(u_i) - R''(0) \chi) }  
\end{equation}
where $\chi$ is a Gaussian random variable of variance $R(0)$ and of mean $V/R(0)$, independent of 
$V''(u_i)$.
So imposing the value of $V_i(u_i)$ amounts to shift the random potential by an independent Gaussian
random variable. This leads to
the following decoupling
\begin{equation}
 \mean{ \mathcal{N}_\mathrm{tot}(H_e,H_d) } 
 =  J({m^2},H_e)\, G({m^2},H_d) 
 \quad , \quad 
 \widetilde{N}(s_e,s_d) = 
 \tilde J({m^2},s_e)\, \tilde G({m^2},s_d) 
\end{equation}
with 
\begin{align}
  \tilde J({m^2},s_e)  =\int  \frac{\D {\bf u}}{(2\pi |R''(0)|)^{K/2}}
   \exp\left\{ 
      - \frac{\left[({m^2}\, \identity_K -\kappa\,\Delta){\bf u}\right]^2}{2  |R''(0)|} 
      - \frac{s_e}{2} {\bf u}^\mathrm{T} ({m^2}\, \identity_K -\kappa\,\Delta) {\bf u}
   \right\}
%   \EXP{-   \frac{s_e}{2} {\bf u}^\mathrm{T} ({m^2}\, \identity_K -\kappa\,\Delta) {\bf u}}
\end{align}
which can be evaluated to be
\begin{equation}
  \tilde J({m^2},s_e)  = \frac{1}{\sqrt{ \det({m^2} \identity_K -\kappa\Delta) 
  \det \left(({m^2}+ s_e |R''(0)|) \identity_K -\kappa\Delta \right) }}
  \:.
\end{equation} 
The second factor is 
\begin{equation}
\tilde G({m^2},s_d) = 
          \mean{
           \left|
              \det \left( {m^2} \, \delta_{ij}- \kappa \Delta_{ij}+V_i''(u_i)\,\delta_{ij} \right)
           \right| \EXP{-s_d \sum_{i=1}^K V_i(u_i)}
            }
  \:.
\end{equation}
Using the property \eqref{property} we obtain     
\begin{align}
& \tilde G({m^2},s_d) = 
          \mean{
           \left|
              \det \left(( {m^2} + |R''(0)| s_d) \, \delta_{ij}-\Delta_{ij}+ U_i \,\delta_{ij} \right)
           \right| 
            } \EXP{\frac{1}{2}  R(0) s_d^2 K}
%              \\
%            & = G({m^2}+ |R''(0)| s_d) \EXP{\frac{1}{2}  R(0) s_d^2 K} 
\end{align}
where the average is over centered independent Gaussian random variables $U_i$ with correlator
$\langle U_i U_j \rangle= \big[R''''(0)/\kappa^2\big]\, \delta_{ij}$.
Hence, $\tilde G({m^2},s_d)$ is equal to the 
average of the absolute value of the determinant studied in this paper, i.e. the numerator 
of \eqref{eq:DetRepres}, up to the simple shift $m^2 \to {m^2}+ |R''(0)| s_d$. 
\\

As a result, the Laplace transforms $\widetilde{N}(s_e,s_d)$ and $\widetilde{N}(s,s)$, defined in \eqref{Ldef} 
are expressed very simply in terms of the determinants studied in this paper
\begin{equation}
\label{eq:DetRepres00}
   \widetilde{N}(s_e,s_d) = \frac{
          \mean{
           \left|
              \det \left( (({m^2} + s_d |R''(0)|)/\kappa)\, \delta_{ij}-\Delta_{ij}+U_i\,\delta_{ij} \right)
           \right|
            } \EXP{\frac{1}{2} R(0) s_d^2 K} 
          }
          { \sqrt{
         \det \left( ({m^2}/\kappa)\, \delta_{ij}-\Delta_{ij}\right) 
             \det \left( (({m^2} + s_e |R''(0)|)/\kappa)\, \delta_{ij}-\Delta_{ij}\right) }
          }
          \:.
\end{equation}
%where now we use the same notations are in the text, i.e. $\langle U_i U_j \rangle= R''''(0) \delta_{ij}$.
Note that since the right hand side is well defined for $s_e,\,s_d> - m^2/|R''(0)|$ we conclude that these
Laplace transforms exist (their defining integrals are convergent in that domain). 
One trivially checks that in the absence of disorder 
one has $\widetilde{N}(s_e,s_d)=1$, i.e. $\mean{\mathcal{N}_\mathrm{tot}({H_e},{H_d})}=\delta({H_e})\,\delta({H_d})$ since there is
then a single equilibrium with {$u_i=0$} and zero elastic energy. 

The formula \eqref{eq:DetRepres} can be similarly written for the continuum model, for which the 
elastic and disorder energies are defined respectively as
\begin{align}
  \mathcal{H}_\mathrm{ela}[u]
  &=\int_0^L \D\tau \, \left[ \frac{\kappa}{2} (\partial_\tau u(\tau))^2 + \frac{{m^2}}{2} u(\tau)^2 \right]  
  \:,
  \\
  \mathcal{H}_\mathrm{dis}[u]
  &=\int_0^L \D\tau \, V(u(\tau), \tau)
  \:.
\end{align}
The joint Laplace transform (defined in the same way as for the discrete model) now reads
\begin{equation}
\label{eq:DetRepres3}
   \widetilde{N}(s_e,s_d) = \frac{
          \mean{
           \left|
              \det \left( ({m^2} + |R''(0)| s_d)/\kappa - \partial^2_\tau + U(\tau) \right)
           \right|
            } \EXP{\frac{1}{2} R(0) s_d^2 L} 
          }
          { \sqrt{
         \det \left( {m^2}/\kappa - \partial^2_\tau \right) 
             \det \left( ({m^2} + |R''(0)| s_e )/\kappa  - \partial^2_\tau \right) }
          }
          \:,
\end{equation}
where we recall that $\langle U(\tau) U(\tau') \rangle=2 D\, \delta(\tau-\tau')$ with $D=R''''(0)/(2 \kappa^2)=1/(2L_c^3)$.
The above product form shows {\it the statistical independence of the elastic and the disorder energy
at a force-free point}.\\

Let us now study some implications of this formula. We first review some of our previous results needed here
\begin{itemize}
\item[(\textit{i})] 
  The determinants in \eqref{eq:DetRepres3} not containing $U(\tau)$ can be read  from Eqs.~\eqref{eq:FreeDetVariousBC} for various boundary conditions, 
all having the same leading behaviour in the large $L$ limit $\det( \gamma - \partial_\tau^2  ) \sim\exp(\sqrt{\gamma}L)$.

\item[(\textit{ii})] The following ratio of determinants was studied for large $L$
\begin{equation}
\label{eq:DetRepres10}
  \mean{ \mathcal{N}_\mathrm{tot} }  %(m^2)
  = \frac{
          \mean{
           \left|
              \det \left( m^2/\kappa -\partial_\tau^2 + U(\tau) \right)
           \right|
            }
          }
          {
            \left| \det \left( m^2/\kappa -\partial_\tau^2 \right) \right|
          }
          \sim \EXP{ \rate(m^2) L} 
\end{equation}
and its rate of growth $\rate(m^2)$ as a function of $m^2$ was shown to be [see Eq.~\eqref{eq:MainResultG}]
\begin{equation} 
  \label{recallg} 
  \rate(m^2) 
  \simeq
  \begin{cases}
    C/L_c - 1/L_m & \mbox{for } L_m\to\infty
    \\[0.125cm]
     \frac{L_m^2}{8\pi\,L_c^3} \exp\Big\{-\frac{8}{3}\left({L_c}/{L_m}\right)^3\Big\} 
                  & \mbox{for } L_m\to0
  \end{cases}
\end{equation}
where $C \simeq 0.46$ and we recall that $L_m=\sqrt{\kappa}/m$ and $L_c=\kappa^{2/3} R''''(0)^{-1/3}$ 
[the first limiting behaviour was given in Eq.~\eqref{eq:RateSmallE}, recalling the correspondence of notations $E=-m^2/\kappa$ and $(2D)^{1/3}=1/L_c$].
\end{itemize}

In view of (\textit{i}) and (\textit{ii}) there are thus two main applications, studied respectively in the
two subsections below~:
\begin{itemize}
\item[(a)] 
we use the information that we know only in the large $L$ limit to study the large deviation
rate functions $\rate(m^2;H)$ and $\rate(m^2;H_e,H_d)$
{controlling the growth of $\mean{ \mathcal{N}_\mathrm{tot}(H) }$ and $\mean{ \mathcal{N}_\mathrm{tot}(H_e,H_d) }$, respectively.}
%, see definitions below.

\item[(b)] 
 The Laplace transform can be written exactly for any $L$, and (in principle) inverted.
Such formula exist in general only for the mean number of equilibria at fixed elastic energy, 
$\int\D H_d\,\mean{ \mathcal{N}_\mathrm{tot}(H_e,H_d) }$,
which is studied below. 
%, and in the particular
%case of the Larkin model ($R''''(0)=0$) for $\mean{\mathcal{N}({H_e,H_d}) }$
%hence for $\mean{\mathcal{N}({H_e,H_d}) }$

\end{itemize}

\subsection{Large deviations and rate functions in the large $L$ limit} 
\label{sec:LDFForFixedEnergy}

We will denote $H_e=h_e L$, $H_d=h_d L$ and $H=h L$, where $h_e$, $h_d$ and $h$ are respectively the  elastic, disorder and total energy densities. 
We expect in the large $L$ limit that
\begin{align}
  & \mean{ \mathcal{N}_\mathrm{tot}(H_e,H_d) } \sim \EXP{\rate(m^2;h_e,h_d) L}  
  \\
  &  \mean{ \mathcal{N}_\mathrm{tot}(H) } \sim \EXP{\rate(m^2;h) L} 
\end{align} 
Hence, from the definition of the Laplace transforms \eqref{Ldef} 
\begin{align}
\label{N1sp} 
  & \widetilde{N}(s_e,s_d)  \sim
  \int \D h_e \D h_d \, \EXP{ (\rate(m^2;h_e,h_d) - s_e h_e - s_d h_d) L }  
  \\
\label{N2sp}
  & 
  \widetilde{N}(s,s) \sim 
  \int \D h \, \EXP{ (\rate(m^2;h) - s h) L }  
\end{align} 

On the other hand from (\textit{i}) and (\textit{ii}) above we can write, in the limit of large $L$
\begin{equation}
  \widetilde{N}(s_e,s_d) \sim \EXP{\Gamma(s_e,s_d) L}
  % \quad , \quad \widetilde{N}(s) \simeq  \EXP{\Gamma(s) L}  \quad , \quad \Gamma(s):=\Gamma(s,s) 
\end{equation} 
with 
\begin{align}
   \label{resGamma} 
   \Gamma(s_e,s_d) 
   = \rate(m^2 + |R''(0)| s_d) 
   + \frac{\sqrt{ m^2 + |R''(0)| s_d}}{\sqrt{\kappa}}  
   + \frac{1}{2} R(0) s_d^2 
   - \frac{m + \sqrt{ m^2 + |R''(0)| s_e}}{2 \sqrt{\kappa}}
\end{align}  

One can use a saddle point to estimate the leading large $L$ behaviour of \eqref{N1sp} and \eqref{N2sp}, and we find that
the rate functions $\rate(m^2;h_e,h_d)$ and 
$\rate(m^2;h)$ are related to $\Gamma(s_e,s_d)$ and $\Gamma(s,s)$ by Legendre transforms. More precisely
one has
\begin{align}
 \Gamma(s_e,s_d) &= 
 \underset{h_e,\,h_d}{\mathrm{max}} 
 \left\{ \rate(m^2;h_e,h_d) - s_e h_e - s_d h_d \right\} 
 \\
 \Gamma(s,s) 
 &= \underset{h}{\mathrm{max}}  \left\{ \rate(m^2;h) - s h \right\}  
\end{align}
Inverting we have 
\begin{align}
  \label{eq:FullRate}
 \rate(m^2;h_e,h_d) &= \underset{s_e,\,s_d}{\mathrm{min}}  
 \left\{ \Gamma(s_e,s_d) + s_e h_e + s_d h_d \right\} \\
 \rate(m^2;h) &= 
  %\CGre
  {
  \underset{s}{\mathrm{min}} \left\{ \Gamma(s,s) + s h \right\}  
  }
\end{align}

Below we will also study the rates associated to fixing the elastic energy alone,
and the disorder energy alone, namely
\begin{align}
 \rate_e(m^2;h_e) &= \underset{s_e}{\mathrm{min}}   \left\{ \Gamma(s_e,0) + s_e h_e  \right\} \\
 \rate_d(m^2;h_d) &= \underset{s_d}{\mathrm{min}}   \left\{ \Gamma(0,s_d) + s_d h_d  \right\} 
\end{align} 

Note the two important general observations, which will be confirmed below
by explicit calculations:
\begin{itemize}
\item
The maximum over $h_e$ of $\rate_e(m^2;h_e)$ occurs at the field $h_e^*$
which corresponds to $s_e=0$ 
and its value is $\Gamma(0,0)= \rate(m^2)$. It is easy
to check that this property holds from the derivative conditions.
The field $h_e^*$ then corresponds to 
the typical (i.e. the most probable)
value of $h_e$. 
Same property holds, respectively, for the rate $\rate_d(m^2;h_d)$
(with typical value $h_d^*$ occurring at $s_d=0$). 

\item
 The form obtained in \eqref{resGamma} satisfies that
\begin{equation}
 \Gamma(s_e,0) + \Gamma(0,s_d) = r(m^2) + \Gamma(s_e,s_d)
\end{equation}
which reflects the independence of the elastic and potential energy noted above.
Hence we have
\begin{equation}
  \rate(m^2;h_e,h_d)-\rate(m^2)
  =
  \big[ \rate_e(m^2;h_e)-\rate(m^2) \big]
   + 
  \big[ \rate_d(m^2;h_d)-\rate(m^2) \big]
  \:.
\end{equation}
In addition one can also show that
\begin{align}
 \rate(m^2;h) &= \underset{h_d}{\mathrm{max}}  \, [\rate_e(m^2;h-h_d) + \rate_d(m^2;h_d)] - r(m^2) \\
 & =  \underset{h_e}{\mathrm{max}} \, [\rate_e(m^2;h_e) + \rate_d(m^2;h-h_e)] - r(m^2)
\end{align}
and finally, $\underset{h}{\mathrm{max}} \, \rate(m^2;h)= r(m^2)$ which is attained
at a field $h^*$, equal to the typical total energy density.
\end{itemize}

To proceed further, it is convenient to introduce dimensionless variables.~\footnote{
  Three main dimensions are involved here~: 
  the energy $[\mathcal{H}]=E$, 
  the length $[\tau]=L$
  and 
  the field $[u]$.
  We deduce the dimensions of the main quantities~:  
  the correlator $[R(u)]=E^2/L$, 
  the mass $[m]=E^{1/2}L^{-1/2}[u]^{-1}$ 
  and 
  the elastic constant $[\kappa]=E\,L[u]^{-2}$.
} 
Because we will be interested in the $m\to0$ limit, we choose to rescale all observables with respect to the length scale $L_c$.
We define the rescaled energy density and its conjugate variable
\begin{align}
  \tilde{h} := \frac{(R''''(0)\kappa)^{1/3}}{|R''(0)|}\, h
  \hspace{1cm}\mbox{and}\hspace{1cm}
  \sigma    := \frac{|R''(0)|\kappa^{1/3}}{R''''(0)^{2/3}}\, s
  \:.
\end{align}
The rates now take the form
\begin{align}
 \rate(m^2;h_e,h_d) &= 
 \frac{1}{L_c}\,\underset{\sigma_e,\,\sigma_d}{\mathrm{min}}  
   \left\{  \widetilde{\Gamma}(\sigma_e,\sigma_d)  + \sigma_e \tilde h_e + \sigma_d \tilde h_d \right\}   
   \\
   \label{eq:RateConstrainedTotal}
 \rate(m^2;h) &= 
  \frac{1}{L_c}\,\underset{\sigma}{\mathrm{min}} 
  \left\{  \widetilde{\Gamma}(\sigma,\sigma)  + \sigma \tilde h \right\}  
  \:,
\end{align}
where   
\begin{align}
  \label{eq:FullGammaTilde}
  \widetilde{\Gamma}(\sigma_e,\sigma_d) 
  = \widetilde{\Lambda}(\rmass+\sigma_d)
  %\tilde{r}(\rmass+\sigma_d) + \sqrt{\rmass+\sigma_d} 
  + \frac{\eta}{2}\sigma_d^2
  -\frac{1}{2} \left( \sqrt{\rmass+\sigma_e}+\sqrt{\rmass} \right)
  \:.
\end{align}
$\widetilde{\Lambda}(\rmass)=L_c\,\Lambda(1)$ is the (rescaled) generalized Lyapunov exponent (cf. Section~\ref{sec:GLE}), defined for $\rmass\in\mathbb{R}$ 
(for reasons which will become clear later, we have preferred to express the rate $\Gamma$ as a function of the GLE $\Lambda(1)=\rate(m^2)+1/L_m$~; a rescaled rate $\tilde{\rate}$ is also defined below).
The expression \eqref{eq:FullGammaTilde} makes clear that the two conjugated parameters belong to different domains~:
$\sigma_e\in[-\rmass,+\infty[$ and $\sigma_d\in\mathbb{R}$, which reflects the fact that $H_e\geq0$ while $H_d\in\mathbb{R}$.
We have also introduced the two dimensionless parameters
\begin{align}
  \rmass &= (L_c/L_m)^2 \propto m^2
  \:,
  \\
  \eta &= \frac{R(0)R''''(0)}{R''(0)^2} \geq1
  \:.
\end{align}
The bound on the parameter $\eta$ will play an important role below and follows from the Gaussian nature~\footnote{
  Since  $R(u)$ is the correlator of a Gaussian field, its Fourier transform is positive, $\widehat{R}(q)\geq0$.
  This property allows to apply the well-known Cauchy-Schwarz inequality 
  $\big[\int\D q\,f^*(q)g(q)\big]^2\leq\int\D q\,|f(q)|^2\int\D q\,|g(q)|^2$ for square integrable functions 
  to the functions $f(q)=\widehat{R}(q)^{1/2}$ and $g(q)=q^2\,\widehat{R}(q)^{1/2}$, leading to
  $\big[\int\D q\,q^2\widehat{R}(q)\big]^2\leq\int\D q\,\widehat{R}(q)\int\D q\,q^4\widehat{R}(q)$, i.e.
  $R''(0)^2\leq R(0)\,R''''(0)$.
  \textsc{Qed}.
} 
of the disordered potential $V(u,\tau)$ in Eq.~\eqref{eq:H}.
This bound is saturated, $\eta=1$, for harmonic correlations $R(u)=\sigma^2\,\cos(u/\rf)$, which corresponds to the charge density wave model studied numerically in Section~\ref{sec:NumericsAlberto}.

\begin{figure}[!ht]
\centering
\includegraphics[width=0.495\textwidth]{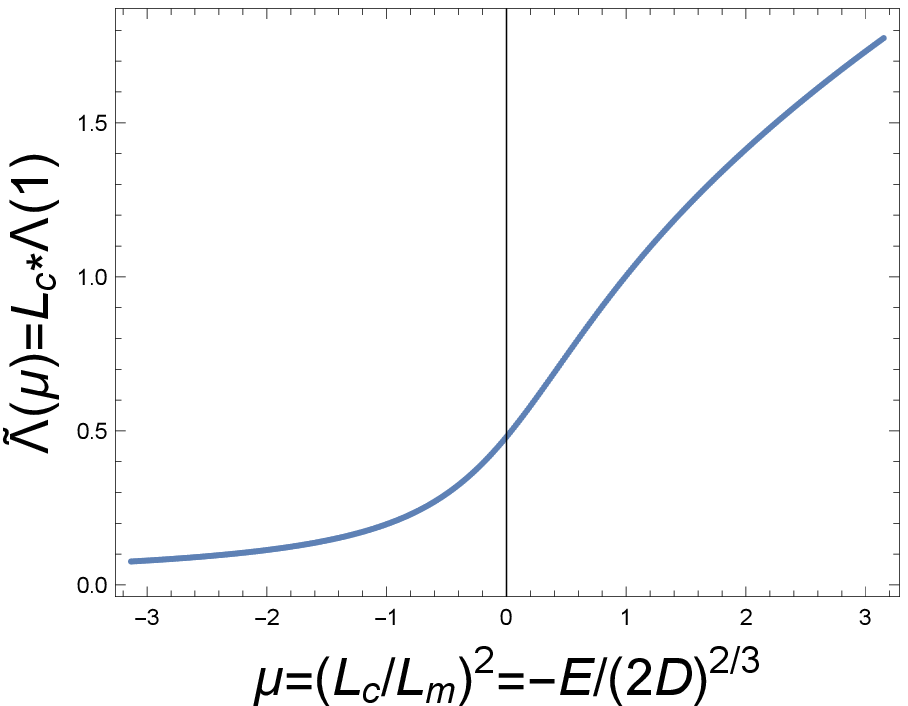}
\includegraphics[width=0.495\textwidth]{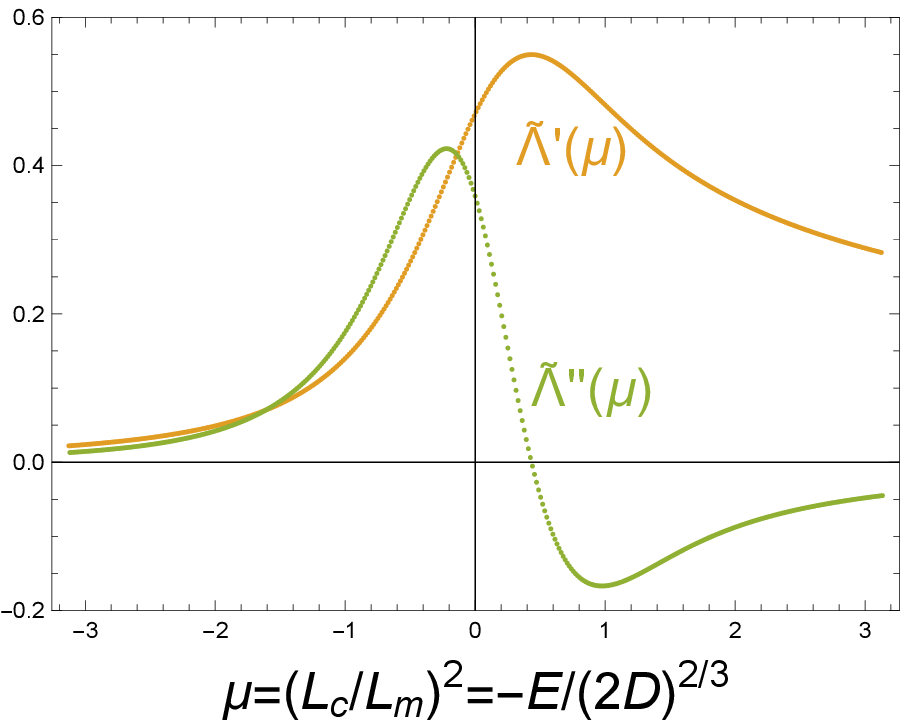}
\caption{\it Rescaled GLE computed numerically according to the method described in Subsection~\ref{subsec:FPDiscrete}.}
\label{fig:LambdaTilde}
\end{figure}

We recall that $\widetilde{\Lambda}(\rmass)$ is a monotonously increasing 
function of $\rmass$, see Fig.~\ref{fig:GLEforPosNegEnergy} and Fig.~\ref{fig:LambdaTilde}, with the three limiting behaviours obtained in the previous sections
\begin{align}
  \label{eq:GLE3LimitingBehaviours}
  \widetilde{\Lambda}(\rmass)
  \simeq 
  \begin{cases}
     -\frac{3}{16\,\rmass}
       & \mbox{for } \rmass\to-\infty
     \\
     C + a_1\, \rmass + \frac{1}{2}a_2\, \rmass^2 + \mathcal{O}(\rmass^3) 
       & \mbox{for } \rmass\to0 
     \\
     \sqrt{\rmass} +  \frac{1}{8\pi\,\rmass}\,\exp\{-(8/3)\rmass^{3/2}\}  
       & \mbox{for } \rmass\to+\infty
  \end{cases}
\end{align}
where the behaviour for $\rmass\to -\infty$ is deduced from \eqref{eq:GLEforLargePositiveEnergy}.
The function $\widetilde{\Lambda}(\rmass)$ is smooth and analytic everywhere, in particular
near $\rmass=0$, a point of special interest in the present study (see below).
The dimensionless coefficient $a_1\simeq0.47$ is positive, see Eq.~\eqref{eq:Lambda1SmallE}.
It is also convenient to introduce the rescaled rate $\tilde{r}(\rmass)=L_c\,\rate(m^2)=\widetilde{\Lambda}(\rmass)-\sqrt{\rmass}$ defined for $\rmass\in\mathbb{R}^+$, with the limiting behaviours 
\begin{equation}
  \label{eq:RateAsymptConvenient}
  \tilde{r}(\rmass) \simeq
  \begin{cases} 
    C - \sqrt{\rmass}  + \, a_1\, \rmass  + \mathcal{O}(\rmass^2)   & \mbox{for } \rmass\to0^+
    \\
    \frac{1}{8\pi\,\rmass}\,\exp\{-(8/3)\rmass^{3/2}\}  & \mbox{for } \rmass\to\infty
  \end{cases}
\end{equation}
corresponding to \eqref{recallg}.

We now study respectively the rates at fixed elastic and disorder energy, and finally at fixed total energy.

\subsubsection{Mean number of equilibria constrained by the elastic energy}

As a warm up simple exercise, we consider first the number of equilibria constrained by the elastic energy 
\begin{equation}
  \int\D H_d\,\mean{ \mathcal{N}_\mathrm{tot}(H_e,H_d) }
  \underset{L\to\infty}{\sim}
  \exp\{ \rate_e(m^2;h_e)\,L \}
\:.
\end{equation}
Here we only compute the large deviation function $r_e(m^2;h_e)$ (a more precise calculation of the distribution of elastic energy will be presented in Subsection~\ref{subsec:DistribHe}).

We consider 
\begin{equation}
  \rate_e(m^2;h_e) = 
 \frac{1}{L_c}\,\underset{\sigma_e}{\mathrm{min}}  
   \left\{ \widetilde{\Gamma}(\sigma_e,0)  + \sigma_e \tilde h_e \right\}
   \hspace{0.5cm}\mbox{for }
   \sigma_e\geq-\rmass
   \:.
\end{equation}
It is straightforward to minimize 
$\widetilde{\Gamma}(\sigma_e,0)  + \sigma_e \tilde h_e=\tilde{\rate}(\rmass)+(1/2)(\sqrt{\rmass}-\sqrt{\sigma_e+\rmass})+\sigma_e \tilde h_e$ over $\sigma_e$~:
the function is minimum for $\sigma_e=-\rmass+1/(4\tilde{h}_e)^2$.
We deduce
\begin{equation}
  L_c \rate_e(m^2;h_e)  
  = \tilde{\rate}(\rmass) - \frac{\rmass}{\tilde h_e} \left( \tilde h_e-\frac{1}{4 \sqrt{\rmass}} \right)^2
  \:,
\end{equation}
i.e. 
\begin{equation}
  \label{eq:reDEhe}
  \rate_e(m^2;h_e) -\rate(m^2) = 
 -\frac{L_c}{L_m^2\,\tilde{h}_e} \left( \tilde{h}_e - \frac{L_m}{4L_c} \right)^2
 \simeq
 -
 \begin{cases}
   \displaystyle
   \frac{|R''(0)|}{16\kappa h_e} & \mbox{as } h_e\to0^+
   \\[0.25cm]
   \displaystyle
   \frac{m^2h_e}{|R''(0)|}       & \mbox{as } h_e\to\infty
 \end{cases}
 \:.
\end{equation}
Writing 
\begin{equation}
  \int\D H_d\,\frac{ \mean{ \mathcal{N}_\mathrm{tot}(H_e,H_d) } }{ \mean{ \mathcal{N}_\mathrm{tot} } }
  \underset{L\to\infty}{\sim}
  \EXP{ [\rate_e(m^2;h_e)-\rate(m^2)]\,L }  
\end{equation}
makes clear that \eqref{eq:reDEhe} is the large deviation function for the elastic energy, controlling the large $L$ behaviour of the distribution of the elastic energy with respect with the annealed measure~\eqref{defAnn}.
We see that the rate is maximum at $\tilde{h}_e = \tilde{h}_e^* := 1/(4\sqrt{\rmass})=L_m/(4L_c)\propto1/m$
(corresponding to $\sigma_e=0$), which coincides with the most probable (typical) value of the elastic energy density for the line at equilibrium.
Coming back to the original variable, 
the typical elastic energy density is
\begin{equation}
  \label{eq:hstar}
  h_e^* = \frac{|R''(0)|}{4m\sqrt{\kappa}}
  \simeq
  \smean{ h_e }_a
  \:,
\end{equation}
which also corresponds with its annealed average value as indicated by $\smean{ \cdots }_a$.
From \eqref{eq:reDEhe}, we deduce the variance of the annealed distribution of
the elastic energy density as
\begin{equation}
  \mathrm{Var}_a( h_e ) 
  \simeq \frac{|R''(0)|^2}{8m^2 \kappa} \frac{L_m}{L} 
  \:.
\end{equation}

We note a peculiarity in the zero mass limit $m \to 0$. In that limit, from \eqref{eq:hstar}
we see that the typical elastic energy density tends to infinity as $h_e^*\propto m^{-1}$, 
as well as the fluctuations $\sqrt{\mathrm{Var}_a( h_e )}\propto m^{-3/2}$. 
This is a genuine feature, understood as follows. We saw in Section~\ref{sec:Configurations} that the
annealed distribution of displacements obeys the Larkin scaling
(see~\ref{app:Larkin}), which, in the massless
case, goes schematically as $u \sim L^{3/2}$. 
The simple dimensional estimate $H_e \sim u^2/L$ 
then leads to an elastic energy growing as $H_e\sim L^2$ instead of $L$ for finite $m$, meaning that the energy density grows with the length as $h_e\sim L$.
This is 
%$H_e$ growing not as $L$, but as $L^2$, 
consistent
with an infinite typical energy density in the thermodynamic limit $L\to\infty$. 
This behaviour of the elastic energy thus arises
because typical stationary configurations wander a lot (with Larkin scaling in $L$), much
further than the ground state (which is known to wander only as $u \sim L^{2/3}$ and to lead to a
finite elastic energy density). Below we provide an exact calculation
of the distribution of $H_e$ for $m=0$, which confirms these qualitative arguments (cf. Subsection~\ref{subsec:TheLastSubsection}).

\subsubsection{Mean number of equilibria constrained by the disorder energy}
\label{sec:disorder}

In a second stage, we consider the number of equilibria constrained by the disorder energy $H_d$~:
\begin{equation}
  \int\D H_e\,\mean{ \mathcal{N}_\mathrm{tot}(H_e,H_d) }
  \underset{L\to\infty}{\sim}
  \exp\{ \rate_d(m^2;h_d)\,L \}
\:,
\end{equation}
where the rate is given by
\begin{equation}
  \label{eq:DefRd}
  \rate_d(m^2;h_d) = 
  \frac{1}{L_c}\,\underset{\sigma_d}{\mathrm{min}} 
  \left\{ \widetilde{\Gamma}(0,\sigma_d)  + \sigma_d \tilde h_d \right\}
  \:,
\end{equation}
where 
\begin{equation}
  \widetilde{\Gamma}(0,\sigma)
  = \widetilde{\Lambda}(\rmass+\sigma) + \frac{\eta}{2}\sigma^2-\sqrt{\rmass}
  \hspace{0.5cm}\mbox{for }
  \sigma\in\mathbb{R}
  \:.
\end{equation}
Although it is not possible to obtain a simple analytical form, as for the elastic energy, one can establish various general features and limiting behaviours.

The minimizer of Eq.~\eqref{eq:DefRd} is given by $\sigma_d=\sigma_*$ solution of
\begin{equation}
  \label{eq:SaddleRd}
  -\tilde{h}_d = \widetilde{\Lambda}'(\rmass+\sigma_*) + \eta\,\sigma_*
  \:.
\end{equation}
We deduce the large deviation function
\begin{equation}
  \label{eq:RateRd}
 L_c \, \rate_d(m^2;h_d) = 
 \widetilde{\Lambda}(\rmass+\sigma_*) - \sigma_*\,\widetilde{\Lambda}'(\rmass+\sigma_*)
 - \frac{\eta}{2}\sigma_*^2 - \sqrt{\rmass}
 \:.
\end{equation}
As discussed above for the elastic energy, from general considerations the maximum of $r_d(m^2;h_d)$ over $h_d$ 
should equal $\rate(m^2)$ and occur at the typical field $h_d^*$ corresponding to the argmin value $\sigma_*=0$. We thus immediately obtain the typical, most probable value of the (dimensionless) disorder energy
density as 
\begin{equation}
  \tilde h_d^* = - \widetilde{\Lambda}'(\rmass) 
 \:.
\end{equation}
%a result which is discussed in more details below.
Because $\widetilde{\Lambda}'(\rmass)>0\ \forall\rmass$ (cf. Fig.~\ref{fig:LambdaTilde}) we see that the typical disorder energy is \textit{negative}
\begin{equation}
  \label{eq:hdstar}
  h_d^*  = - \frac{|R''(0)|}{(R''''(0)\kappa)^{1/3}}\,\widetilde{\Lambda}'(\rmass) <0
  \:.
\end{equation}
We recall that $\rmass= (L_c/L_m)^2 \propto m^2$. 
In the limit of zero mass, $m \to 0$ (absence of external confinement), we can use the value obtained from the numerics
$\widetilde{\Lambda}'(0)=a_1\simeq0.47$ (cf. Fig.~\ref{fig:LambdaTilde}) to get a good estimate for the typical disorder energy. 
In the other limit of strong confinement ($L_m\ll L_c$), using \eqref{eq:GLE3LimitingBehaviours}, we deduce that the typical disorder energy takes the form 
\begin{equation}
  h_d^* \simeq - \frac{ |R''(0)| }{ 2m\sqrt{\kappa} } 
  \hspace{1cm}\mbox{for }
  m\to\infty
  \:,
\end{equation} 
i.e. is twice the elastic energy \eqref{eq:hstar}, with the opposite sign.
Note that our numerics indicate
that $\widetilde{\Lambda}'(\rmass)$ reaches its maximum at $\rmass \approx 0.4$
(when the curvature $\widetilde{\Lambda}''(\rmass)$ changes in sign). 
Intuition about ground states would suggest that the lower the mass, the larger the available space for the elastic line to explore better locations in the random potential, hence the lower the typical disorder energy. However here we are dealing with all stationary points, which seems, from our numerics, to behave differently (with the minimum $h_d^*$ occurring at a non zero value of $\rmass$).

The study of the typical fluctuations is rather simple~:
the expansion of \eqref{eq:SaddleRd} at first order in $\sigma_*$ gives
$
\sigma_*\simeq -(\tilde h_d-\tilde h_d^*)/(\widetilde{\Lambda}''(\rmass)+\eta)
$
and, similarly, the expansion of the rate \eqref{eq:RateRd} takes the form 
$
L_c \, \rate_d(m^2;h_d)
= \tilde{\rate}(\rmass) - (1/2)(\widetilde{\Lambda}''(\rmass)+\eta)\,\sigma_*^2+\mathcal{O}(\sigma_*^3)
$.
We deduce the quadratic behaviour 
\begin{equation}
  \rate_d(m^2;h_d) \simeq \rate(m^2)
   -\frac{1}{2L_c\,(\widetilde{\Lambda}''(\rmass)+\eta)} \left( \tilde h_d-\tilde h_d^* \right)^2
   \hspace{1cm}\mbox{for }
   \tilde h_d \sim \tilde h_d^*
   \:.
\end{equation}
Our numerics also indicates that $\widetilde{\Lambda}''(\rmass)$ is always above $\sim-0.17$ 
(Fig.~\ref{fig:LambdaTilde}).
Thus the combination 
$\eta+\widetilde{\Lambda}''(\rmass)>0$ always remain positive, as needed.
This combination enters the variance of the disorder energy, from
the annealed distribution, as one finds, from \eqref{eq:rdDEhd},
\begin{equation}
    \mathrm{Var}_a( \tilde h_d )
  = (\eta+\widetilde{\Lambda}''(\rmass)) \frac{L_c}{L} 
  \quad \Leftrightarrow \quad 
    \mathrm{Var}_a( h_d ) 
  = \left( R(0) + \frac{R''(0)^2}{R''''(0)}    \widetilde{\Lambda}''(\rmass) \right) \frac{1}{L}
\end{equation}
Because $\eta\geq1$ and $|\widetilde{\Lambda}''(\rmass)|$ is bounded by a number of order one, the first expression makes clear that the first term $\eta\,L_c/L$ gives a good estimate of the variance.
Moreover, in the strong confinement limit ($L_m\ll L_c$), the variance is independent of the mass, 
$\mathrm{Var}_a( h_d )\simeq R(0)/L$, contrary to the typical value which vanishes as $h_d^*\sim1/m$.

\begin{figure}[!ht]
\centering
\includegraphics[width=8cm]{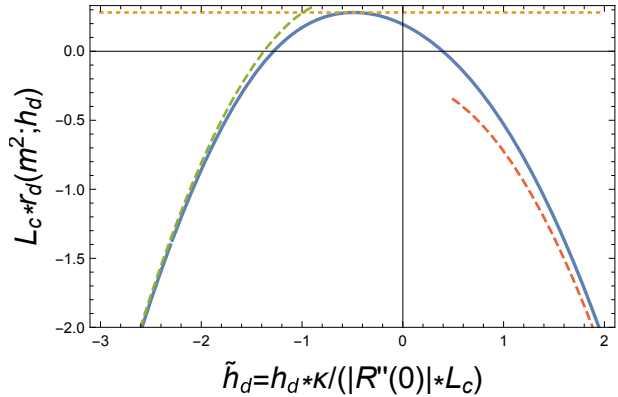}
\caption{\it Rate $\rate_d(m^2;h_d)$ for $\rmass=(L_c/L_m)^2=0.05$ and $\eta=1$. The dashed lines correspond to the limiting behaviours \eqref{eq:rdDEhd}~; the dotted horizontal line is $\rate(m^2)$.
The rate was computed by making use of the data for the GLE obtained above and represented in Fig.~\ref{fig:LambdaTilde}.
}
\label{fig:rddeh}
\end{figure}

Let us now analyse the large deviations, i.e. the \textit{atypical} fluctuations.
The limiting behaviours \eqref{eq:GLE3LimitingBehaviours} show that $\widetilde{\Lambda}'(\rmass+\sigma_*)$ is bounded and decays at $\sigma_*\to\pm\infty$.
This is supported by numerics (Fig.~\ref{fig:LambdaTilde}) 
As a result we get $\sigma_*\simeq-\tilde{h}_d/\eta$ in the tails for $\tilde{h}_d\to\pm\infty$.
We have perform a more detailed analysis of the limiting behaviours of $\sigma_*$ at large $h_d$.
Putting all together, we obtain the respective behaviours in the three regions~:
\begin{equation}
  \sigma_* \simeq
  \begin{cases}
     -\frac{1}{\eta}\tilde{h}_d - \frac{1}{2 \sqrt{\eta |\tilde{h}_d|}} + \mathcal{O}(|\tilde{h}_d|^{-3/2})
     & \mbox{for } \tilde{h}_d\to-\infty
     \\[0.25cm]
     -\frac{1}{\eta+\widetilde{\Lambda}''(\rmass)}(\tilde{h}_d+\widetilde{\Lambda}'(\rmass))
     & \mbox{for } \tilde{h}_d+\widetilde{\Lambda}'(\rmass)\to0
     \\[0.25cm]
     -\frac{1}{\eta}\tilde{h}_d + \mathcal{O}(\tilde{h}_d^{-2})
     %\blue{- \frac{3 \eta}{16 \tilde h_d^2} }
     & \mbox{for } \tilde{h}_d\to+\infty
  \end{cases}
\end{equation}
Correspondingly, we deduce the following behaviours for the large deviation rate function~\footnote{
  The rate can be deduced from Eq.~\eqref{eq:RateRd}.
  More directly, one can use the general property
  $\sigma_*(\tilde{h}_d)=\deriv{}{\tilde{h}_d}L_c \rate_d(m^2;h_d)$, i.e.
  $
  L_c \, \left[ \rate_d(m^2;h_d) - \rate(m^2) \right]
  =\int_{\tilde{h}_d^*}^{\tilde{h}_d}\D t\, \sigma_*(t)
  $.
  A similar relation was recently used in Refs.~\cite{GraTex15,GraTex16b} in order to study some properties of random matrices.  
}
\begin{equation}
  \label{eq:rdDEhd}
  L_c \, \left[ \rate_d(m^2;h_d)  - \rate(m^2) \right]
  \simeq  
  \begin{cases}
    \displaystyle
    - \frac{1}{2\eta}\,\tilde{h}_d^2 + \sqrt{|\tilde{h}_d|/\eta}  - \widetilde{\Lambda}(\rmass)
    + \mathcal{O}(|\tilde{h}_d|^{-1/2})
     & \mbox{for } \tilde{h}_d\to-\infty
    \\[0.25cm]
    \displaystyle
    -\frac{1}{2\big[\eta+\widetilde{\Lambda}''(\rmass)\big]}
     \left(\tilde{h}_d-\tilde{h}_d^*\right)^2 
     & \mbox{for } \tilde{h}_d\sim\tilde{h}_d^*
    \\[0.25cm]
    \displaystyle
    - \frac{1}{2\eta}\,\tilde{h}_d^2 - \widetilde{\Lambda}(\rmass)  
    + \mathcal{O}(\tilde{h}_d^{-1})  
     & \mbox{for } \tilde{h}_d\to+\infty
  \end{cases}
\end{equation}
We compare in Fig.~\ref{fig:rddeh} these limiting behaviours with the result of a numerical resolution of Eqs.~(\ref{eq:SaddleRd},\ref{eq:RateRd})~: the agreement is excellent.

%prg NtotDeH.nb

An interesting feature is that the decay of the rate at large positive energy is not controlled by the mass, as it was the case for the elastic energy, cf. Eq.~\eqref{eq:reDEhe}: instead it has a parabolic shape controlled by $R(0)$~:
\begin{equation}
\label{eq:rdDEhd2}
  \rate_d(m^2;h_d) \simeq -\frac{h_d^2}{2R(0)}
  \hspace{1cm}\mbox{for }
  h_d\to\pm\infty
  \:.
\end{equation}

We have used the numerical data (Sections~\ref{subsec:FPDiscrete} and \ref{subsec:DiffEq}) to compute the rate $r_d(m^2;h_d)$ in order to check our analysis~: the agreement is good, as one can see in Fig.~\ref{fig:rddeh}.

Finally, it is interesting for the following to study the lowest point $h_d^\mathrm{min}$ where the rate $\rate_d(m^2;h_d)$ vanishes~:
in the small mass limit, $\rmass\to0$, we obtain numerically
$\tilde{h}_d^\mathrm{min}\simeq-1.503$ for $\eta=1$, and using \eqref{eq:rdDEhd}, 
$\tilde{h}_d^\mathrm{min}\simeq-2^{2/3}\eta^{1/3}\simeq-1.587\,\eta^{1/3}$ for $\eta\gg1$,
so that setting $\eta=1$ in the approximate expression obtained for $\eta\gg1$ is close to the correct result.

\subsubsection{Mean number of equilibria constrained by the total energy}

We now study the rate \eqref{eq:RateConstrainedTotal} at fixed total energy $H$. 
We recall that the total energy can have both signs, contrary to the elastic energy $H_e$, which is positive. Here also 
%is not possible to derive a simple analytic form, however, 
we can obtain the asymptotic behaviours easily.
We have now to consider 
\begin{align}
  \widetilde{\Gamma}(\sigma,\sigma) 
  = \widetilde{\Lambda}(\rmass+\sigma) 
  - \frac{1}{2} \left( \sqrt{\rmass+\sigma} + \sqrt{\rmass} \right)+ \frac{\eta}{2}\sigma^2  
   \hspace{0.5cm}\mbox{for }
   \sigma\geq-\rmass
   \:.
\end{align}
%using again that $\tilde r(\rmass)=\widetilde{\Lambda}(\rmass)-\sqrt{\rmass}$.
The sum $\widetilde{\Gamma}(\sigma,\sigma)  + \sigma \tilde h$ is minimum for $\sigma=\sigma_*$ solution of 
\begin{equation}
  \label{eq:SaddleEq}
  -\tilde{h} = \widetilde{\Lambda}'(\rmass+\sigma_*) + \eta\,\sigma_* - \frac{1}{4\sqrt{\rmass+\sigma_*}}
  \:.
\end{equation}
As a result
\begin{align}
  \label{eq:10.42}
  \rate(m^2;h) = \frac{1}{L_c}
  \left\{
    \widetilde{\Lambda}(\rmass+\sigma_*) - \sigma_*\,\widetilde{\Lambda}'(\rmass+\sigma_*) 
    - \frac{1}{4}\sqrt{\rmass+\sigma_*} 
    - \frac{1}{2} \sqrt{\rmass} 
    - \frac{\rmass}{4\sqrt{\rmass+\sigma_*}} 
    - \frac{\eta}{2}\sigma_*^2 
  \right\}
  \:.
\end{align}
As discussed above, the maximum of $\rate(m^2;h)$ over $h$ equals $r(m^2)$ and is
attained for the typical field $\tilde h^*$ corresponding to $\sigma^*=0$, hence
from \eqref{eq:SaddleEq}
\begin{equation}
  \tilde h^*  = - \widetilde{\Lambda}'(\rmass) + \frac{1}{4\sqrt{\rmass}} = \tilde h_d^* + \tilde h_e^*
\end{equation}
hence, not surprisingly, the typical total energy is the sum $h^*=h_e^*+h_d^*$ of the typical
disorder and elastic energies obtained in the previous sections. In the original
units, the typical total energy density reads
\begin{equation}
  h^*  
  =  \frac{|R''(0)|}{4 m \sqrt{\kappa}} \left(1 - 4\sqrt{\rmass}\, \widetilde{\Lambda}'(\rmass) \right)
  \hspace{0.5cm}\mbox{where } \rmass=\left(\frac{L_c}{L_m}\right)^2
  \:.
\end{equation}
From \eqref{eq:GLE3LimitingBehaviours} we see that the last factor 
$(1 - 4\sqrt{\rmass}\, \widetilde{\Lambda}'(\rmass))$ changes sign as the mass increases and 
varies from $1$ (positive typical energy, elastic energy dominates) for $m \to 0$, 
corresponding to weak confinement,
to $-1$ (negative typical energy, disorder energy dominates) for $m \to +\infty$, corresponding to strong confinement. 

The expansion of \eqref{eq:SaddleEq} for $\sigma_*\to0$ allows to study the typical fluctuations, in the same way as for the disorder energy.
As expected, the variance is given by the sum of variances of elastic and disorder energy computed above~:
\begin{equation}
  \mathrm{Var}_a( h ) =\mathrm{Var}_a( h_e ) + \mathrm{Var}_a( h_d )
=  \frac{R''(0)^2}{R''''(0) L} \left(\eta+\widetilde{\Lambda}''(\rmass) + \frac{1}{8\rmass^{3/2}}  \right)
  \:.
\end{equation}
We deduce that the fluctuations are dominated by the elastic energy in the weak confinement regime ($L_m\gg L_c$), $\mathrm{Var}_a( h ) \simeq \mathrm{Var}_a( h_e )\propto m^{-3}$, while in the strong confinement case ($L_m\ll L_c$) they are dominated by the disorder energy 
$\mathrm{Var}_a( h ) \simeq \mathrm{Var}_a( h_d )\simeq R(0)/L$.

\begin{figure}[!ht]
\centering
\includegraphics[width=8cm]{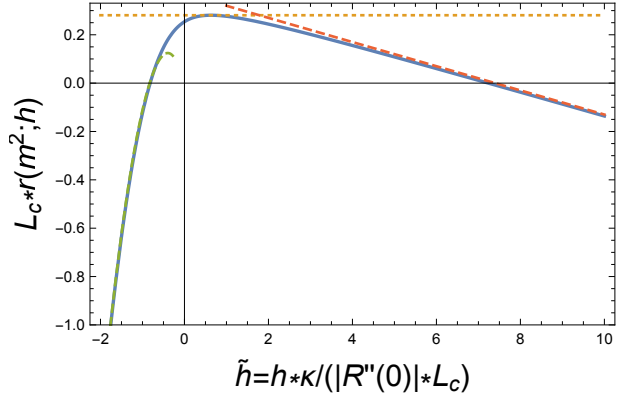}
\caption{\it Rate $\rate(m^2;h)$ for $\rmass=(L_c/L_m)^2=0.05$ and $\eta=1$. The dashed lines correspond to the limiting behaviours \eqref{eq:rDEh}~; the dotted horizontal line is $\rate(m^2)$.
The rate was computed by making use of the data for the GLE obtained above and represented in Fig.~\ref{fig:LambdaTilde}.
}
\label{fig:rdeh}
\end{figure}

We now derive limiting behaviours for the rate.
Asymptotic
analysis of \eqref{eq:SaddleEq} using \eqref{eq:GLE3LimitingBehaviours} 
%and \eqref{eq:RateAsymptConvenient} and 
leads to (the analysis is quite similar to the two previous sections)
\begin{equation}
  \sigma_* \simeq
  \begin{cases}
    - \frac{1}{\eta} \tilde{h}  - \frac{1}{4 \sqrt{\eta |\tilde{h}|}} +\mathcal{O}(|\tilde{h}|^{-3/2})
     & \mbox{for } \tilde{h} \to-\infty
     \\[0.25cm]
     -\frac{1}{\eta+\widetilde{\Lambda}''(\rmass) + 1/(8 \rmass^{3/2}) }(\tilde{h} - \tilde h^*)
     & \mbox{for } \tilde{h} \sim \tilde h^*
     \\[0.25cm]
     - \rmass + \frac{1}{16 (\tilde h + \widetilde{\Lambda}'(0) - \eta \rmass)^2} + \mathcal{O}(\tilde h^{-5}) 
     & \mbox{for } \tilde{h} \to+\infty
  \end{cases}
\end{equation}
Correspondingly, using \eqref{eq:10.42} we deduce the following behaviours for the large deviation rate function 
\begin{align}
  \label{eq:rDEh}
  &L_c \, \left[ \rate(m^2;h)  - \rate(m^2) \right] 
  \\
  \nonumber
  &
  \simeq   
  \begin{cases}
    \displaystyle
    - \frac{1}{2\eta}\,\tilde{h}^2 + \frac{1}{2} \sqrt{|\tilde{h}|/\eta} 
    + \frac{1}{2}\sqrt{\rmass} -\widetilde{\Lambda}(\rmass) + \mathcal{O}(|\tilde{h}|^{-1/2}) 
     & \mbox{for } \tilde{h} \to-\infty
    \\[0.25cm]
    \displaystyle
    -\frac{1}{2\big[\eta+\widetilde{\Lambda}''(\rmass) + 1/(8 \rmass^{3/2})\big]}
    \left(\tilde{h}    - \tilde h^*\right)^2 
     & \mbox{for } \tilde{h} \sim \tilde h^* \nonumber
    \\[0.25cm]
    \displaystyle
     - \rmass \tilde h
        +\frac{1}{2}\sqrt{\rmass} 
        - \widetilde{\Lambda}(\rmass) + \frac{1}{2}\eta \rmass^2 + \widetilde{\Lambda}(0) 
      % - \frac{1}{16 \tilde h} 
        + \mathcal{O}(\tilde h^{-1})
     & \mbox{for } \tilde{h} \to +\infty 
  \end{cases}
\end{align}
Using the data of the numerical calculation (Sections~\ref{subsec:FPDiscrete} and \ref{subsec:DiffEq}), we have solved numerically \eqref{eq:SaddleEq} and computed the rate \eqref{eq:10.42}~: the result is plotted in Fig.~\ref{fig:rdeh}. We see that the agreement with the limiting behaviours discussed in the text is excellent.

%prg NtotDeH.nb

Let us discuss some main features of this result. In the limit of large total negative energy density we thus obtain the dominant term [corresponding to $-(\eta/2)\sigma_*^2$ in Eq.~\eqref{eq:10.42}]
\begin{equation}
  \rate(m^2;h) 
  \simeq - \frac{h^2}{2R(0)}
  \hspace{0.5cm}\mbox{for }
  h\to-\infty
  \:.
\end{equation}
which is identical to the leading behaviour for large negative disorder energy from \eqref{eq:rdDEhd2}.
Since the elastic energy is positive, it is reasonable to expect that the two tails coincide to
leading order. Notice by comparing \eqref{eq:rDEh} and \eqref{eq:rdDEhd} that they differ
however by the prefactor of the next (subleading) order. 

Consider now the limit of large positive total energy density.
From \eqref{eq:rDEh} we obtain
\begin{equation}
  \label{eq:ldfLargePositiveH}
  \rate(m^2;h) \simeq - \frac{\rmass}{L_c}\,\tilde{h}
  = - \frac{m^2}{|R''(0)|}\,h
  \hspace{0.5cm}\mbox{for }
  h\to+\infty
  \:.
\end{equation}
This is the same behaviour as \eqref{eq:reDEhe}, obtained for the elastic energy.
Again it is quite reasonable that these two tails coincide, since the disorder energy rate function was found to
decay much faster at large positive disorder energy density than the elastic energy
one. Hence the elastic energy dominates in that regime. Note however that the
next (subleading) order term, which is $\mathcal{O}(1)$, 
is different for the total and the elastic energy rate functions.

As was already discussed in the context of the elastic energy, the 
decay of the large deviation function with $h$ is controlled by the mass~:
when the mass vanishes, $\rate(0;h)$ is a monotonous function which saturates  
\begin{equation}
  \label{eq:ldfLimit2}
  \rate(0;h) \simeq \frac{C}{L_c} - \frac{|R''(0)|}{16\kappa\,h}
  \hspace{0.5cm}\mbox{for }
  h\to+\infty
  \:.
\end{equation}
%(see Fig.~\ref{fig:LDFzeroMASS}).
Because $\mean{ \mathcal{N}_\mathrm{tot}(H) }/\mean{ \mathcal{N}_\mathrm{tot} }$ must be a decreasing function of $H$ for $H\to\pm\infty$, the saturation of the large deviation function \eqref{eq:ldfLimit2} for $h\to+\infty$ needs some explanation, provided at the end of the Subsection~\ref{subsec:DistribHe}.
It has the same physical origin as discussed above for the elastic energy (large wandering
of the elastic line at equilibria in the massless case leading to $H \sim L^2$).

As can be seen on Fig. \ref{fig:rdeh}, the rate $\rate(m^2;h)$ is positive for 
$h \in [h_\mathrm{min},h_\mathrm{max}]$.
%\color{black}
%and becomes negative below $\tilde h_0$. 
The value $h_\mathrm{min}$ provides \textit{a lower bound} on the energy density
of the ground state $h_\mathrm{GS}$. Indeed one has the convexity exact bound
\begin{equation} 
  \ln \, \mean{ \mathcal{N}_\mathrm{tot}(H) }
  \geq \mean{ \ln \mathcal{N}_\mathrm{tot}(H) }
\end{equation}
Since when decreasing $H= h L$ the l.h.s vanishes for $h = h_\mathrm{min}$, it implies that 
the r.h.s has already vanished at that point. Since the r.h.s. vanishes when $h$ 
reaches the typical ground state energy density $h_\mathrm{GS}$ we deduce that
$h_\mathrm{min}\leq h_\mathrm{GS}$. 
We now assume small mass $m\to0$. 
For $\eta=1$, we get numerically $\tilde{h}_\mathrm{min}\simeq-0.97$ for $\rmass\to0$ (Fig.~\ref{fig:rdeh}). 
If we assume $\eta\gg1$, we can use the asymptotic \eqref{eq:rDEh} (for $\tilde{h}\to-\infty$) and deduce $\tilde{h}_\mathrm{min} \simeq - \eta^{1/3}$. 
Reintroducing the dimensional factors, we conclude that the lower bound on the ground state energy density is
\begin{equation}
  h_\mathrm{min} \simeq 
  \begin{cases}
     -0.97\, \frac{|R''(0)|}{(R''''(0)\kappa)^{1/3}}   & \mbox{ and }  \eta=1
     \\[0.25cm]
     - \left( \frac{R(0)|R''(0)|}{\kappa} \right)^{1/3} & \mbox{ and }  \eta\gg1
  \end{cases}  
  \hspace{1cm}\mbox{for }
  L_c\ll L_m
  \:.
\end{equation}
Similarly, $h_\mathrm{max}$ provides an \textit{upper bound} on the highest equilibrium energy density, $h_\mathrm{max}\geq h_\mathrm{HS}$.
For small mass, we can use the asymptotic \eqref{eq:rDEh} (for $\tilde{h}\to+\infty$) to get $\tilde h_\mathrm{max} \simeq C/\rmass\gg1$, corresponding to 
\begin{equation}
  h_\mathrm{max} \simeq +C\,\frac{|R''(0)|\,R''''(0)^{1/3}}{m^2 \kappa^{2/3}}
  \hspace{1cm}\mbox{for }
  L_c\ll L_m
  \:.
\end{equation}

It is interesting to compare the bound $\tilde{h}_\mathrm{min}$ ($\simeq-0.97$ for $\eta=1$ and $\simeq - \eta^{1/3}$, for $\rmass=0$) for the total ground state energy density with the similar bound obtained above for the disorder energy density, 
$\tilde{h}_d^\mathrm{min}\simeq-1.503$ for $\eta=1$ and  
$\tilde{h}_d^\mathrm{min}\simeq-1.587\,\eta^{1/3}$ for $\eta\gg1$.
We have $\tilde{h}_\mathrm{min}>\tilde{h}_d^\mathrm{min}$ as it should.
However the comparison of the two bounds suggest that the elastic energy and the disorder energy are comparable for the \textit{atypical} configurations of the line corresponding to the ground state.
This is very different for the \textit{typical} configurations, which are characterized by typical disorder energy density $\tilde{h}_d^*=- \widetilde{\Lambda}'(0)\simeq-0.47$ while we show below that the typical elastic energy density is very large, growing with the length $\tilde{h}_e^*\sim L/L_c$, cf.~Eq.~\eqref{eq:TypicalElasticEnergyAtZeroMass}.

%{\red here it would be nice to give the numerical values for $h_\mathrm{min}$ and $h_\mathrm{max}$
%from the figure.}
%\blue{Finally note that the same arguments can be made for the elastic and disorder energy, separately,
%providing also bounds for the states with lowest elastic and disorder energy} \red{ I had that idea
%but is that correct? Can you make it precise?}
%
%\red{Pierre says: I am not too clear about what is written below. Christophe, can you rewrite what you want
%to keep in view of what is above? I left all the material below. I do not understand at all
%your estimates for $h_\mathrm{min}$ and $h_\mathrm{max}$, they look
%interesting but I dont understand how you get them, how you control them, which limit etc.. and
%if they agree with the numerics.}

\subsection{Mean number of equilibria at fixed value of the elastic energy for any $L$} 
\label{subsec:DistribHe}

Let us study the mean number of equilibria at fixed value of the elastic energy $H_e$.
As discussed in~\ref{app:Larkin},
the annealed distribution of elastic energy $H_e$ for our model, exactly coincides with the one of the Larkin model.
The present calculation, however, which treats carefully the boundary conditions,
has not, to our knowledge, been reported previously, even in the context of the Larkin model.

Setting $s_d=0$ in \eqref{eq:DetRepres00}, we obtain
\begin{equation}
  \label{eq:DetRepres4}
  {
   \int\D H_e\D H_d \,  
   \mean{ \mathcal{N}_\mathrm{tot}(H_e,H_d) }
   \,  
   \EXP{-s_e H_e}  
   }
   =  \mean{ \mathcal{N}_\mathrm{tot} }
   \,
   \sqrt{ 
      \frac{ \det \left( {m^2}/\kappa - \partial^2_\tau \right) }
           {  \det \left( ({m^2} + s_e |R''(0)|)/\kappa  - \partial^2_\tau \right)}
         }
\end{equation}
Note that since the elastic energy is positive, these are bona-fide Laplace transforms. 
We can now use the {expressions of these determinants, Eq.~\eqref{eq:FreeDetVariousBC}.}

\subsubsection{Finite mass $m$}

Let us focus on periodic boundary conditions, 
which leads to the simplest formula, and indicate the results for others. 
Then we have, with $L_m=\sqrt{\kappa}/m$ 
\begin{equation}
%\label{eq:DetRepres3}  %%%% DOUBLE LABEL...
  {
   \int\D H_e\D H_d \,  
   \mean{ \mathcal{N}_\mathrm{tot}(H_e,H_d) }
   \,  
   \EXP{-s_e H_e}  
   }
 =  \mean{\mathcal{N}_\mathrm{tot} }
 \,\frac{\sinh L/2L_m}{ \sinh ( \sqrt{ 1 + \frac{s_e |R''(0)|}{m^2}} L/2L_m )}
 \end{equation}
The natural unit of ${H_e}$ is ${H_e}_m={|R''(0)|}/{m^2}$.
The quantity of most interest is the annealed (i.e. over samples) probability distribution, $\mathcal{P}_{L,m}(H_e)$,
that an equilibrium chosen at random has elastic energy $H_e$,
%mean number of equilibria with value $H_e$ of the elastic energy 
takes the form (for all classes of boundary conditions) 
\begin{equation}
  \mathcal{P}_{L,m}(H_e)
  =
   \int\D H_d \,  
   \frac{ \mean{ \mathcal{N}_\mathrm{tot}(H_e,H_d) } }{ \mean{\mathcal{N}_\mathrm{tot} } }
  = \frac{1}{{H_e}_m} f_{L/L_m}({H_e}/{H_e}_m)  
\end{equation}
For the periodic case, we have to compute the inverse Laplace transform
\begin{equation}
  f^\mathrm{Per}_\ell (\reshe) = 
  \mathscr{L}^{-1}_{s \to \reshe}\left[   \frac{\sinh (\ell/2)}{ \sinh ( \sqrt{ 1 + s}\, \ell/2 )} \right]
  \:.
\end{equation}
Note that $f_\ell(\reshe)$ integrates to unity since it is a probability distribution for 
the scaled elastic energy variable $\reshe$.
%can thus 
%be interpreted as a probability distribution for $\reshe$: similarly for ${H_e}$
%the ratio  
%%$\mean{\mathcal{N}({H_e}) }/\mean{\mathcal{N}_\mathrm{tot} }$ 
%\CGre{
%$\int\D H_d \,\mean{ \mathcal{N}_\mathrm{tot}(H_e,H_d) }/\mean{\mathcal{N}_\mathrm{tot} } $ 
%}
%is the
%(annealed-i.e. over samples) probability density that if one picks one equilibrium at random, its
%elastic energy is $H_e$. 

Let us also indicate the result for Dir/Dir 
\begin{equation}
f^\mathrm{D/D}_\ell(\reshe) = \mathscr{L}^{-1}_{s \to \reshe}\left[
  \sqrt{ \frac{\sqrt{1+s}\,\sinh \ell}{ \sinh ( \sqrt{ 1 + s}\, \ell )}}
 \right]
\end{equation}
The above formulae can be inverted in principle to obtain  $\mathcal{P}_{L,m}(H_e)$
%\CGre{
%$\int\D H_d \,\mean{ \mathcal{N}_\mathrm{tot}(H_e,H_d) }/\mean{\mathcal{N}_\mathrm{tot} } $ 
%}
for any $L$. 
{For periodic boundary condition,} %instance, 
using 
$$
\frac{1}{\sinh(z)} = \sum_{n=-\infty}^{+\infty} \frac{(-1)^n}{z+ \I n \pi}
$$
we obtain 
%\\
%\red{\textbf{Pierre:} I leave all 3 lines so you can check, but one
%can comment later (dont erase it!) the second line.. note that I tripled checked that formula}
\begin{align}
 f^\mathrm{Per}_\ell(\reshe) 
 &=  \sinh (\ell/2) \int_0^{+\infty} \D t 
 \sum_{n=-\infty}^{+\infty} (-1)^n \EXP{- \I n \pi t} 
  \mathscr{L}^{-1}_{s \to \reshe}\left[  \EXP{- \frac{1}{2}\ell t \sqrt{1+s}} \right]
  \nonumber 
  \\
  & =  \sinh (\ell/2) \frac{\ell}{4 \sqrt{\pi} \reshe^{3/2}} \EXP{-\reshe} 
  \int_0^{+\infty} \D t \sum_{n=-\infty}^{+\infty} 
  (-1)^n \EXP{- \I n \pi t}  t \EXP{- \ell^2t^2/(16 \reshe) }
  \nonumber 
  \\
  & = \sinh (\ell/2)  \sum_{n=-\infty}^{+\infty} (-1)^n
  \frac{2 \EXP{-\reshe} \left(\ell-4 \pi  n \sqrt{\reshe} \,
  F\left(\frac{2 n \pi \sqrt{\reshe}}{\ell}\right)\right)}{\sqrt{\pi } \ell^2 \sqrt{\reshe}}
\end{align}
where $F(x)=\EXP{-x^2} \int_0^x\D y\, \EXP{-y^2}$ the Dawson integral (FDawson in {\tt Mathematica}). 
Let us further discuss two limits:\\

In the limit of small $\ell=L/L_m \to 0$ one sees that the result for the Dirichlet (D/D) boundary conditions converges to $\lim_{\ell \to 0} f^\mathrm{D/D}_\ell(\reshe) =\delta(\reshe)$ while for the periodic (Per) converges to a non trivial fixed distribution, $\lim_{\ell \to 0} f^\mathrm{Per}_\ell(\reshe) = \EXP{-\reshe}/\sqrt{\pi \reshe}$. Presumably this is related to the fact that in the later case the center of mass degree of freedom is
still free even in that limit, and effectively reduces to a zero dimensional model. \\
%\\
%\red{\textbf{Pierre:} understand this point better}.
%\\

In the limit of large $\ell = L/L_m \gg 1$ one finds
\begin{equation}
\label{ff} 
  f^\mathrm{Per}_\ell(\reshe) 
  \underset{\ell \gg 1}{\simeq}
  \mathscr{L}^{-1}_{s \to \reshe} \left[  \EXP{(1- \sqrt{ 1 + s}) \ell/2} \right]
  = \frac{1}{4 \sqrt{\pi} \reshe^{3/2}} \EXP{- \frac{(\reshe-\ell/4)^2}{\reshe}}  
\end{equation}
%which, incidentally, is the same distribution as the avalanche size distribution in the mean field ABBM model.
From this we conclude that for $L/L_m \gg 1$ the leading (i.e. {\it typical}) fluctuations of the elastic energy
are Gaussian 
\begin{equation}
  H_e \simeq \Hem \left( \frac{L}{4 L_m} + \sqrt{ \frac{L}{8 L_m}  } \chi \right)  
  \quad , \quad \Hem = \frac{|R''(0)|}{m^2}
\end{equation}
where $\chi$ is a Gaussian random variable of unit variance. 
The form \eqref{ff} is furthermore consistent with the {\it large deviation} rate function
(in the notations of the previous section) associated to $H_e= h_e L$, that is
\begin{equation}
  { \rate_e(m^2;h_e)  }
  = \rate(m^2) - \frac{\Hem}{h_e} \left( \frac{h_e}{\Hem} - \frac{1}{4 L_m}\right)^2
  \:, 
\end{equation}
which indeed coincides with the Legendre transform w.r.t $s_e$ of $\Gamma(s_e,s_d=0)$ in 
Eq.~\eqref{resGamma}, {cf. Eq.~\eqref{eq:reDEhe}.}

\subsubsection{Case $m=0$}
\label{subsec:TheLastSubsection}

In this final paragraph, we consider the distribution of the elastic energy in the zero mass limit. 
One of our aim is here to clarify the origin of the saturation of the rate $\rate(m^2;h)$, Eq.~\eqref{eq:ldfLimit2}, or equivalently of the rate $\rate_e(m^2;h_e)$, as the tail of the distribution of the total energy is controlled by the distribution of the elastic energy.
We consider the case of Neumann/Dirichlet boundary conditions, which describes the case where the line is attached by one end, the other end being free.
We choose these boundary conditions for the simplicity of the determinant, Eq.~\eqref{eq:FreeDetVariousBC}.
From \eqref{eq:DetRepres4}, we see that the distribution then takes the form
\begin{equation}
  \label{eq:ILTwithNoName}
  \mathcal{P}_{L,0} (H_e)
  = \frac{\kappa}{|R''(0)|L^2} \, 
  \varphi\!\left(\frac{\kappa\,H_e}{|R''(0)|L^2} \right)
  \hspace{0.5cm}\mbox{where}\hspace{0.5cm}
  \varphi(\reshe)=\mathscr{L}^{-1}_{s \to \reshe} \left[  \frac{1}{\sqrt{\cosh\sqrt{s}}} \right]
  \:.
\end{equation}
This expression shows that the typical elastic energy \textit{density} grows with the length
\begin{equation}
  \label{eq:TypicalElasticEnergyAtZeroMass}
  h_e^* \sim \frac{|R''(0)|\,L}{\kappa}
  \:,
  \hspace{1cm}\mbox{i.e.}\hspace{1cm}
  \tilde{h}_e^* \sim \frac{L}{L_c}
\end{equation}
in terms of the dimensionless variable introduced above.

In~\ref{app:AnInvLT}, we show that the limiting behaviours of the inverse Laplace transform~\eqref{eq:ILTwithNoName} are 
\begin{align}
  \varphi(\reshe) \sim 
  \begin{cases}
    \frac{1}{\reshe} \, \exp\left\{-1/(16\reshe)\right\} 
      & \mbox{for } \reshe \to 0 
    \\
    \frac{1}{\sqrt{\reshe}} \, \exp\left\{ -(\pi/2)^2\reshe \right\} 
      & \mbox{for } \reshe \to \infty
  \end{cases}
  \:.
\end{align}
If the distribution is rewritten under the form 
$\mathcal{P}_{L,0}(H_e)\sim\exp\{L\,\big[\rate_e(m^2;h_e)-\rate(m^2)\big]\}$ for $L\to\infty$, we have 
\begin{align}
  \rate_e(m^2;h_e) - \rate(m^2)
  \simeq
  \begin{cases}
    - \frac{ |R''(0)| }{16\kappa\,h_e}      & \mbox{for } h_e \ll |R''(0)|L/\kappa
    \\[0.2cm]
    - \frac{\pi^2\kappa\,h_e}{4|R''(0)|L^2} & \mbox{for }  h_e \gg |R''(0)|L/\kappa
  \end{cases}  
\end{align}
The first behaviour is in exact correspondence with Eq.~\eqref{eq:ldfLimit2} or Eq.~\eqref{eq:reDEhe} for $m=0$.
The second behaviour ensures the decay of $\mathcal{P}_{L,0}(H_e)$ for large $H_e$, as it should. 
This makes clear that the saturation of the rate Eq.~\eqref{eq:ldfLimit2} only reflects a subtlety concerning the order of the two limits $L\to\infty$ and $h_e\to\infty$ when $m=0$.

%%%%%%%%%%%%%%%%%%%%%%%%%%%%%%%%%%%%%%%%%%%%%%%%%%%%%%%%%%%%%%%%%%%%%%%%%%%%%%%%%%%%%%%%%%
%%%%%%%%%%%%%%%%%%%%%%%%%%%%%%%%%%%%%%%%%%%%%%%%%%%%%%%%%%%%%%%%%%%%%%%%%%%%%%%%%%%%%%%%%%

\section{Conclusion}

By extending the Kac-Rice approach to manifolds of finite internal dimension, 
we have shown that the mean number of equilibria of an elastic line in
a random potential in presence of a parabolic confinement, 
grows exponentially with its length, and developed a theory
to calculate the rate of growth $\rate$. The mean number equilibria is shown to be related to 
the expectation value of the modulus of the determinant of the Laplace operator in
presence of diagonal disorder, the same operator which features in the Anderson 
localization problem. The relation is valid for elastic interface of arbitrary internal dimension 
$d$. In $d=1$ (elastic line) using the Gelfand-Yaglom theorem, 
the rate of growth can be related with the large deviation
function of the Lyapunov exponent fluctuations associated to a 1D random Schr\"odinger
problem, but in the negative energy range.
This large deviation function (generalized Lyapunov exponent GLE) is 
given by the lowest eigenvalue of an associated Fokker-Planck operator
which is analysed in detail by several complementary techniques.

From these methods, we found that the rate $\rate$ is described by a universal function of
the disorder strength for which we obtained analytical and numerical results.
The disorder strength is naturally parameterized using the Larkin length $L_c$ and
the dimensionless control parameter is the ratio of $L_c$ to the length
$L_m$ imposed by the parabolic confining potential. We extract
analytically the asymptotic behaviours of this scaling function for
small and large value of the argument. For strong confinement, the rate $\rate$ 
is small and given by a non-perturbative (instanton, Lifshitz tail-like) contribution to GLE. 
For weak confinement, the rate $\rate$ is found to be proportional to the inverse Larkin length of the pinning theory. We have also discussed the question of counting of stable equilibria.
Finally, we have shown how to extend the method to calculate the asymptotic number of equilibria at
fixed energy~:
we have obtained large deviation rate functions controlling the number of equilibria constrained by either the total, the elastic or the disorder energy.
These large deviation functions control the distribution of the (total, elastic or disorder) energy density of the line at equilibrium. Some connections with the Larkin model have been discussed.

One of the most interesting application is to study the evolution of the mean number of
equilibria as an external force $f$ is turned on. We find that a ``topology trivialization'' phenomenon
occurs, and the rate associated to the mean number of equilibria decreases to zero at
some threshold force $f=f_c^\mathrm{tot}$ which can be calculated from
the above methods. This is related to the so-called depinning
transition, which occurs at $f=f_c$, at which all stable equilibria disappear. 
As a result we showed that $f_c^\mathrm{tot}$ provides an exact upper bound to $f_c$,
differing only by a factor of order unity. This prediction is confirmed numerically.

From the point of view of Anderson localization, we have also discovered several interesting properties of the generalized Lyapunov exponent (GLE) $\Lambda(q)$.
We have shown a crucial difference between the GLE for $q=1$, which is a non-analytic function of the disorder strength $D$ (for $E\to-\infty$) while $\Lambda(q)$ is analytic when $q$ is an even integer.
It would be interesting to see if the difference persists for larger odd integer arguments.
We were able to obtain few exact results for even integer $q$.
The limit of large positive argument $q\to+\infty$ was analysed, which is related to the large deviations of the wave function (solution of the Cauchy initial value problem) for large values.
The analysis of the limit $q\to-\infty$ of the GLE has remained an open question, which can be related to the large deviations of the wave function for small values.

The connection discussed here with the generalized Lyapunov exponents of a localization
problem opens a bridge between pinning theory and localization theory
that should inspire further works.

%%%%%%%%%%%%%%%%%%%%%%%%%%%%%%%%%%%%%%%%%%%%%%%%%%%%%%%%%%%%%%%%%%%%%%%%%%%%%%%%%%%%%%%%%%
%%%%%%%%%%%%%%%%%%%%%%%%%%%%%%%%%%%%%%%%%%%%%%%%%%%%%%%%%%%%%%%%%%%%%%%%%%%%%%%%%%%%%%%%%%

\section*{Acknowledgements}

CT acknowledges stimulating discussions and suggestions from Philippe Bougerol, Alain Comtet, Aur\'elien Grabsch, Jean-Marc Luck, Nicolas Pavloff, Yves Tourigny and Denis Ullmo.
We thank David Saykin for sharing his results and writing~\ref{app:Saykin}.
We thank the PCMI Summer School 2017, where some of this work was performed.
The research at King's College London was supported by EPSRC grant EP/N009436/1 The many faces of random characteristic polynomials. 
This research was also supported by ANR grant ANR-17-CE30-0027-01 RaMaTraF.
We are grateful to an anonymous referee for numerous useful remarks.

%%%%%%%%%%%%%%%%%%%%%%%%%%%%%%%%%%%%%%%%%%%%%%%%%%%%%%%%%%%%%%%%%%%%%%%%%%%%%%%%%%%%%%%%%%
%%%%%%%%%%%%%%%%%%%%%%%%%%%%%%%%%%%%%%%%%%%%%%%%%%%%%%%%%%%%%%%%%%%%%%%%%%%%%%%%%%%%%%%%%%
%%%%%%%%%%%%%%%%%%%%%%%%%%%%%%%%%%%%%%%%%%%%%%%%%%%%%%%%%%%%%%%%%%%%%%%%%%%%%%%%%%%%%%%%%%

\appendix

%%%%%%%%%%%%%%%%%%%%%%%%%%%%%%%%%%%%%%%%%%%%%%%%%%%%%%%%%%%%%%%%%%%%%%%%%%%%%%%%%%%%%%%%%%
\section{Boundary conditions and determinants}
\label{app:bc}

In this section, we discuss several formulae for the determinant $\det (H-E)$ of the Schr\"odinger operator
to justify and make more precise the formula of the main text and of \ref{app:PinningDirichlet} below.

Consider the discrete Hamiltonian which appears in the text, $H_{i,j} = - \Delta_{i,j} + U_{i} \delta_{i,j}$
and the associated Schr\"odinger equation
\begin{equation}
  \label{eq:DiscreteSchrod}
  \sum_j H_{i,j} \psi_j =   -\psi_{i+1} + (2 + U_i) \psi_i - \psi_{i-1} =E \, \psi_i
\end{equation}
for $i\in\{1,\cdots,K\}$~;
boundary conditions set the values for $\psi_0$ and~$\psi_{K+1}$.

\subsection{Boundary conditions}

There are three natural boundary conditions for the elastic line problem
studied in this paper.
We detail them here in the discrete and continuum settings.

\subsubsection{Dirichlet boundary conditions}

The pinned boundary conditions correspond to setting $\psi_0=\psi_{K+1}=0$ in \eqref{eq:DiscreteSchrod}
($u_0=u_{K+1}=0$ with the notations of the body of the paper).
The associated $K \times K$ Laplacian matrix $\Delta = \Delta_\mathrm{Dir}$ then reads~:
\begin{equation}
\Delta_\mathrm{Dir} =
\begin{pmatrix}
  -2 & 1  & 0 & \cdots & 0 \\
  1  & -2 &  1 & 0 & \cdots \\
 % 0  & 1  & -2 & 1 &  \\
  0 &  1 & \ddots & \ddots & \\
  \vdots & \ddots & \ddots &  -2 & 1  \\
 0  &  \cdots & 0  & 1 & -2
\end{pmatrix}
\end{equation}
In the continuum limit it corresponds to the usual Laplacian with Dirichlet boundary conditions
$\psi(0)=\psi(L)=0$.

\subsubsection{Neumann boundary conditions}

  The free boundary conditions~:
  set $\psi_0=\psi_1$ and $\psi_K=\psi_{K+1}$ in \eqref{eq:DiscreteSchrod}, i.e.  the associated Laplacian $\Delta = \Delta_\mathrm{Neu}$ is
\begin{equation}
\Delta_\mathrm{Neu} =
\begin{pmatrix}
  -1 & 1  & 0 & \cdots & 0 \\
  1  & -2 &  1 & 0 & \cdots \\
 % 0  & 1  & -2 & 1 &  \\
  0 &  1 & \ddots & \ddots \\
  \vdots & \ddots & \ddots &  -2 & 1  \\
 0  &  \cdots & 0  & 1 & -1
\end{pmatrix}
\end{equation}
In the continuum limit it
corresponds to usual Laplacian with Neumann boundary conditions
$\psi'(0)=\psi'(L)=0$.

\subsubsection{Periodic boundary conditions}

  The periodic boundary conditions, $\psi_0=\psi_K$, $\psi_{K+1}=\psi_1$, with the associated
Laplacian $\Delta = \Delta_\mathrm{Per}$ given by
\begin{equation}
\Delta_\mathrm{Per} =
\begin{pmatrix}
  -2 & 1  & 0 & \cdots & 1 \\
  1  & -2 &  1 & 0 & \cdots \\
 % 0  & 1  & -2 & 1 &  \\
  0 &  1 & \ddots & \ddots \\
  \vdots & \ddots & \ddots &  -2 & 1  \\
 1  &  \cdots & 0  & 1 & -2
\end{pmatrix}
\end{equation}
In the continuum limit this corresponds to usual Laplacian for
$\psi(0)=\psi(L)$ and $\psi'(0)=\psi'(L)$.

\subsection{Determinants}

We now discuss separately the formulae for the determinant $\det (H-E)$ for the three types of boundary conditions.

\subsubsection{Pinned boundary conditions (Dirichlet)}

We denote by $y_i(E)$ the solution of the initial value problem with
\begin{equation}
  y_0=0
  \hspace{1cm}\mbox{and}\hspace{1cm}
  y_1=1
  \:.
\end{equation}
The $K$ eigenvalues $\{E_\alpha\}_{\alpha=1,\cdots,K}$ of the $K\times K$ Hamiltonian $H$ are the solutions of 
the quantization condition
\begin{equation}
    y_{K+1}(E)=0
    \:.
\end{equation}
We see by recursion that $y_2(E)=2+ U_1-E$,  $y_3(E)=(2+U_2-E)(2+U_1-E)-1$, etc.
By induction, it is straightforward to prove that $y_{j+1}(E)$ is a polynomial of degree $j$ in $E$ with higher degree term $(-E)^j$.
Using these remarks we 
can write
\begin{equation}
  \label{eq:DisDirDet}
  \det(H - E) = \prod_{\alpha=1}^K(E_\alpha - E) = y_{K+1}(E)
\end{equation}
%where the $E_\alpha$ are the eigenvalues of $H$.
This can now be used for the discrete polymer model, formula \eqref{eq:DetRepres} in the text, with the correspondence $E=-m^2/\kappa$.
As a simple illustration consider the free case with $U_n=-2$. We deduce $y_n=\sin qn/\sin q$ where $E=-2\cos q$.
As a consequence the Dirichlet determinant is
$\det(H-E)=\sin q(K+1)/\sin q$, corresponding to the spectrum $q_n=n\pi/(K+1)$ with $n=1,\cdots,K$.

Denoting $y(\tau)$ the solution of $(H-E) y(\tau)=0$ with initial conditions
\begin{equation}
 y(0)=0
  \hspace{1cm}\mbox{and}\hspace{1cm}
 y'(0)=1
 \:,
\end{equation}
we see that Eq.~\eqref{eq:DisDirDet} is obviously the discrete version of the Gelfand-Yaglom formula
\begin{equation}
  \det(H - E) = 2 y(L)
  \:.
\end{equation}
up to a $E$ independent multiplicative factor which depends on the ultraviolet regularization (the factor chosen here and formulae given below correspond to zeta regularization \cite{Tex10,HarKirTex12}).
Let us illustrate the formula in the free case where $H=-\partial_\tau^2$~:
setting $\gamma=-E$ for convenience for the following, we find $y(\tau)=\sinh(\sqrt{\gamma}\tau)/\sqrt{\gamma}$ thus
$\det(\gamma-\partial_\tau^2)=2\sinh(\sqrt{\gamma}L)/\sqrt{\gamma}$.
Rewriting the hyperbolic function as an infinite product
\begin{equation}
  \det(\gamma-\partial_\tau^2)
  =
  2L \prod_{n=1}^\infty \bigg( 1 + \frac{\gamma}{q_n^2}\bigg)
  \,,
  \hspace{0.5cm}\mbox{where }
  q_n = \frac{n\pi}{L}
  \hspace{0.25cm}\mbox{for }
  n\in\mathbb{N}^*
  \:.
\end{equation}
We recognize the eigenvalues of the Laplacian.
In this case we recall the eigenfunctions, $-\partial_\tau^2\psi_n(\tau)=q_n^2\psi_n(\tau)$,
\begin{equation}
  \psi_n(\tau) = \sqrt{\frac{2}{L}}\, \sin\left(q_n\tau\right)
  \mbox{ with }
  n\in\mathbb{N}^*
  \:.
\end{equation}

\subsubsection{Free boundary conditions (Neumann)}

 $\psi_0=\psi_1$ and $\psi_K = \psi_{K+1}$.
We now denote by $x_i(E)$ the solution of the initial value problem with
\begin{equation}
 x_0= x_1=1
\end{equation}
Now the quantization condition is
\begin{equation}
  x_{K+1}(E)=x_K(E)
\end{equation}
Similar argument as above gives
\begin{equation}
  \label{eq:DisNeuDet}
  \det(H - E) = x_{K+1}(E)-x_K(E)
  \:.
\end{equation}

In the continuum limit, we consider the solution $x(\tau)$ of the
differential equation $(H-E) x(\tau)=0$, where $H= - \deriv{^2}{\tau^2} + U(\tau)$,
with initial conditions
\begin{equation}
 x(0)=1
  \hspace{1cm}\mbox{and}\hspace{1cm}
 x'(0)=0
\end{equation}
The determinant is then given by
\begin{equation}
  \det(H - E) = 2\,x'(L)
\end{equation}
%up to a $E$ independent multiplicative factor which depends on the ultraviolet regularization (the factor chosen here and corresponds to zeta regularization \cite{HarKirTex12}).
As in the text, a regularization independent formula is obtained by forming ratio of two such determinants.
We illustrate the formula in the free case~:
setting again $E=-\gamma$ we find $x(\tau)=\cosh(\sqrt{\gamma}\tau)/\sqrt{\gamma}$, thus
$\det(\gamma-\partial_\tau^2)=2\sqrt{\gamma}\sinh(\sqrt{\gamma}L)$, which can also be expressed in terms of the spectrum of eigenvalues $\{q_n^2\}_{n\in\mathbb{N}}$ of the Laplacian:
\begin{equation}
  \begin{cases}
  \psi_0(\tau) = \frac{1}{\sqrt{L}} \\
  \psi_n(\tau) = \sqrt{\frac{2}{L}}\, \cos\left(q_n\tau\right)
  \end{cases}
  \,,
 \hspace{0.5cm}\mbox{where }
   q_n = \frac{n\pi}{L}
  \mbox{ with }
  n\in\mathbb{N}
\end{equation}

\subsubsection{Periodic boundary conditions}

For the sake of completeness, let us add a phase and now write the periodic boundary conditions as
\begin{equation}
  \label{eq:PeriodicBCflux}
  \psi_K=\psi_0\,\EXP{\I\theta}
  \hspace{1cm}\mbox{and}\hspace{1cm}
  \psi_{K+1} = \psi_{1}\,\,\EXP{\I\theta}
  \:,
\end{equation}
(this corresponds to a ring pierced by a magnetic flux).
The spectral analysis can be formulated conveniently by introducing two specific solutions $y_i(E)$ and $\tilde y_i(E)$ for two different initial value problems corresponding to
\begin{align}
  \begin{cases}
     y_0=0
     \\
     y_1=1
  \end{cases}
    \hspace{1cm}\mbox{and}\hspace{1cm}
  \begin{cases}
     \tilde y_{K+1}=0
     \\
     \tilde y_{K}=1
  \end{cases}
\end{align}
We have pointed out above that $y_n(E)$ is a polynomial of degree $n-1$ in $E$~; similarly $\tilde y_n(E)$ a polynomial of degree $K-n-1$.
The Wronskian $W_j = y_j \tilde y_{j+1} - y_{j+1} \tilde y_j $ of the two solutions is constant (starting from \eqref{eq:DiscreteSchrod}, one finds $W_{j+1}=W_j$), and is simply denoted $W=-\tilde y_0 =- y_{K+1}$ (which is the Dirichlet determinant).

The solution of the spectral problem can then be decomposed as $\psi_j= a\, y_j + b\, \tilde y_j$, where $a$ and $b$ are two constants.
Imposing the boundary conditions \eqref{eq:PeriodicBCflux} leads to the quantization equation
\begin{align}
  \det
  \begin{pmatrix}
    y_K                    & 1- \tilde y_0\,\EXP{\I\theta}
    \\
    \EXP{\I\theta}-y_{K+1} &  \tilde y_1 \,\EXP{\I\theta}
  \end{pmatrix} = 0
  \:.
\end{align}
Expanding this equation gives
\begin{equation}
  \label{eq:DisPerDet}
  \det(H-E)
   = \frac{1}{\tilde y_0} + \tilde y_0 - \frac{\tilde y_1\, y_{K}}{\tilde y_0}  - 2\,\cos\theta
   \:.
\end{equation}
As an example of application, we consider the free case ($U_n=-2$) where $y_n=\sin qn/\sin q=\tilde{y}_{K+1-n}$ for energy $E=-2\cos q$.
Some algebra gives $\det(H-E)=2\big(\cos qK-\cos\theta\big)=2(\,T_K(-E/2)-\cos\theta\big)$, where $T_K(x)$ is a Chebyshev polynomial.

We can proceed along the same lines in the continuum~:
we consider the two solutions $y(\tau)$ and $\tilde y(\tau)$
of the differential equation
with initial conditions
\begin{equation}
\label{bcper}
  \begin{cases}
     y(0)=0
     \\
     y'(0)=1
  \end{cases}
    \hspace{1cm}\mbox{and}\hspace{1cm}
  \begin{cases}
     \tilde y(L)=0
     \\
     \tilde y'(L)=-1
  \end{cases}
\end{equation}
The determinant is then given by
\begin{equation}
 % \boxed{
  \det(H - E) = y'(L) - \tilde y'(0) - 2\,\cos \theta
 %  }
\end{equation}
%(the precise $E$-independent factor corresponds to zeta regularization \cite{HarKirTex12}).
In the free case we have simply $\tilde{y}(\tau)=y(L-\tau)$ and we find
$\det(\gamma-\partial_\tau^2)=2\big[\cosh(\sqrt{\gamma}L) - \cos\theta\big]$,
which can again be expanded over the spectrum of eigenvalues %$(q_n-\theta/L)^2$:
\begin{equation}
  \det(\gamma-\partial_\tau^2)
  =
  4\sin^2(\theta/2) \prod_{n\in\mathbb{Z}} \bigg( 1 + \frac{\gamma}{(q_n-\theta/L)^2}\bigg)
\end{equation}
We recall also the related eigenfunctions for completeness
\begin{equation}
  \psi_n(\tau) = \frac{1}{\sqrt{L}}\, \EXP{\I q_n\tau}
  \,,
 \hspace{0.5cm}\mbox{where }
  q_n = \frac{2n\pi}{L}
  \mbox{ with }
  n\in\mathbb{Z}
  \:.
\end{equation}

For references on functional determinants, cf. Refs.~\cite{Des00,Tex10,HarKirTex12} and the review~\cite{ComDesTex05}.

%%%%%%%%%%%%%%%%%%%%%%%%%%%%%%%%%%%%%%%%%%%%%%%%%%%%%%%%%%%%%%%%%%%%%%%%%%%%%%%%%%%%%%%%%%
%%%%%%%%%%%%%%%%%%%%%%%%%%%%%%%%%%%%%%%%%%%%%%%%%%%%%%%%%%%%%%%%%%%%%%%%%%%%%%%%%%%%%%%%%%

\section{Pinning of the line for fixed boundary conditions}
\label{app:PinningDirichlet}

%\paragraph{Dirichlet boundary conditions.---}

We provide some additional information for the analysis of \S~\ref{subsec:ElasticLineContinuous}~: we show how the discussion can be extended to the case of a line with fixed endpoints.
We can also perform the calculation for an elastic line pinned at its two ends
(Dirichlet boundary conditions).
\footnote{
  although it is not the standard one for the depinning problem, where the elastic line is usually free to move, it is a usual setting in the related sandpile problem~\cite{Dha99,Pru12,LeDWie15}.
  }
The Fourier coefficients are
\begin{align}
  \tilde f_n &= \frac{2  f \sqrt{2 L} }{n \pi}\,, &  \quad n=2 p+1
    \\
             &= 0 \,,                             & n=2 p+2
\end{align}
for $p\in\mathbb{N}$.
Performing the sum over the modes, $q_n=n\pi/L$ with $n\in\mathbb{N}^*$, one finds that (minus) the argument of the
exponential in \eqref{eq:NiceFormula3} becomes
%\bea
%\frac{ L f^2}{2 |R''(0)|} \phi_L
%\eea
\begin{equation}
  \frac{ L f^2}{2 |R''(0)|}  \sum_{n\ \mathrm{odd}}
  \frac{8}{(n\pi)^2}
  \frac{1}{1+(L_w/L)^4[(n\pi)^2+(L/L_m)^2]^2}
  \:.
\end{equation}
In the limit $L_m/L\to\infty$ (vanishing mass) and $L/L_w\to\infty$, we find
\begin{equation}
  \frac{ L f^2}{2 |R''(0)|}  \sum_{n\ \mathrm{odd}}
  \frac{8}{(n\pi)^2}
  =  \frac{ L f^2}{2 |R''(0)|}
\end{equation}
%with, in the small mass limit, and $L \gg L_w$,
%\bea
%\phi_L \to \sum_{p=0}^{+\infty} \frac{8}{\pi^2 (2p+1)^2} =1
%\frac{1}{1+ \cot^2(2\theta)
%(1+ (\frac{L_m}{L} \pi (2 p+1))^2)^2} \nonumber
%\eea
leading to a value for the second term in \eqref{rr} {\it identical} to
the case of periodic and Neumann boundary conditions (despite the summation over
an infinite set of modes). In addition, we also expect that
$\Lambda(1)$, the first term in \eqref{rr} is independent of the
boundary conditions. For instance, in the framework of the Ricatti equation \eqref{Riccati}
defined in the text, the initial condition is $z(0)=-\infty$ for Dirichlet, and
$z(0)=0$ for Neumann boundary conditions. However, it is known that
the rate of growth, i.e. the GLE, does not depend on the
initial condition. Hence $\Lambda(1)$, and the rate $\rate_\mathrm{tot}(f)$ in
\eqref{rr} is the same for all three types of boundary conditions, leading to the same value
of $f_c^\mathrm{tot}$.

%%%%%%%%%%%%%%%%%%%%%%%%%%%%%%%%%%%%%%%%%%%%%%%%%%%%%%%%%%%%%%%%%%%%%%%%%%%%%%%%%%%%%%%%%%
%%%%%%%%%%%%%%%%%%%%%%%%%%%%%%%%%%%%%%%%%%%%%%%%%%%%%%%%%%%%%%%%%%%%%%%%%%%%%%%%%%%%%%%%%%

\section{The Larkin model}
\label{app:Larkin}

The Larkin model is a simplified model for pinning. Let us recall it here for an elastic line
in the continuum (discrete versions, and extensions to higher $d$ are immediate,
for reviews see Refs.~\cite{BlaFeiGesLarVin94,LeD11}). In the
Larkin model the non-linear random potential term in Eq.~\eqref{eq:H} is replaced 
simply by a linear random force term
\begin{equation}
  \label{eq:HLarkin}
  \mathcal{H}_\mathrm{Lar}[u(\tau)]
  =\int_0^L
   \D\tau
  \left[
      \frac{\kappa}{2}
      \left(
         \derivp{u(\tau)}{\tau}
       \right)^2
       + \frac{m^2}{2}u^2(\tau)
       - \phi(\tau)\, u(\tau)
  \right]
\end{equation}
which is centered Gaussian with correlator 
$\mean{\phi(\tau) \phi(\tau')} = - R''(0)\, \delta(\tau-\tau')$.
It amounts to expand $V(u,\tau)=V(0,\tau) + V'(0,\tau)\, u + \cdots$ and discard higher order terms. 
Being quadratic, there is a unique energy minimum (i.e. a single equilibrium)
\begin{equation}
  u_0(\tau) = \int_0^L \D\tau'\, 
  \bra{\tau}(m^2 - \kappa\, \partial_\tau^2)^{-1} \ket{\tau'}\,
  \phi(\tau') 
\end{equation}
which is distributed as a Gaussian field with correlator
%\begin{equation}
%  \overline{u_0(\tau) u_0(\tau')} = (m^2 - \kappa \partial_\tau^2)_{\tau \tau'}^{-2} \quad , \quad   \overline{(u_0(\tau) - u_0(\tau'))^2} =  \int \frac{dq}{ \pi} \frac{1-\cos q(\tau-\tau')}{(\kappa q^2 + m^2)^2} 
%\end{equation}
\begin{align}
  &\mean{u_0(\tau) u_0(\tau')} 
  = |R''(0)|\,\bra{\tau}(m^2 - \kappa\, \partial_\tau^2)^{-2} \ket{\tau'} 
  = |R''(0)|\,\int \frac{\D q}{2\pi} \frac{\cos q(\tau-\tau')}{( m^2+\kappa q^2 )^2} 
  \:,    
  \\
  &\mean{(u_0(\tau) - u_0(\tau'))^2} 
  = |R''(0)|\, \int \frac{\D q}{ \pi} \frac{1-\cos q(\tau-\tau')}{( m^2+\kappa q^2 )^2} 
\end{align}
where in the second equation we assumed $L$ large enough to ignore boundary conditions:
this integral behaves as $L_m^{2 \zeta_L}$ where $\zeta_L=3/2$ is the Larkin roughness exponent.

It is shown that at zero temperature the Larkin model,  
a zero dimensional version of the elastic line model \eqref{eq:H},
has the same correlation functions for the field $u(\tau)$, to all orders
in perturbation theory in the disorder, as 
the original model \eqref{eq:H}
(the so-called dimensional reduction phenomenon).
However it lacks the (non-perturbative) feature of multiple equilibria. Indeed, one notes that it depends on a single parameter, $R''(0)$, and lacks the physics of pinning arising from a non vanishing $R''''(0)$.
However, it does provide a good approximate description of the correlation functions of the true model (in its ground state) at scales smaller than $L_c$, or for $m \gg m_c$ (such that $L_{m_c}=L_c$), although even in this regime it lacks the non-perturbative corrections originating from rare events, such as the one
obtained in this paper. 

The elastic energy at the minimum is easily obtained as
\begin{equation} 
\label{HL} 
  H_e^\mathrm{Lar} = \frac{1}{2} \int_0^L\D\tau   \int_0^L \D\tau'   \,
  \phi(\tau)\,  \bra{\tau}(m^2 - \kappa\, \partial_\tau^2)^{-1} \ket{\tau'} \, \phi(\tau')
\end{equation}
and its Laplace transform is simply obtained by integration over the Gaussian field 
$\phi(\tau)$ as
\begin{equation}
  \big\langle \EXP{- s_e H_e^\mathrm{Lar}} \big\rangle
  =  \sqrt{ 
      \frac{ \det \left( {m^2}/\kappa - \partial^2_\tau \right) }
           {  \det \left( ({m^2} + s_e |R''(0)|)/\kappa  - \partial^2_\tau \right)}
         }
\end{equation}
which is exactly the same factor which occurs in \eqref{eq:DetRepres4}.
Thus the prediction of the Larkin model (which has a single equilibrium)
coincides {\it exactly} with the {\it annealed} probability distribution of
the elastic energy of the true model over all equilibria. 
As an example, by averaging \eqref{HL} over $\phi(\tau)$ 
one recovers its average (or most probable)
value as
\begin{equation}
  \lim_{L \to \infty} \frac{1}{L} \mean{H_e^\mathrm{Lar}} 
  \simeq \frac{|R''(0)|}{2} \int \frac{\D q}{2 \pi} \,
  \frac{1}{m^2+\kappa q^2 } = \frac{|R''(0)|}{4 m \sqrt{\kappa}} 
\end{equation}
which is exactly Eq. \eqref{eq:hstar}.

In the Larkin model as written in \eqref{HL} the disorder energy is simply $H_d=-2 H_e$ and 
total energy $H=-H_e$, so $H_d$ and $H_e$ are not independent as in the true model 
(in Section~\ref{sec:LDFForFixedEnergy}, the relation $H_d=-2 H_e$ was shown to hold only for the typical values in the strong confinement regime, $L_m\ll L_c$).
Trying to improve on the model (as adding the $V(0,\tau)$ term from the expansion of
the potential) does not lead to a consistent description of $H_d$. That no such simple
approximation exist for $H_d$ and $H$ is corroborated by the results of Section \ref{sec:disorder}
where the typical disorder energy density is obtained in terms of the (quite non-trivial) GLE of a
1D Anderson localization problem. 

%%%%%%%%%%%%%%%%%%%%%%%%%%%%%%%%%%%%%%%%%%%%%%%%%%%%%%%%%%%%%%%%%%%%%%%%%%%%%%%%%%%%%%%%%%
%%%%%%%%%%%%%%%%%%%%%%%%%%%%%%%%%%%%%%%%%%%%%%%%%%%%%%%%%%%%%%%%%%%%%%%%%%%%%%%%%%%%%%%%%%

\section{Spectrum of the Fokker-Planck generator $\mathscr{G}^\dagger$ in the presence of a non equilibrium stationary state -- Spectrum of $\mathscr{O}_q$}
\label{app:SpectrumGdagger}

\subsection{Generator of the FPE}

We make several remarks on the spectrum of the forward generator
\begin{equation}
  \label{eq:GeneFP}
  \mathscr{G}^\dagger = D\deriv{^2}{z^2} + \deriv{}{z}\mathscr{U}'(z)
  = D \deriv{}{z} \EXP{-\mathscr{U}(z)/D} \deriv{}{z} \EXP{\mathscr{U}(z)/D}
\end{equation}
when the potential is such that the diffusion is characterized by a non-equilibrium stationary state (NESS) on the full real axis.
This situation requires that $\mathscr{U}(z)\to\pm\infty$ for $z\to\pm\infty$, and the drift $|\mathscr{U}'(z)|$ grows sufficiently fast at infinity so that the particle is driven in a finite time from $+\infty$ to $-\infty$~:
\begin{equation}
  \label{eq:ConditionNESS}
  \int^\infty\D z\, \left|\potRic'(z)\right|^{-1} < \infty
  \hspace{0.5cm}\mbox{and} \hspace{0.5cm}
  \int_{-\infty}\D z\, \left|\potRic'(z)\right|^{-1} < \infty
  \:.
\end{equation}
This is the case for the potential $\potRic(z)=Ez+z^3/3$ considered in the paper.

In order to get some insight, we perform the following non-unitary transformation relating the generator to the Hermitian operator
\begin{align}
  \label{eq:Hsusy}
  \mathscr{H}_0
  &= - \EXP{\potRic(z)/(2D)} \mathscr{G}^\dagger \EXP{-\potRic(z)/(2D)}
  = - D\,\EXP{\potRic(z)/(2D)}\deriv{}{z}\EXP{-\potRic(z)/D}\deriv{}{z}\EXP{\potRic(z)/(2D)}
  \\
  \label{eq:Hsusy2}
  &= -D\deriv{^2}{z^2}
   +\frac{[\potRic'(z)]^2}{4D}
   -\frac{\potRic''(z)}{2}
   \:.
\end{align}
The transformation is well-known in the context of the Fokker-Planck equation (FPE), see for instance Ref.~\cite{Ris89} (see also the recent article \cite{GraTexTou14} and references therein).
The first equation emphasizes a particular symmetry of the operator, known as ``supersymmetry'' \cite{Jun96}, which rewrites
\begin{equation}
  \mathscr{H}_0=\mathscr{Q}^\dagger\mathscr{Q}
  \hspace{0.5cm}\mbox{where}\hspace{0.5cm}
  \mathscr{Q}=-\sqrt{D}\,\EXP{-\potRic(z)/(2D)}\deriv{}{z}\EXP{\potRic(z)/(2D)}
%  = \sqrt{D}\left(-\deriv{}{z} -\frac{\potRic'(z)}{2D}\right)
  \:.
\end{equation}
This makes clear that $\mathscr{H}_0$ has a positive spectrum, $\mathrm{Spec}(\mathscr{H}_0)\subset\mathbb{R}^+$, as well as $\mathrm{Spec}(\mathscr{G}^\dagger)$.

The conditions \eqref{eq:ConditionNESS} imply that the drift grows at infinity so that the potential in the Hamiltonian \eqref{eq:Hsusy2} is a confining potential (for the case studied in the paper, the potential behaves as $z^4/(4D)$ at infinity).
As a consequence the spectrum of the Hamiltonian and of the generator is \textit{discrete}~:
we have denoted $\mathscr{E}_n(0)$, $\Phi^\mathrm{R}_n(z;0)$ and $\Phi^\mathrm{L}_n(z;0)$ the eigenvalues and eigenvectors of the generator, defined by (\ref{eq:EqForPhiR},\ref{eq:EqForPhiL}) for $q=0$.
We can also associate to the eigenvalue $\mathscr{E}_n(0)$ the eigenfunction $\Psi_n(z)$ of $\mathscr{H}_0$.
Making use of
$\mathscr{H}_0\Psi_n=\mathscr{E}_n\Psi_n$,
$\mathscr{G}^\dagger\Phi^\mathrm{R}_n=-\mathscr{E}_n\Phi^\mathrm{R}_n$ and
$\mathscr{G}\Phi^\mathrm{L}_n=-\mathscr{E}_n\Phi^\mathrm{L}_n$,
if $\mathscr{E}_n(0)>0$ we get
\begin{equation}
  \label{eq:DetailedBalance}
  \Psi_n(z) = \Phi^\mathrm{R}_n(z;0)\,\EXP{\potRic(z)/(2D)} = \Phi^\mathrm{L}_n(z;0)\,\EXP{-\potRic(z)/(2D)}
\end{equation}
which shows that the strictly positive parts of the spectra of the two operators $\mathscr{G}^\dagger$ and  $\mathscr{H}_0$ coincide.

The conditions \eqref{eq:ConditionNESS} also ensure the existence of null right and left eigenvectors
\begin{equation}
  \Phi_0^\mathrm{R}(z;0)=\frac{N(E)}{D}\,\EXP{-\potRic(z)/D}\int_{-\infty}^z\D z'\,\EXP{\potRic(z')/D}
  \hspace{0.25cm}\mbox{and}\hspace{0.25cm}
  \Phi_0^\mathrm{L}(z;0)=1
  \:,
\end{equation}
where $N(E)$ is given by the normalization.
The right eigenvector is the stationary state for constant current~:
using the relation between the probability density and the current
\begin{equation}
  J_\tau(z) = -D\, \EXP{-\mathscr{U}(z)/D} \derivp{}{z} \left[\EXP{\mathscr{U}(z)/D} P_\tau(z)\right]
 \:,
\end{equation}
we deduce that $P(z)=\Phi_0^\mathrm{R}(z;0)$ is the stationary solution of the FPE for a uniform current $J=-N(E)$, where $N(E)$ is the integrated DoS of the random Schr\"odinger operator \eqref{eq:DisorderedHamiltonian} introduced in Subsection~\ref{subsec:EstimateRate} (see also Refs.~\cite{LifGrePas88,Luc92,ComTexTou13,ComLucTexTou13,GraTexTou14}).
The inspection of its asymptotic behaviour
\begin{equation}
  \label{eq:AsympPhi0Rq0}
  \Phi_0^\mathrm{R}(z;0)
  \simeq N(E)/\left|\potRic'(z)\right|\simeq N(E)\,z^{-2}
  \quad\mbox{for}\quad z\to\pm\infty
  \:,
\end{equation}
illustrates that the drift $\potRic'(z)\simeq-z^2$ dominates the diffusion at large $z$.
This also makes clear that the function $\Psi_0(z)=\Phi_0^\mathrm{R}(z;0)\,\EXP{\potRic(z)/(2D)}$, solution of $\mathscr{H}_0\Psi_0=0$, is \textit{not normalizable}, thus $\mathscr{E}_0(0)=0$ is not in the spectrum of $\mathscr{H}_0$ and one says that the ``supersymmetry is broken''~\cite{Jun96}~:~\footnote{
When the potential $\mathscr{U}(z)$ confines the particle such that an equilibrium state (with zero current) exists, the two null eigenvectors are
$\Phi_0^\mathrm{R}(z;0)=c\,\exp\big\{-\potRic(z)/D\big\}$ and $\Phi_0^\mathrm{L}(z;0)=1$ so that the wave function $\Psi_0(z)=c\,\exp\big\{-\potRic(z)/(2D)\big\}$ is normalizable and the zero mode also belongs to the spectrum of the Hamiltonian. This is the case of ``good susy'' where the spectra exactly coincide,
$\mathrm{Spec}(-\mathscr{G}^\dagger)=\mathrm{Spec}(\mathscr{H}_0)$.
Note that the condition \eqref{eq:DetailedBalance} ensures the detailed balance property in this case.
}
\begin{equation}
  \label{eq:BrokenSUSY}
  \mathrm{Spec}(-\mathscr{G}^\dagger) = \mathrm{Spec}(\mathscr{H}_0) \cup \{\mathscr{E}_0(0)=0\}
  \hspace{1cm} \mbox{(broken SUSY).}
\end{equation}
A consequence of importance for the article is that both
\begin{equation}
  \label{eq:Phi0Phi1NoNode}
  \Phi_0^\mathrm{R}(z;0)>0
  \hspace{0.5cm}\mbox{and}\hspace{0.5cm}
  \Phi_1^\mathrm{R}(z;0)>0
\end{equation}
i.e. have no node, where the second condition follows from the fact that $\Phi_1^\mathrm{R}(z;0)$ is related to the ground state $\Psi_1$ of $\mathscr{H}_0$.
Since $\Psi_n$ for $n\geq1$ is the $(n-1)$-th excited state of the Hamiltonian, it has $n-1$ nodes, like  $\Phi_n^\mathrm{R}(z;0)$.

We now briefly discuss the asymptotic behaviour of the eigenvectors. We assume that $\mathscr{U}(z)\to\pm\infty$ for $z\to\pm\infty$ like in the paper.
The asymptotic behaviour of the eigenfunctions of $\mathscr{H}_0$ is given by the WKB form
$
  \Psi_n(z)\simeq
  \big[p(z)\big]^{-1/2}\, \exp\pm\int\D z\, p(z)
$
where
$
p(z)= \sqrt{[\mathscr{U}'(z)/2D]^2-\mathscr{U}''(z)/2D -\mathscr{E}_n/D }
$.
With the condition that $\mathscr{U}'(z)$ grows at infinity, we can expand its integral as
$
  \int\D z\,p(z) \simeq \pm\mathscr{U}(z)/(2D) \mp \frac{1}{2}\ln|\mathscr{U}'(z)|
$ for $z\to\pm\infty$.
As a consequence, the two asymptotic behaviours for the wave function are
\begin{equation}
  \Psi_n(z)\sim
  \left\{
  \begin{array}{ll}
     \EXP{-\mathcal{U}(z)/(2D)}
       & \mbox{ for } z\to +\infty \\
     \frac{1}{|\mathcal{U}'(z)|}\,\EXP{\mathcal{U}(z)/(2D)}
       & \mbox{ for } z\to -\infty
  \end{array}
  \right.
\end{equation}
which is obviously normalizable. 
% as argued above since the potential in \eqref{eq:Hsusy} is confining.
Correspondingly, the right and left eigenvectors of the generator behave asymptotically as
\begin{equation}
  \label{eq:AsimptoticPhiRn}
  \Phi_n^\mathrm{R}(z;0)
  \sim
  \left\{
  \begin{array}{ll}
     \EXP{-\mathcal{U}(z)/D}
%       & \mbox{ for } z\to +\infty 
  \\
     \frac{1}{|\mathcal{U}'(z)|}
%       & \mbox{ for } z\to -\infty
  \end{array}
  \right.
%\end{equation}
%and
%\begin{equation}
%  \label{eq:AsimptoticPhiLn}
\hspace{0.5cm}\mbox{and}\hspace{0.5cm}
  \Phi_n^\mathrm{L}(z;0)
  \sim
  \left\{
  \begin{array}{ll}
     1
       & \mbox{ for } z\to +\infty \\
     \frac{1}{|\mathcal{U}'(z)|} \EXP{\mathcal{U}(z)/D}
       & \mbox{ for } z\to -\infty
  \end{array}
  \right.
\end{equation}

\subsection{The spectrum of $\mathscr{O}_q$}

We now make some remark on the spectrum of the operator $\mathscr{O}_q=\mathscr{G}^\dagger+q\,z$ of importance for the paper.
The similar non-unitary transformation is possible
\begin{equation}
  \label{eq:Hq}
  \mathscr{H}_q
  = - \EXP{\potRic(z)/(2D)} \mathscr{O}_q \EXP{-\potRic(z)/(2D)}
  = \mathscr{H}_0 - q\,z
  \:,
\end{equation}
which leads to a Hermitian operator. 
$\mathscr{H}_q$ also describes a quantum particle trapped in a confining potential, hence has a discrete spectrum.
However the presence of the additional potential term $-q\,z$ breaks the supersymmetry.
It is not possible to construct explicitly the state $\Phi_0^\mathrm{R}(z;q)$ like for $q=0$, which makes the discussion more complicated. From a continuity argument, we expect also the property
\begin{equation}
  \label{eq:Spectra}
    \mathrm{Spec}(-\mathscr{O}_q) = \mathrm{Spec}(\mathscr{H}_q) \cup \{\mathscr{E}_0(q)\}
    \:,
\end{equation}
what we have verified by diagonalisation of the discretised operators for several values of~$q$.

\begin{figure}[!ht]
\centering
\includegraphics[width=6.5cm]{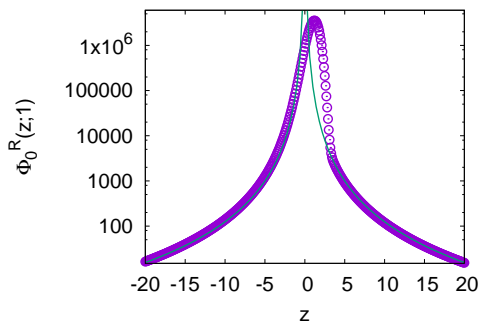}
\caption{\it Function $\Phi_0^\mathrm{R}(z;q)$ (not normalized) for $k=1$ and $q=1$.
  The line is $\propto|z|^{-3}$.}
  \label{fig:Phi0R}
\end{figure}

Some asymptotic analysis is also possible. Starting from the differential equation $\mathscr{O}_q\Phi_n^\mathrm{R}(z;q)=-\mathscr{E}_n(q)\Phi_n^\mathrm{R}(z;q)$, we obtain that the two independent asymptotic behaviours are
\begin{equation}
|z|^{-2-q}
\hspace{0.5cm}\mbox{and}\hspace{0.5cm}
|z|^q\,\EXP{-\potRic(z)/D}
\:,
\end{equation}
which corresponds to \eqref{eq:AsimptoticPhiRn} for $q=0$.
All ``excited state'' $n\geq1$ are in correspondence with eigenfunctions of \eqref{eq:Hq} and thus present the exponential behaviour $z^q\,\EXP{-\potRic(z)/D}$ for $z\to+\infty$.
Like for $q=0$, the first right eigenvector is the only one with algebraic behaviour on both sides $\Phi_0^\mathrm{R}(z;q)\sim|z|^{-2-q}$ for $z\to\pm\infty$ (see Fig.~\ref{fig:Phi0R}).

%%%%%%%%%%%%%%%%%%%%%%%%%%%%%%%%%%%%%%%%%%%%%%%%%%%%%%%%%%%%%%%%%%%%%%%%%%%%%%%%%%%%%%%%%%%%%%%%%%%%%%%%%

\section{WKB calculation for the ground state of the Fokker-Planck operator (written by David Saykin)}
\label{app:Saykin}

We consider the spectral problem defined by Eq.~\eqref{eq:EqForPhiR} for $q=1$.
A closely related problem was considered recently in Ref.~\cite{SayTikRod18}.
In order to make the equation of the Schr\"{o}dinger type, one performs the transformation~:
$\Phi(z)=\psi(z)\,\exp[\frac{1}{2}(|E|z-\frac{z^3}{3})]$  
%Let me change $\psi(z) \mapsto \exp[\frac{1}{2}(|E|z-\frac{z^3}{3})]\psi(z)$,  i.e. do the transformation given by \Cred{Eq.~(102)} from the paper, 
\begin{equation}
	-\psi'' + \left[  \frac{1}{4}(z^2-|E|)^2-2z   - \mathscr{E}  \right]\psi=0
	\:.
\end{equation}
%How it is noted in paper, % (See \Cred{Eq.~(107)}) 
%the sought solution would not be the eigenfunction of transformed operator because now it has to blow up at $z\to+\infty$, however it doesn't hinder from solving the problem using conventional WKB analysis. 
As we have seen in Section~\ref{sec:GLEspectral}, one seeks for a solution $\Phi_0^\mathrm{R}(z;q=1)$ presenting a power law decay at infinity, implying an exponential blow up of $\psi(z)$ as $z\to+\infty$, however this does not preclude us from solving the problem using conventional WKB analysis.

Let us rescale the coordinate as $z\mapsto|E|^{-1/4}y$ and introduce $g \equiv (4|E|^{3/2})^{-1}$ and $2\nu \equiv 1 + \mathscr{E} /\sqrt{|E|}$. One recovers the shifted double--well potential in a more standard form~:
\begin{equation}
  \label{eq:double--well}
	-\psi''+\left[g\left(y^2-\frac{1}{4g}\right)^2-4\sqrt{g}y+1-2\nu\right]\psi = 0
	\:,
\end{equation}
The desired eigenfunction satisfies the boundary conditions
\begin{align}
	\psi &\sim A_-\frac{\exp\int_0^y |k(x)|dx}{\sqrt{|k(y)|/g}},	\quad	
	\mbox{as }  y\to-\infty
	\:,	\qquad
	\\
	\psi &\sim A_+\frac{\exp\int_0^y |k(x)|dx}{\sqrt{|k(y)|/g}} + A_+'\frac{\exp-\int_0^y |k(x)|dx}{\sqrt{|k(y)|/g}},	\quad	
	\mbox{as }  y\to+\infty
	\:,
\end{align}
where
\begin{equation}
	|k(y)|^2 = g\left(y^2-\frac{1}{4g}\right)^2-4\sqrt{g}y+1-2\nu
	\:.
\end{equation}
The energy parameter $\nu$ is expected to behave asymptotically as $\nu \sim -2\#g\exp(-\frac{1}{3g})$ for $g \to +0$.

\paragraph{Hermite equation}

Let us introduce the notations $y_\pm = y \mp \frac{1}{2\sqrt{g}}$, $\nu_+=\nu$ and $\nu_-=\nu-2$. Near the minima of the potential $|y_\pm|\ll\frac{1}{\sqrt{g}}$, Eq.~\eqref{eq:double--well} reduces to the Hermite equation
\begin{equation}
\label{eq:approx_hermite}
	\left[ -\partial^{2} +  y_\pm^2 - (2\nu_\pm + 1) \right] \chi(y_\pm) \approx 0,	\quad	|y_\pm| \ll \frac{1}{\sqrt{g}}.
\end{equation}
The solution of this differential equation is known as the Hermite function used in Section~\ref{sec:WKBE1}, or, equivalently, the parabolic cylinder functions $\psi^{\text{osc}}_\nu(y) = \sqrt{2}^\nu D_\nu(\sqrt{2}y)$ (see also Section~\ref{sec:WKBE1}). We deduce the asymptotic
\begin{equation}
\label{eq:hermite_asymptotics}
	\psi_\nu^{\text{osc}}(y) \sim
	\begin{cases}
		(2y)^\nu \EXP{-\frac{y^2}{2}},		&	y \to +\infty	\\
		\cos{\pi\nu}(-2y)^\nu \EXP{-\frac{y^2}{2}} + \frac{\sqrt{\pi}}{\Gamma(-\nu)}	(-y)^{-\nu-1} \EXP{\frac{y^2}{2}},	
											&	y \to -\infty	\\
	\end{cases}
\end{equation}
The second linearly independent solution may be chosen as $\bar{\psi}^{\text{osc}}_\nu(y) = -\cos\pi\nu \,\psi^{\text{osc}}_\nu(y) + \psi^{\text{osc}}_\nu(-y)$, with asymptotic behaviour
\begin{equation}\label{eq:hermite_asymptotics_2}
	\bar{\psi}_\nu^{\text{osc}}(y) \sim
	\begin{cases}
		\frac{\sqrt{\pi}}{\Gamma(-\nu)}y^{-\nu-1} \EXP{\frac{y^2}{2}},		&	y \to +\infty	\\
		\sin^2{\pi\nu}(-2y)^\nu \EXP{-\frac{y^2}{2}} - \frac{\cos\pi\nu\sqrt{\pi}}{\Gamma(-\nu)}(-y)^{-\nu-1} \EXP{\frac{y^2}{2}},	
		&	y \to -\infty	\\
	\end{cases}	
\end{equation}
Note that, when $\nu\in\N_0$, the solution becomes $\psi_n^{\text{osc}}(y)=(2y)^n \EXP{-\frac{y^2}{2}}$, $\Gamma(-n)\bar{\psi}_n^{\text{osc}}(y) = \sqrt{\pi}y^{-n-1} \EXP{\frac{y^2}{2}}$.

\paragraph{Matching}  

In order to match the global WKB solutions with Hermite-like solutions approximating the solution in the neighbourhood of the points $\frac{\pm1}{2\sqrt{g}}$, one expands the semiclassical exponent 
$S(y)=\int_{0}^{y}k(x)dx - \frac{1}{2}\ln(k(y)/g)$ in the vicinity of these points~:
\begin{align}
	\text{I}: S(y_-<0) &\sim \frac{1}{12g}-\frac{y_-^2}{2}-\ln(-4\sqrt{g}y_+)+\left(\nu-\frac{1}{2}\right)\ln(\sqrt{g}y_+)+\bigo{g}+\bigo{y_-^3}\\
	\text{II}: S(y_->0) &\sim -\frac{1}{12g}+\frac{y_-^2}{2}+\ln(4\sqrt{g}y_+)-\left(\nu-\frac{1}{2}\right)\ln(\sqrt{g}y_+)+\bigo{g}+\bigo{y_-^3}\\
	\text{III}: S(y_+<0) &\sim \frac{1}{12g}-\frac{y_+^2}{2}+\ln(-4\sqrt{g}y_+)+\left(\nu-\frac{1}{2}\right)\ln(\sqrt{g}y_+)+\bigo{g}+\bigo{y_+^3}\\
	\text{IV}: S(y_+>0) &\sim -\frac{1}{12g}+\frac{y_+^2}{2}-\ln(4\sqrt{g}y_+)-\left(\nu-\frac{1}{2}\right)\ln(\sqrt{g}y_+)+\bigo{g}+\bigo{y_+^3}
\end{align} 
Starting from $y=-\infty$, one connects $A_-$ to $A_+$, $A_+'$ passing the turning points (of second order) with the help of Eqs.~\eqref{eq:hermite_asymptotics}, \eqref{eq:hermite_asymptotics_2}. 
Here one omits lower indices $y_\pm\mapsto y$ to make equations more readable~:
\begin{align}
	\text{I}: \psi(y_-<0) 	&\sim A_-\EXP{\frac{1}{12g}}(4\sqrt{g}^{\frac{3}{2}-\nu})^{-1}(-y)^{\nu-2} \EXP{-y^2/2} \\
	\text{II}: \psi(y_->0) 	&\sim \cos{\pi\nu_-}A_-\EXP{\frac{1}{12g}}(4\sqrt{g}^{\frac{3}{2}-\nu})^{-1}y^{\nu_-} \EXP{-\frac{y^2}{2}} 
	  \nonumber\\
							&+ {\textstyle\frac{2^{-\nu_-}\sqrt{\pi}}{\Gamma(-\nu_-)}}A_-\EXP{\frac{1}{12g}}(4\sqrt{g}^{\frac{3}{2}-\nu})^{-1} y^{-\nu_--1}\EXP{\frac{y^2}{2}} \\
							&\sim C_-\EXP{\frac{1}{12g}}(4\sqrt{g}^{\frac{3}{2}-\nu})^{-1}y^{\nu-2}\EXP{-y^2/2}
								+ C_+\EXP{-\frac{1}{12g}}4\sqrt{g}^{\frac{3}{2}-\nu}y^{-\nu+1}\EXP{y^2/2}	\\
	\text{III}: \psi(y_+<0) &\sim C_+\EXP{\frac{1}{12g}}4\sqrt{g}^{\nu+\frac{1}{2}}(-y)^{\nu}\EXP{-y^2/2}
								+ C_-\EXP{-\frac{1}{12g}}(4\sqrt{g}^{\nu+\frac{1}{2}})^{-1}(-y)^{-\nu-1}\EXP{y^2/2}	\\
	\text{IV}: \psi(y_+>0) 	
	\nonumber\\ 
	&\hspace{-1cm}
	\sim \textstyle\left[\cos\pi\nu\,C_+\EXP{\frac{1}{12g}}4\sqrt{g}^{\nu+\frac{1}{2}} 
		+ \frac{\Gamma(-\nu)}{\sqrt{\pi}}2^\nu\sin^2\pi\nu C_-\EXP{-\frac{1}{12g}}(4\sqrt{g}^{\nu+\frac{1}{2}})^{-1}\right]y^\nu \EXP{-y^2/2} 
    \nonumber\\ 
	&
	\hspace{-1cm}
    + \textstyle\left[\frac{2^{-\nu}\sqrt{\pi}}{\Gamma(-\nu)}C_+\EXP{\frac{1}{12g}}4\sqrt{g}^{\nu+\frac{1}{2}}
		 - \cos\pi\nu\,C_-\EXP{-\frac{1}{12g}}(4\sqrt{g}^{\nu+\frac{1}{2}})^{-1}\right]	y^{-\nu-1}\EXP{y^2/2}	\\
    &\hspace{-1cm}
    \sim A_+\EXP{-\frac{1}{12g}}(4\sqrt{g}^{\nu+\frac{1}{2}})^{-1}y^{-\nu-1}\EXP{y^2/2} +
		 A_+'\EXP{\frac{1}{12g}}4\sqrt{g}^{\nu+\frac{1}{2}}y^\nu \EXP{-y^2/2}.
\end{align}
$C_\pm$ are the coefficients of the semicalssical expansion under the potential barrier.
\begin{equation}\label{eq:semiclassics}
	\psi\left(|y|\lesssim \frac{1}{2\sqrt{g}}\right) \sim C_+\frac{\exp\int_{0}^y |k(x)|dx}{\sqrt{|k(y)|/g}} + C_-\frac{\exp-\int_{0}^y |k(x)|dx}{\sqrt{|k(y)|/g}}.
\end{equation}
\begin{equation}
	C_+ = \frac{\EXP{\frac{1}{6g}}}{4\sqrt{g}^{3-2\nu}}\frac{2^{-\nu}\sqrt{\pi}}{\Gamma(2-\nu)}A_-,	\quad
	C_- = \cos\pi\nu\,A_-.
\end{equation}
Finally, the coefficients $A_+$, $A_-$ are given by
\begin{align}
	A_+/A_- &= 
	\frac{\EXP{\frac{1}{3g}}}{\Gamma(-\nu)\Gamma(2-\nu)}\frac{4\pi}{g}
	\left(\frac{g}{2}\right)^{2\nu} - \cos^2\pi\nu,	
	\\
	A_+'/A_- &= \frac{\cos\pi\nu\,\EXP{\frac{1}{6g}}}{\Gamma(2-\nu)}\frac{\sqrt{g}^{2\nu-3}}{2^{\nu+2}} 
			- \cos\pi\nu\sin^2\pi\nu\frac{\EXP{-\frac{1}{6g}}\Gamma(-\nu)2^{\nu-4}}{\sqrt{\pi}\sqrt{g}^{2\nu-1}}.
\end{align}
One could expect that the correct boundary condition at $+\infty$ is $A_+'=0$, however this is not the case. 
Instead, let us exploit $A_+=A_-$, cf. Eq.~\eqref{eq:QuantifForE0} of the paper. 
This leads to 
\begin{equation}
  \label{eq:SaykinResult}
	\nu \simeq -\frac{g}{2\pi}\exp\left(-\frac{1}{3g}\right),	\quad g\ll 1
	\:.
\end{equation}
Substitution of this result leads to $A_+'/A_- \sim \frac{1}{4g}\sqrt{\pi/g}\EXP{\frac{1}{6g}} \gg 1$. 
This means that the solution of the initial problem at $+\infty$ has corrections of order 
$\psi(z)\sim A_+|z|^{-3} + \dots + \#\exp(|E|z-z^3/3)$, $z\to+\infty$, as well as monomial corrections.

Using $g=s/4=1/(4|E|^{3/2})$, we can relate \eqref{eq:SaykinResult} with the quantity of interest in the paper~:
\begin{equation}
  \Lambda(1) = -\mathscr{E}_0(1) = \sqrt{|E|}(1 - 2\nu) 
  \simeq  \sqrt{|E|}  + \frac{1}{4\pi|E|}\EXP{-4|E|^{3/2}/3}
  \:.
\end{equation}
This coincides with the energy dependence of the pre-exponential function obtained by a neat numerical analysis (Section~\ref{sec:GLEspectral}), and demonstrates the statement below Eq.~\eqref{eq:ConclusionSDE1} about the dimensionless factor $1/(4\pi)\simeq0.079577$.
%\Cred{
%\paragraph{Conclusion.---} Analytical solution of that connection problem strongly depends on fact that $A_+=A_-$ and cannot be reformulated to conventional boundary problem. Is there any other way to see that $A_-=A_+$  besides numerical calculation? 
%}

%%%%%%%%%%%%%%%%%%%%%%%%%%%%%%%%%%%%%%%%%%%%%%%%%%%%%%%%%%%%%%%%%%%%%%%%%%%%%%%%%%%%%%%%%%%%%%%%%%%%%%%%%%

\section{Action $S_q$ of the WKB treatment in the $s\to0$ limit}
\label{app:cq}

In this appendix, we analyse the action \eqref{eq:DefActionWKB} of the WKB treatment presented in Section~\ref{sec:WKBE1} and prove that its $s\to0$ behaviour is given by Eqs.~(\ref{eq:ActionWKB},\ref{eq:cq}).
For this purpose, we introduce the function 
\begin{equation}
  \label{eq:DefFq}
  F_q(s) := 2s\,S_q
  =\int_{-1}^{1-\sqrt{s}}
  \D\zeta\,
  \sqrt{(1-\zeta^2)^2-4s(q+1)\zeta+4sq\,(1-\sqrt{s})-s^2}
  \:.
\end{equation}
As it is quite obvious that $F_q(0)=4/3$, this allows one to extract easily the leading term of the action $S_q\simeq2/(3s)$ by considering
\begin{equation}
  F_q'(s)
  =
  2
  \int_{-1}^{1-\sqrt{s}}
  \D\zeta\,
  \frac{ 
      q - (q+1)\zeta - (3/2)q\sqrt{s} - s/2
    }{
      \sqrt{(1-\zeta^2)^2-4s(q+1)\zeta+4sq\,(1-\sqrt{s})-s^2}
    }
    \:.
\end{equation} 
The integral is logarithmically divergent in the $s\to0$ limit. 
This makes clear that we can safely neglect the two terms $- (3/2)q\sqrt{s} - s/2$ in the numerator, which contribute as $\mathcal{O}(\sqrt{s}\ln s)$ to $F_q'(s)$.

In the limit $s\to0$, the integral is dominated by the two boundaries.
Because the logarithmic divergences are cut off in two different manners, it is convenient to split the integral into two parts, which we analyse separately~:
\begin{equation}
  F_q'(s) = G_-(s) + G_+(s) + \mathcal{O}(\sqrt{s}\ln s)
\end{equation}
where 
\begin{equation}
   G_-(s):= 
  2\int_{-1}^{0}
  \D\zeta\,
  \frac{ q - (q+1)\zeta }{ \sqrt{ f(\zeta) }  }
  \hspace{0.5cm}\mbox{and}\hspace{0.5cm}
  G_+(s):= 
  2\int_{0}^{1-\sqrt{s}}
  \D\zeta\,
  \frac{ q - (q+1)\zeta }{ \sqrt{ f(\zeta) } }
\end{equation}
and
\begin{equation}
  f(\zeta) = (1-\zeta^2)^2-4s(q+1)\zeta+4sq\,(1-\sqrt{s})-s^2
  \:.
\end{equation}

\subsubsection*{Term $G_-$}

Inspection of the Taylor expansion of the function $f$ near $\zeta=-1$
\begin{equation}
  f(\zeta) \underset{\zeta\sim-1}{\simeq}
  4s(2q+1) -s\,(4q\sqrt{s}+s) -4s(q+1)(\zeta+1) + 4(\zeta+1)^2
\end{equation}
makes clear that the constant term cut off the logarithmic divergence of the integral at a scale $\delta\zeta=\mathcal{O}(\sqrt{s})$, so that the linear term can be neglected~:
\begin{align}
  G_-(s) &\underset{s\to0}{\simeq}
  2\int_{-1} 
  \D\zeta\,
  \frac{ q - (q+1)\zeta  }{ \sqrt{ s(2q+1) +  (\zeta+1)^2 }  }
  \\
  &\simeq 
  (2q+1)
  \int_{\sqrt{s(2q+1)}} 
  \frac{\D y}{y}
  \simeq -\left(q+\frac12\right)\ln s
\end{align}
In order to obtain the subleading constant term, we consider the integral 
\begin{align}
  &A_-(s) := 
  \int_{-1}^0 
  \D\zeta\,
  \frac{ q - (q+1)\zeta  }{ \sqrt{ s(2q+1) +  (\zeta+1)^2 }  }
  \\
  &=(2q+1)\ln\left(\frac{1}{\sqrt{s(2q+1)}}+\sqrt{\frac{1}{s(2q+1) }+1}\right)
  \nonumber
%  \\
%  &\hspace{1cm}
  -(q+1) \left[
    \sqrt{s(2q+1)+1}-\sqrt{s(2q+1)}
  \right]
\end{align}
It is straightforward to show that 
\begin{equation}
  \lim_{s\to0} \left[ G_-(s) - A_-(s) \right] = q+1 - \ln2
\end{equation}
from which we conclude that 
\begin{equation}
  \label{eq:Gminus}
  G_-(s) \underset{s\to0}{\simeq}
  -\left(q+\frac12\right)\ln\left[ s (2q+1)\right] + 2q\ln2 +  \mathcal{O}(\sqrt{s})
\end{equation}

\subsubsection*{Term $G_+$}

We proceed in a similar manner for the second term. 
The starting point is the Taylor expansion
\begin{equation}
  f(\zeta) \underset{\zeta\sim1-\sqrt{s}}{\simeq}
  -\left[8\sqrt{s}+\mathcal{O}(s)\right] (\zeta-1+\sqrt{s}) + 4(\zeta-1+\sqrt{s})^2
\end{equation}
which shows that we have now to consider the integral 
\begin{align}
  A_+(s) &:= 
  \int_{0}^{1-\sqrt{s}}
  \D\zeta\,
  \frac{ q - (q+1)\zeta  }{ \sqrt{ 2\sqrt{s}\,(1-\sqrt{s}-\zeta) + (\zeta-1+\sqrt{s})^2 } }
  \\
  \nonumber
  &=-\ln\left(\frac{1}{\sqrt{s}}+\sqrt{\frac{1}{s}+1}\right) 
  + (q+1)\sqrt{1-s}
\end{align}
Using that 
\begin{equation}
  \lim_{s\to0} \left[ G_+(s) - A_+(s) \right] 
  = -(q+1) +(2q+1) \ln2
\end{equation}
we conclude that 
\begin{equation}
  \label{eq:Gplus}
  G_+(s) \underset{s\to0}{\simeq}  \frac12 \ln  s + 2q\ln2 +  \mathcal{O}(s)
\end{equation}

\subsubsection*{Conclusion}

Gathering the two 
expressions (\ref{eq:Gminus},\ref{eq:Gplus}),
we finally obtain 
\begin{equation}
  F'(s) \underset{s\to0}{=}
  - q\,\ln s + 4q\ln 2 -\left(q+\frac12\right)\ln(2q+1) + \mathcal{O}(\sqrt{s}\ln s)
\end{equation}
An integration, with $F(0)=4/3$, leads to (\ref{eq:ActionWKB},\ref{eq:cq}).

%%%%%%%%%%%%%%%%%%%%%%%%%%%%%%%%%%%%%%%%%%%%%%%%%%%%%%%%%%%%%%%%%%%%%%%%%%%%%%%%%%%%%%%%%%%%%%%%%%%%%%%%%%

\section{An inverse Laplace transform}
\label{app:AnInvLT}

We study in this appendix the function defined by its Laplace transform
\begin{equation}
  \varphi(\reshe) = \mathscr{L}^{-1}_{s \to \reshe} \left[  \frac{1}{ \sqrt{\cosh\sqrt{s}} } \right]
  \hspace{0.5cm}\mbox{for }\reshe\geq0
  \:.
\end{equation}
A direct calculation of the inverse Laplace is quite tricky due to the complicate analytic structure of the function $\sqrt{\cosh\sqrt{s}}$, which presents numbers of branch cuts in the complex plane of the variable $s=x+\I\,y$~:
a succession of segments on the real axis, $[-\pi^2(2n+3/2)^2,-\pi^2(2n+1/2)^2]$ for $n\in\mathbb{N}$, and parabolas intersecting those segments, $-x+[y/(2m\pi)]^2=(m\pi)^2$ for non zero odd integers~$m$.

Instead, we consider 
\begin{equation}
  (\varphi*\varphi)(\reshe) 
  = \mathscr{L}^{-1}_{s \to \reshe} \left[  \frac{1}{ \cosh\sqrt{s} } \right]
\end{equation}
which can be easily computed thanks to residue's theorem
\begin{equation}
  (\varphi*\varphi)(\reshe) = \pi
  \sum_{n=0}^\infty (-1)^n (2n+1) \, \exp\left\{ - \frac{\pi^2}{4}(2n+1)^2 \reshe \right\}
  \:.
\end{equation}
This first representation is useful to analyse the $\reshe\to\infty$ limit.
Using a Poisson formula given in an appendix of Ref.~\cite{TexHag10}, we obtain the other useful representation
\begin{equation}
  (\varphi*\varphi)(\reshe) = \frac{1}{\sqrt{\pi}\,\reshe^{3/2}}
   \sum_{n=0}^\infty (-1)^n (2n+1) \, \exp\left\{ - \frac{(n+1/2)^2}{ \reshe} \right\}
   \:,
\end{equation}
appropriate to describe the $\reshe\to0$ limit.

We now relate the two limiting behaviours 
\begin{align}
  \label{eq:LimitsVarphi2}
   (\varphi*\varphi)(\reshe)
   \simeq
   \begin{cases}
     \frac{1}{\sqrt{\pi}\,\reshe^{3/2}}\, \exp\left\{ - 1/(4\reshe) \right\} 
       & \mbox{for } \reshe \to 0 
    \\[0.2cm]
     \pi \, \exp\left\{ - \pi^2 \reshe / 4 \right\} 
       & \mbox{for } \reshe \to \infty
  \end{cases}
\end{align}
to the corresponding one for $\varphi(\reshe)$.
We use the two following remarks~:
\begin{itemize}
\item
  We introduce $w(x)=c\,x^{-\alpha}\,\EXP{-x-1/(16x)}$ for $x>0$. Some algebra gives 
  $$
    (w*w)(x)=c^2\,(x/2)^{-2\alpha+1}\,\EXP{-x}
    \int_1^\infty\D u\frac{u^{\alpha-3/2}}{\sqrt{u-1}}\,\EXP{-u/(4x)}
  $$
  with $(w*w)(x)\sim x^{-2\alpha+1/2}\EXP{-1/(4x)}$ for $x\to0$.
\item
  Now considering the function $p(x)=\big[b^{\alpha+1}/\Gamma(\alpha+1)\big]x^{\alpha}\,\EXP{-bx}$ for $x>0$, it is easy to get 
  $$
  (p*p)(x)=\frac{b^{2\alpha+2}}{\Gamma(2\alpha+2)} \, x^{2\alpha+1}\, \EXP{-bx}
  $$
\end{itemize}
Using the first remark for $\alpha=1$ and the second remark for $\alpha=-1/2$, we can relate the limiting behaviours \eqref{eq:LimitsVarphi2} to
\begin{align}
  \varphi(\reshe) \sim 
  \begin{cases}
    \frac{1}{\reshe} \, \EXP{-1/(16\reshe)} 
      & \mbox{for } \reshe \to 0 
    \\
    \frac{1}{\sqrt{\reshe}} \, \EXP{-\pi^2\reshe/4} & \mbox{for } \reshe \to \infty
  \end{cases}
  \:.
\end{align}

%%%%%%%%%%%%%%%%%%%%%%%%%%%%%%%%%%%%%%%%%%%%%%%%%%%%%%%%%%%%%%%%%%%%%%%%%%%%%%%%%%%%%%%%%%%%%%%%%%%

%%\bibliographystyle{elsarticle-harv}
%\bibliographystyle{phreport}
%\bibliography{biblio}

\addcontentsline{toc}{section}{\protect\bibname}
\begin{thebibliography}{100}

\bibitem{BlaFeiGesLarVin94}
G.~Blatter, M.~V. Feigel'man, V.~B. Geshkenbein, A.~I. Larkin, and V.~M.
  Vinokur,
 Vortices in high-temperature superconductors,
 Rev. Mod. Phys. {\bf 66}, 1125--1388 (1994).

\bibitem{LeD11}
P.~Le~Doussal,
 Novel phases of vortices in superconductors,
 Int. J. Mod. Phys. B {\bf 24}, 3855--3914 (2010),
 in \textit{BCS: 50 years}, L. N. Cooper and D. Feldman (eds.), World
  Scientic, 2011.

\bibitem{HalZha95}
T.~Halpin-Healy and Y.-C. Zhang,
 Kinetic roughening phenomena, stochastic growth, directed polymers
  and all that. Aspects of multidisciplinary statistical mechanics,
 Phys. Rep. {\bf 254}(4–6), 215--414 (1995).

\bibitem{CalLeDRos10}
P.~Calabrese, P.~L. Doussal, and A.~Rosso,
 Free-energy distribution of the directed polymer at high temperature,
 Europhys. Lett. {\bf 90}, 20002 (2010).

\bibitem{Dot10a}
V.~Dotsenko,
 Bethe ansatz derivation of the Tracy-Widom distribution for
  one-dimensional directed polymers,
 Europhys. Lett. {\bf 90}, 20003 (2010).

\bibitem{Dot10b}
V.~Dotsenko,
 Replica Bethe ansatz derivation of the Tracy-Widom distribution of
  the free energy fluctuations in one-dimensional directed polymers,
 J. Stat. Mech. , P07010 (2010).

\bibitem{SasSpo10}
T.~Sasamoto and H.~Spohn,
 One-dimensional Kardar-Parisi-Zhang equation: an exact solution and
  its universality,
 Phys. Rev. Lett. {\bf 104}, 230602 (2010).

\bibitem{AmiCorQua11}
G.~Amir, I.~Corwin, and J.~Quastel,
 Probability distribution of the free energy of the continuum directed
  random polymer in $1+1$ dimensions,
 Commun. Pure Appl. Math. {\bf 64}, 466--537 (2011).

\bibitem{Joh00}
K.~Johansson,
 Shape Fluctuations and Random Matrices,
 Commun. Math. Phys. {\bf 209}(2), 437--476 (2000).

\bibitem{AzaWsc09}
J.-M. Azais and M.~Wschebor,
 {\em Level Sets and Extrema of Random Processes and Fields},
 John Wiley \& Sons, 2009.

\bibitem{Fyo15}
Y.~V. Fyodorov,
 High-Dimensional Random Fields and Random Matrix Theory,
 Markov Processes Related Fields {\bf 21}, 483--518 (2015).

\bibitem{AufBenCer13}
A.~Auffinger, G.~{Ben Arous}, and J.~Cerny,
 Random matrices and complexity of spin glasses,
 Commun. Pure. Appl. Math. {\bf 66}, 165 (2013).

\bibitem{AufBen13}
A.~Auffinger and G.~{Ben Arous},
 Complexity of random smooth functions on the high-dimensional sphere,
 Ann. Probab. {\bf 41}(6), 4214--4247 (2013).

\bibitem{Nic14}
L.~I. Nicolaescu,
 Complexity of random smooth functions on compact manifolds,
 Indiana Univ. Math. J. {\bf 63}, 1037 (2014).

\bibitem{SubZei17}
E.~Subag and O.~Zeitouni,
 The extremal process of critical points of the pure p-spin spherical
  spin glass model,
 Probab. Theory Related Fields {\bf 168}(3), 773--820 (2017).

\bibitem{CamWig17}
V.~Cammarota and I.~Wigman,
 Fluctuations of the total number of critical points of random
  spherical harmonics,
 Stochastic Processes and their Applications {\bf 127}(12), 3825--3869
  (2017).

\bibitem{FyoKho16}
Y.~V. Fyodorov and B.~A. Khoruzhenko,
 Nonlinear analogue of the May−Wigner instability transition,
 Proc. Natl. Acad. Sci. (USA) {\bf 113}, 6827 (2016).

\bibitem{Lon60}
M.~S. {Longuet-Higgins},
 Reflection and refraction at a random moving surface. II. Number of
  specular points in a Gaussian surface,
 J. Opt. Soc. Am. {\bf 50}, 845 (1960).

\bibitem{HalLax66}
B.~I. Halperin and M.~Lax,
 Impurity-Band Tails in the High-Density Limit. I. Minimum Counting
  Methods,
 Phys. Rev. {\bf 148}, 722--740 (1966).

\bibitem{WeiHal82}
A.~Weinrib and B.~I. Halperin,
 Distribution of maxima, minima, and saddle points of the intensity of
  laser speckle patterns,
 Phys. Rev. B {\bf 26}, 1362--1368 (1982).

\bibitem{Fre95}
I.~Freund,
 Saddles, singularities, and extrema in random phase fields,
 Phys. Rev. E {\bf 52}, 2348--2360 (1995).

\bibitem{Fyo04}
Y.~V. Fyodorov,
 Complexity of Random Energy Landscapes, Glass Transition, and
  Absolute Value of the Spectral Determinant of Random Matrices,
 Phys. Rev. Lett. {\bf 92}, 240601 (2004).

\bibitem{BraDea07}
A.~J. Bray and D.~S. Dean,
 Statistics of Critical Points of Gaussian Fields on Large-Dimensional
  Spaces,
 Phys. Rev. Lett. {\bf 98}, 150201 (2007).

\bibitem{FyoWil07}
Y.~V. Fyodorov and I.~Williams,
 Replica Symmetry Breaking Condition Exposed by Random Matrix
  Calculation of Landscape Complexity,
 J. Stat. Phys. {\bf 129}(5), 1081--1116 (2007).

\bibitem{FyoNad12}
Y.~V. Fyodorov and C.~Nadal,
 Critical Behavior of the Number of Minima of a Random Landscape at
  the Glass Transition Point and the Tracy-Widom Distribution,
 Phys. Rev. Lett. {\bf 109}, 167203 (2012).

\bibitem{Par05}
G.~Parisi,
 Computing the number of metastable states in infinite-range models,
 in {\em Les Houches summer school, Session LXXXIII}, edited by
  A.~Bovier and \textit{et al}, volume~83, pages 295--329, Amsterdam, 2005,
  Elsevier.

\bibitem{AnnCavGiaPar03}
A.~Annibale, A.~Cavagna, I.~Giardina, and G.~Parisi,
 Supersymmetric complexity in the Sherrington-Kirkpatrick model,
 Phys. Rev. E {\bf 68}, 061103 (2003).

\bibitem{FyoLeD14}
Y.~V. Fyodorov and P.~Le~Doussal,
 Topology Trivialization and Large Deviations for the Minimum in the
  Simplest Random Optimization,
 J. Stat. Phys. {\bf 154}(1), 466--490 (2014).

\bibitem{WaiTou13}
G.~Wainrib and J.~Touboul,
 Topological and Dynamical Complexity of Random Neural Networks,
 Phys. Rev. Lett. {\bf 110}, 118101 (2013).

\bibitem{Fyo16}
Y.~V. Fyodorov,
 Topology trivialization transition in random non-gradient autonomous
  ODEs on a sphere,
 J. Stat. Mechanics: Theor. Exp. {\bf 2016}, 124003 (2016).

\bibitem{DouShiZel04}
M.~R. Douglas, B.~Shiffman, and S.~Zelditch,
 Critical Points and supersymmetric vacua, I,
 Commun. Math. Phys. {\bf 252}, 325 (2004).

\bibitem{DouShiZel06}
M.~R. Douglas, B.~Shiffman, and S.~Zelditch,
 Critical Points and supersymmetric vacua, III: String/M models,
 Commun. Math. Phys. {\bf 265}, 617 (2006).

\bibitem{EasGutMas16}
R.~Easther, A.~H. Guth, and A.~Masoumi,
 Counting Vacua in Random Landscapes,
 preprint hep-th arXiv:1612.05224 (2016).

\bibitem{Fyo05}
Y.~V. Fyodorov,
 Counting stationary points of random landscapes as a random matrix
  problem,
 Acta Phys. Pol. B {\bf 36}(9), 2699--2707 (2005).

\bibitem{LeDMulWie08}
P.~Le~Doussal, M.~M\"uller, and K.~J. Wiese,
 Cusps and shocks in the renormalized potential of glassy random
  manifolds: How functional renormalization group and replica symmetry breaking
  fit together,
 Phys. Rev. B {\bf 77}, 064203 (2008).

\bibitem{BalBouMez96}
L.~Balents, J.-P. Bouchaud, and M.~Mezard,
 The large scale energy landscape of randomly pinned objects,
 J. Phys. I (France) {\bf 6}, 1007 (1996).

\bibitem{MezPar91}
M.~M\'ezard and G.~Parisi,
 Replica field theory for random manifolds,
 J. Phys. I (France) {\bf 1}, 809 (1991).

\bibitem{Fis86}
D.~S. Fisher,
 Interface Fluctuations in Disordered Systems:
  $5\ensuremath{-}\ensuremath{\epsilon}$ Expansion and Failure of Dimensional
  Reduction,
 Phys. Rev. Lett. {\bf 56}, 1964--1967 (1986).

\bibitem{LeD09AnnPhys}
P.~Le~Doussal,
 Exact results and open questions in first principle functional RG,
 Ann. Phys. {\bf 325}, 49 (2009).

\bibitem{LeDWie09b}
P.~Le~Doussal and K.~J. Wiese,
 Size distributions of shocks and static avalanches from the
  functional renormalization group,
 Phys. Rev. E {\bf 79}, 051106 (2009).

\bibitem{Fis85}
D.~S. Fisher,
 Sliding charge-density waves as a dynamic critical phenomenon,
 Phys. Rev. B {\bf 31}, 1396--1427 (1985).

\bibitem{RosKra02}
A.~Rosso and W.~Krauth,
 Roughness at the depinning threshold for a long-range elastic string,
 Phys. Rev. E {\bf 65}, 025101 (2002).

\bibitem{KolBusFerRos13}
A.~B. Kolton, S.~Bustingorry, E.~E. Ferrero, and A.~Rosso,
 Uniqueness of the thermodynamic limit for driven disordered elastic
  interfaces,
 J. Stat. Mech.: Theor. Exp. {\bf 2013}, P12004 (2013).

\bibitem{DemLecRos14}
V.~D\'emery, V.~Lecomte, and A.~Rosso,
 The effect of disorder geometry on the critical force in disordered
  elastic systems,
 J. Stat. Mech.: Theor. Exp. {\bf 2014}, P03009 (2014).

\bibitem{DemRosPon14}
V.~D\'emery, A.~Rosso, and L.~Ponson,
 From microstructural features to effective toughness in disordered
  brittle solids,
 Europhys. Lett. {\bf 105}(3), 34003 (2014).

\bibitem{LeDWie09a}
P.~Le~Doussal and K.~J. Wiese,
 Driven particle in a random landscape: Disorder correlator, avalanche
  distribution, and extreme value statistics of records,
 Phys. Rev. E {\bf 79}, 051105 (2009).

\bibitem{FedLeDWie06}
A.~A. Fedorenko, P.~Le~Doussal, and K.~J. Wiese,
 Universal distribution of threshold forces at the depinning
  transition,
 Phys. Rev. E {\bf 74}, 041110 (2006).

\bibitem{RosLeDWie09}
A.~Rosso, P.~Le~Doussal, and K.~J. Wiese,
 Avalanche-size distribution at the depinning transition: A numerical
  test of the theory,
 Phys. Rev. B {\bf 80}, 144204 (2009).

\bibitem{LeDWieCha02}
P.~Le~Doussal, K.~J. Wiese, and P.~Chauve,
 Two-loop functional renormalization group theory of the depinning
  transition,
 Phys. Rev. B {\bf 66}, 174201 (2002).

\bibitem{HubPraPumWil18}
G.~Huber, M.~Pradas, A.~Pumir, and M.~Wilkinson,
 Persistent stability of a chaotic system,
 Physica A {\bf 492}, 517--523 (2018).

\bibitem{EfeLar77}
K.~B. Efetov and A.~I. Larkin,
 Charge-density wave in a random potential,
 Sov. Phys. JETP {\bf 45}(6), 1236--1241 (1977).

\bibitem{ParSou79}
G.~Parisi and N.~Sourlas,
 Random Magnetic Fields, Supersymmetry, and Negative Dimensions,
 Phys. Rev. Lett. {\bf 43}, 744--745 (1979).

\bibitem{GiaLeD94}
T.~Giamarchi and P.~Le~Doussal,
 Elastic theory of pinned flux lattices,
 Phys. Rev. Lett. {\bf 72}, 1530--1533 (1994).

\bibitem{Fis85b}
D.~S. Fisher,
 Random fields, random anisotropies, nonlinear \ensuremath{\sigma}
  models, and dimensional reduction,
 Phys. Rev. B {\bf 31}, 7233--7251 (1985).

\bibitem{Fel00}
D.~E. Feldman,
 Quasi-long-range order in the random anisotropy Heisenberg model:
  Functional renormalization group in $4\ensuremath{-}\ensuremath{\epsilon}$
  dimensions,
 Phys. Rev. B {\bf 61}, 382--390 (2000).

\bibitem{Fel01}
D.~E. Feldman,
 Quasi-long range order in glass states of impure liquid crystals,
  magnets, and superconductors,
 Int. J. Mod. Phys. B {\bf 15}(22), 2945--2976 (2001).

\bibitem{WieLeD07}
K.~J. Wiese and P.~L. Doussal,
 Functional Renormalization for Disordered Systems, Basic Recipes and
  Gourmet Dishes,
 Markov Processes Relat. Fields {\bf 13}, 777--818 (2007),
 preprint arXiv:cond-mat/0611346.

\bibitem{Mid92}
A.~A. Middleton,
 Asymptotic uniqueness of the sliding state for charge-density waves,
 Phys. Rev. Lett. {\bf 68}(5), 670 (1992).

\bibitem{BaeMac98}
C.~Baesens and R.~S. MacKay,
 Gradient dynamics of tilted Frenkel-Kontorova models,
 Nonlinearity {\bf 11}, 949 (1998).

\bibitem{Hir85}
M.~W. Hirsch,
 Systems of Differential Equations that are Competitive or Cooperative
  II: Convergence Almost Everywhere,
 SIAM J. Math. Anal. {\bf 16}(3), 423 (1985).

\bibitem{gragra}
I.~S. Gradshteyn and I.~M. Ryzhik,
 {\em Table of integrals, series and products},
 Academic Press, fifth edition, 1994.

\bibitem{Tex10}
C.~Texier,
 $\zeta$-regularised spectral determinants for metric graphs,
 J.~Phys.~A: Math. Theor. {\bf 43}, 425203 (2010).

\bibitem{HerJon71}
D.~C. Herbert and R.~Jones,
 Localized states in disordered systems,
 J.~Phys.~C: Solid St. Phys. {\bf 4}(10), 1145 (1971).

\bibitem{Tho72}
D.~J. Thouless,
 A relation between the density of states and range of localization
  for one-dimensional random systems,
 J.~Phys.~C: Solid St. Phys. {\bf 5}, 77 (1972).

\bibitem{Luc92}
J.-M. Luck,
 {\em Syst\`emes d\'esordonn\'es unidimensionnels},
 CEA, collection Al\'ea Saclay, Saclay, 1992.

\bibitem{GraTexTou14}
A.~Grabsch, C.~Texier, and Y.~Tourigny,
 One-dimensional disordered quantum mechanics and Sinai diffusion with
  random absorbers,
 J. Stat. Phys. {\bf 155}(2), 237--276 (2014).

\bibitem{GelYag60}
I.~M. {Gel'fand} and A.~M. Yaglom,
 Integration in functional spaces and its applications in quantum
  physics,
 J. Math. Phys. {\bf 1}(1), 48--69 (1960).

\bibitem{LevSmi77}
S.~Levit and U.~Smilansky,
 A theorem on infinite products of eigenvalues of Sturm-Liouville type
  operators,
 Proc. Am. Math. Soc. {\bf 65}, 299 (1977).

\bibitem{Kol10}
I.~V. Kolokolov,
 Statistical geometry of chaotic two-dimensional transport,
 JETP Lett. {\bf 92}, 107 (2010).

\bibitem{ComLucTexTou13}
A.~Comtet, J.-M. Luck, C.~Texier, and Y.~Tourigny,
 The Lyapunov exponent of products of random $2\times2$ matrices close
  to the identity,
 J. Stat. Phys. {\bf 150}(1), 13--65 (2013).

\bibitem{ComTexTou13}
A.~Comtet, C.~Texier, and Y.~Tourigny,
 Lyapunov exponents, one-dimensional Anderson localisation and
  products of random matrices,
 J.~Phys.~A: Math. Theor. {\bf 46}, 254003 (2013),
 Special issue \og Lyapunov analysis: from dynamical systems theory to
  applications \fg{}.

\bibitem{Hal65}
B.~I. Halperin,
 Green's Functions for a Particle in a One-Dimensional Random
  Potential,
 Phys. Rev. {\bf 139}(1A), A104--A117 (1965).

\bibitem{LifGrePas88}
I.~M. Lifshits, S.~A. Gredeskul, and L.~A. Pastur,
 {\em Introduction to the theory of disordered systems},
 John Wiley \& Sons, 1988.

\bibitem{PalVul87}
G.~Paladin and A.~Vulpiani,
 Anomalous scaling in multifractal objects,
 Phys. Rep. {\bf 156}(4), 147--225 (1987).

\bibitem{PalVul87b}
G.~Paladin and A.~Vulpiani,
 Anomalous scaling and generalized Lyapunov exponents of the
  one-dimensional Anderson model,
 Phys. Rev. B {\bf 35}, 2015--2020 (1987).

\bibitem{CriPalVul93}
A.~Crisanti, G.~Paladin, and A.~Vulpiani,
 {\em Products of random matrices in statistical physics},
 Springer-Verlag, 1993,
 Springer Series in Solid-State Sciences vol.~{\bf104}.

\bibitem{AntPasSly81}
T.~N. Antsygina, L.~A. Pastur, and V.~A. Slyusarev,
 Localization of states and kinetic properties of one-dimensional
  disordered systems,
 Sov. J. Low Temp. Phys. {\bf 7}(1), 1--21 (1981).

\bibitem{SchTit02}
H.~Schomerus and M.~Titov,
 Statistics of finite-time Lyapunov exponents in a random
  time-dependent potential,
 Phys. Rev. E {\bf 66}, 066207 (2002).

\bibitem{CohRotSha88}
A.~Cohen, Y.~Roth, and B.~Shapiro,
 Universal distributions and scaling in disordered systems,
 Phys. Rev. B {\bf 38}(17), 12125--12132 (1988).

\bibitem{RamTex14}
K.~Ramola and C.~Texier,
 Fluctuations of random matrix products and 1D Dirac equation with
  random mass,
 J. Stat. Phys. {\bf 157}(3), 497--514 (2014).

\bibitem{BouGeoHanLeDMai86}
J.-P. Bouchaud, A.~Georges, D.~Hansel, P.~{Le Doussal}, and J.-M. Maillard,
 Rigorous bounds and the replica method for products of random
  matrices,
 J. Phys. A: Math. Gen. {\bf 19}, L1145--L1152 (1986).

\bibitem{BouGeoLeD86}
J.-P. Bouchaud, A.~Georges, and P.~{Le Doussal},
 Fluctuations of the Lyapunov exponent and intermittency in dynamical
  and disordered systems: the example of 1D localization,
 in {\em Proceedings of the Meeting on Dynamical Systems}, Rome,
  Italy, 1986,
 Preprint LPTENS 86/34.

\bibitem{Pen94}
J.~B. Pendry,
 Symmetry and transport of waves in one-dimensional disordered
  systems,
 Adv. Phys. {\bf 43}(4), 461--542 (1994).

\bibitem{Tex00}
C.~Texier,
 Individual energy level distributions for one-dimensional diagonal
  and off-diagonal disorder,
 J.~Phys.~A: Math. Gen. {\bf 33}, 6095--6128 (2000).

\bibitem{HagTex08}
C.~Hagendorf and C.~Texier,
 Breaking supersymmetry in a one-dimensional random Hamiltonian,
 J.~Phys.~A: Math. Theor. {\bf 41}, 405302 (2008).

\bibitem{LanLif66c}
L.~D. Landau and E.~Lifchitz,
 {\em M\'ecanique quantique},
 Mir, 1966,
 tome 3.

\bibitem{Gar00}
A.~Garg,
 Tunnel splittings for one dimensional potential wells revisited,
 Am. J. Phys. {\bf 68}, 430 (2000).

\bibitem{Son08}
D.-Y. Song,
 Tunneling and energy splitting in an asymmetric double-well
  potential,
 Ann. Phys. {\bf 323}(12), 2991--2999 (2008).

\bibitem{Tex15book}
C.~Texier,
 {\em M\'ecanique quantique},
 Dunod, Paris, second edition, 2015.

\bibitem{NikOuv83}
A.~Nikiforov and V.~Ouvarov,
 {\em Fonctions sp\'eciales de la physique math\'ematique},
 Mir, Moscou, 1983.

\bibitem{Gar89}
C.~W. Gardiner,
 {\em Handbook of stochastic methods for physics, chemistry and the
  natural sciences},
 Springer, 1989.

\bibitem{ZilPik03}
R.~Zillmer and A.~Pikovsky,
 Multiscaling of noise-induced parametric instability,
 Phys. Rev. E {\bf 67}, 061117 (2003).

\bibitem{MalMar02}
K.~Mallick and P.~Marcq,
 Anomalous diffusion in nonlinear oscillators with multiplicative
  noise,
 Phys. Rev. E {\bf 66}, 041113 (2002).

\bibitem{GraTex15}
A.~Grabsch and C.~Texier,
 Capacitance and charge relaxation resistance of chaotic cavities --
  Joint distribution of two linear statistics in the Laguerre ensemble of
  random matrices,
 Europhys. Lett. {\bf 109}, 50004 (2015).

\bibitem{GraTex16b}
A.~Grabsch and C.~Texier,
 Distribution of spectral linear statistics on random matrices beyond
  the large deviation function -- Wigner time delay in multichannel disordered
  wires,
 J. Phys. A: Math. Theor. {\bf 49}, 465002 (2016).

\bibitem{HarKirTex12}
J.~M. Harrison, K.~Kirsten, and C.~Texier,
 Spectral determinants and Zeta functions of Schr\"odinger operators
  on metric graphs,
 J.~Phys.~A: Math. Theor. {\bf 45}, 125206 (2012).

\bibitem{Des00}
J.~Desbois,
 Spectral determinant of Schr\"odinger operators on graphs,
 J.~Phys.~A: Math. Gen. {\bf 33}, L63 (2000).

\bibitem{ComDesTex05}
A.~Comtet, J.~Desbois, and C.~Texier,
 Functionals of the Brownian motion, localization and metric graphs,
 J.~Phys.~A: Math. Gen. {\bf 38}, R341--R383 (2005).

\bibitem{Dha99}
D.~Dhar,
 The Abelian sandpile and related models,
 Physica A {\bf 263}(1-4), 4--25 (1999),
 Proceedings of the 20th IUPAP International Conference on Statistical
  Physics.

\bibitem{Pru12}
G.~Pruessner,
 {\em Self-organised criticality: theory, models and
  characterisation},
 Cambridge University Press, 2012.

\bibitem{LeDWie15}
P.~Le~Doussal and K.~J. Wiese,
 Exact Mapping of the Stochastic Field Theory for Manna Sandpiles to
  Interfaces in Random Media,
 Phys. Rev. Lett. {\bf 114}, 110601 (2015).

\bibitem{Ris89}
H.~Risken,
 {\em The Fokker-Planck Equation: Methods of Solution and
  Applications},
 Springer, Berlin, 1989.

\bibitem{Jun96}
G.~Junker,
 {\em Supersymmetric methods in quantum and statistical physics},
 Springer, 1996.

\bibitem{SayTikRod18}
D.~R. Saykin, K.~S. Tikhonov, and Y.~I. Rodionov,
 Landau levels with magnetic tunneling in a Weyl semimetal and
  magnetoconductance of a ballistic $p\text{\ensuremath{-}}n$ junction,
 Phys. Rev. B {\bf 97}, 041202 (2018).

\bibitem{TexHag10}
C.~Texier and C.~Hagendorf,
 Effect of boundaries on the spectrum of a one-dimensional random mass
  Dirac Hamiltonian,
 J.~Phys.~A: Math. Theor. {\bf 43}, 025002 (2010).

\end{thebibliography}
%\end{document}

\end{document}